\documentclass[
reprint,
superscriptaddress,
amsmath,amssymb,
aps,
floatfix
]{revtex4-2}

\usepackage{booktabs}
\usepackage{graphicx}
\usepackage{dcolumn}
\usepackage{bm}

\usepackage{hyperref}
\usepackage[dvipsnames]{xcolor}
\hypersetup{
    colorlinks=true,
    linkcolor=Blue,
    citecolor=Blue,   
   urlcolor=Blue
   }

\def\bea{\begin{eqnarray}}
\def\eea{\end{eqnarray}}
\def\beq{\begin{equation}}
\def\eeq{\end{equation}}
\def\f{\frac}

\def\s{\sigma}

\def\la{\langle}
\def\ra{\rangle}
\def\nn{\nonumber}

\def\d{\delta}
\def\p{\partial}

\def\Pe{\mathrm{Pe}}

\def\uv{ {\hat{\mathbf{u}}}}
\def\rv{ {\bf r}}
\def\vv{ {\bf v}}

\def\g{\gamma}

\def\l{\mathrm{Pe}}

\begin{document}

\title{Controlling inertial active Brownian motion via stochastic resetting}

\author{Manish Patel}%
\email[contact author:~]{manish.patel@iopb.res.in}
\affiliation{Institute of Physics, Sachivalaya Marg, Bhubaneswar, Odisha 751005, India}
\affiliation{Homi Bhabha National Institute, Anushakti Nagar, Mumbai, Maharashtra 400094, India}

\author{Amir Shee}
\email[contact author:~]{amir.shee@uvm.edu}
\affiliation{Department of Physics, University of Vermont, Burlington, Vermont 05405, USA}

\date{\today}

\begin{abstract}
Inertia is intrinsic to many living and synthetic active systems, from
animals and robotic agents to colloidal swimmers, and it strongly shapes
transport.
Many such systems employ intermittent restart protocols to regulate
exploration.
Stochastic resetting provides a theoretical framework for these strategies
and a route to control nonequilibrium steady states, yet the role of inertia
in reset-controlled active dynamics remains poorly understood. Here we study an inertial active Brownian particle in two dimensions under complete and partial stochastic resetting. We derive exact stationary second- and fourth-order scalar moments for the different protocols wherever their stationary limits exist, and obtain the full transient evolution of the mean-squared velocity and displacement under complete resetting. Under complete resetting, inertia strongly enhances localization: at large reset rates, the steady-state mean-squared displacement decays as $r^{-3}$, rather than
following the overdamped scaling $r^{-1}$.
Kurtosis-based phase diagrams reveal strongly non-Gaussian position and
velocity statistics, including activity-induced re-entrant behavior as a
function of reset rate and inertia.
The transient solutions show that resetting shifts the velocity-relaxation
rates, whereas the terminal relaxation of the displacement is governed
solely by the mean resetting time $1/r$, with inertia controlling the
faster transient modes.
Comparing complete resetting with the partial protocols, we find that
resetting the velocity together with the position substantially enhances
localization.
Resetting the orientation to a fixed or a random direction drives
the mean-squared velocity toward opposite limits.
These results establish inertia, reset frequency, and the choice of reset
variables as coupled control parameters for localization, relaxation, and
non-Gaussian fluctuations in active systems.
\end{abstract}

\maketitle

\section{Introduction}

Efficient navigation is a fundamental challenge for active agents, both living and synthetic, operating in noisy and structured environments~\cite{Romanczuk2012, Marchetti2013, Bechinger2016, Basu2024}. From foraging organisms to microrobots, effective motion requires balancing extensive exploration with reliable return, while operating under finite time and energetic constraints~\cite{Viswanathan2011, Volpe2017, Palagi2018}. A powerful and widely observed strategy to regulate this balance is to intermittently interrupt motion and reset key state variables, thereby limiting unproductive excursions and maintaining control over spatial exploration. This restart mechanism, formalized as stochastic resetting, provides a general framework for controlling search dynamics and has emerged as a unifying principle across physical, chemical, and biological systems~\cite{Evans2011PRL, Evans2011JPA, Reuveni2014SubstrateUnbinding, Rotbart2015, Roldan2016, Pal2017, Robin2018SingleMolecule, Bressloff2020}.

Stochastic resetting provides a theoretical framework for restart protocols~\cite{Evans2011PRL, Evans2011JPA}, in which the dynamics is interrupted at random times and returned to a prescribed state, thereby preventing indefinite wandering and producing well-defined nonequilibrium steady states~\cite{Evans2020, Gupta2019, Gupta2022}. 
Resetting can reduce first-passage times, enhance search efficiency, and tune steady-state fluctuations in both passive and active systems~\cite{ Evans2018, Santra2020, Bressloff2020PRE, Kumar2020, Boyer2024, Ghosh2026,Debasish2025, Andrey2025}. 
For active particles, however, most existing work has focused on overdamped dynamics, where inertia is neglected and the velocity is effectively slaved to the propulsion direction~\cite{ Kumar2020,Baouche2024, PSPal2024, Baouche2025}.  The overdamped limit may not appropriate for describing many experimentally relevant settings, including low-damping colloids, vibrated granular particles, self-propelled robots, and animal locomotion, where inertia plays an essential dynamical role~\cite{ Koumakis2016, Scholz2018, Deblais2018, Dombrowski2019, Dauchot2019, Loewen2020Inertial, Sandoval2020, Nguyen2022InertialAOUP, Altshuler2024, Paramanick2024, Olsen2025Optimal, Sinha2025,Antonov2024}. 
Finite inertia introduces a momentum-relaxation timescale that delays the response of the velocity to changes in orientation and can qualitatively reshape transport and fluctuation statistics~\cite{GutierrezMartinez2020, Caprini2021, Martins2022, Patel2023, Patel2024, Patel2025, Lisin2025, Smith_2022,Santra_2025}. 
Despite this, steady-state statistics of inertial active particles under stochastic resetting, including the emergence of non-Gaussian fluctuations that control exploration and localization, remain largely unexplored.

Here, we systematically analyze complete resetting and all six partial resetting protocols obtained by resetting a nonempty proper subset of the position, velocity, and orientation of an inertial active Brownian particle. For protocols that reset orientation, we consider both fixed and uniformly random reset directions.
 We additionally analyze the passive limit to clearly delineate the role of activity and its interplay with inertia under stochastic resetting.
We focus on translational inertia, assuming that rotational inertia relaxes rapidly and can be neglected, although it may play an important role in active colloids and related systems~\cite{Palacci2013, Caprini2022, Lisin_26}. Using a moment-generating framework~\cite{Hermans1952, Shee2020, Chaudhuri2021, Patel2023, Pattanayak2024, Patel2024} combined with the Final-Value Theorem, we derive exact stationary second- and fourth-order scalar moments across the resetting protocols wherever stationary limits exist, as functions of inertia $M$, Péclet number $\Pe$ (activity), and reset rate $r$. For complete resetting, we also obtain the full transient MSV and MSD. We use the excess kurtosis as our primary metric to systematically characterize the roles of inertia and reset rate as control parameters for tuning the nonequilibrium steady state.

The steady-state mean-squared velocity (MSV) $\langle \vv^2 \rangle^{\rm st}$ decreases with increasing inertia, scaling as $M^{-1}$ in the absence of resetting~\cite{Patel2023}. With complete stochastic resetting, the MSV exhibits the same scaling at low inertia but is more strongly suppressed at large inertia, where $\langle \vv^2 \rangle^{\rm st}_{r} \sim M^{-2}$. As a function of the reset rate, the MSV remains approximately constant at low $r$ and decays at large $r$ as $\langle \vv^2 \rangle^{\rm st}_{r} \sim r^{-1}$.  
The steady-state mean-squared displacement (MSD) $\langle \rv^2 \rangle^{\rm st}_{r}$ is nearly independent of inertia at low $M$ but is strongly suppressed at high inertia, scaling as $\langle \rv^2 \rangle^{\rm st}_{r} \sim M^{-2}$. In the overdamped limit, the MSD decreases with the reset rate as $\langle \rv^2 \rangle^{\rm st}_{r} \sim r^{-1}$~\cite{Shee2025}, a scaling that persists at low reset rates for inertial active Brownian particles. In contrast, at large reset rates the suppression is much stronger, with $\langle \rv^2 \rangle^{\rm st}_{r} \sim r^{-3}$. Notably, all these asymptotic exponents are activity-independent: passive particles exhibit the same asymptotic behavior, differing only in quantitative amplitudes.

The steady-state second-order moments exhibit clear qualitative and quantitative differences in the presence of inertia, reflecting a pronounced suppression of steady-state fluctuations. To further quantify the emergence of heavy-tailed statistics, we use the excess kurtosis as a primary metric and construct phase diagrams in the $(M,r)$ plane. 
 The phase diagram of the velocity excess kurtosis shows re-entrant behavior as a function of inertia and reset rate, which is absent for the passive case. Importantly, the excess kurtosis in position space shows re-entrant behavior as a function of reset rate, which is absent for the passive case.
The excess kurtosis thus reveals broad regimes of strong non-Gaussianity and re-entrant behavior arising from the competition between persistence, momentum relaxation, and reset frequency.
 The re-entrant behavior of the position excess kurtosis reveals regimes where displacements of both large and small amplitudes are more probable than in the passive case.
Together, these results show that inertia is a key control variable in reset-controlled active systems: it sharpens typical localization at large reset rates while amplifying rare excursions, thereby enriching steady-state fluctuation statistics and producing tunable non-Gaussian steady states.
The analytical structure uncovered here provides a compact baseline for experiments in colloidal, granular, and robotic platforms and opens pathways toward reset-informed navigation and sampling protocols in autonomous active agents~\cite{Paramanick2024, Altshuler2024, Olsen2025Optimal, Sinha2025}.

The paper is organized as follows. In Sec.~\ref{sec:model} we introduce the inertial active Brownian particle model, define complete resetting as the reference protocol, specify the extension to partial resetting, and define the dimensionless control parameters $(M,\Pe,r)$. In Sec.~\ref{sec:second_moments} we present exact steady-state second moments of velocity and displacement and summarize their limiting behavior. In Sec.~\ref{sec:phase_diagrams} we present kurtosis-based phase diagrams in the $(M,r)$ plane and relate them to the underlying steady-state distributions, highlighting inertial re-entrant behavior and heavy-tailed regimes. In Sec.~\ref{sec:dynamics} we give closed-form expressions for the transient approach of both second moments to the steady state under complete resetting and identify the relaxation time scales.
In Sec.~\ref{sec:partial} we compare the complete protocol with the six partial
resetting protocols obtained from the state variables $(\rv,\vv,\uv)$, and give the steady-state mean-squared velocity of all seven protocols, and the mean-squared displacement of the four position-resetting protocols, in closed form. Section~\ref{sec:conclusions} concludes with a discussion of implications and outlook.

%====== Figure 1 ========
\begin{figure*}[!t]
\begin{center}
\includegraphics[width=\linewidth]{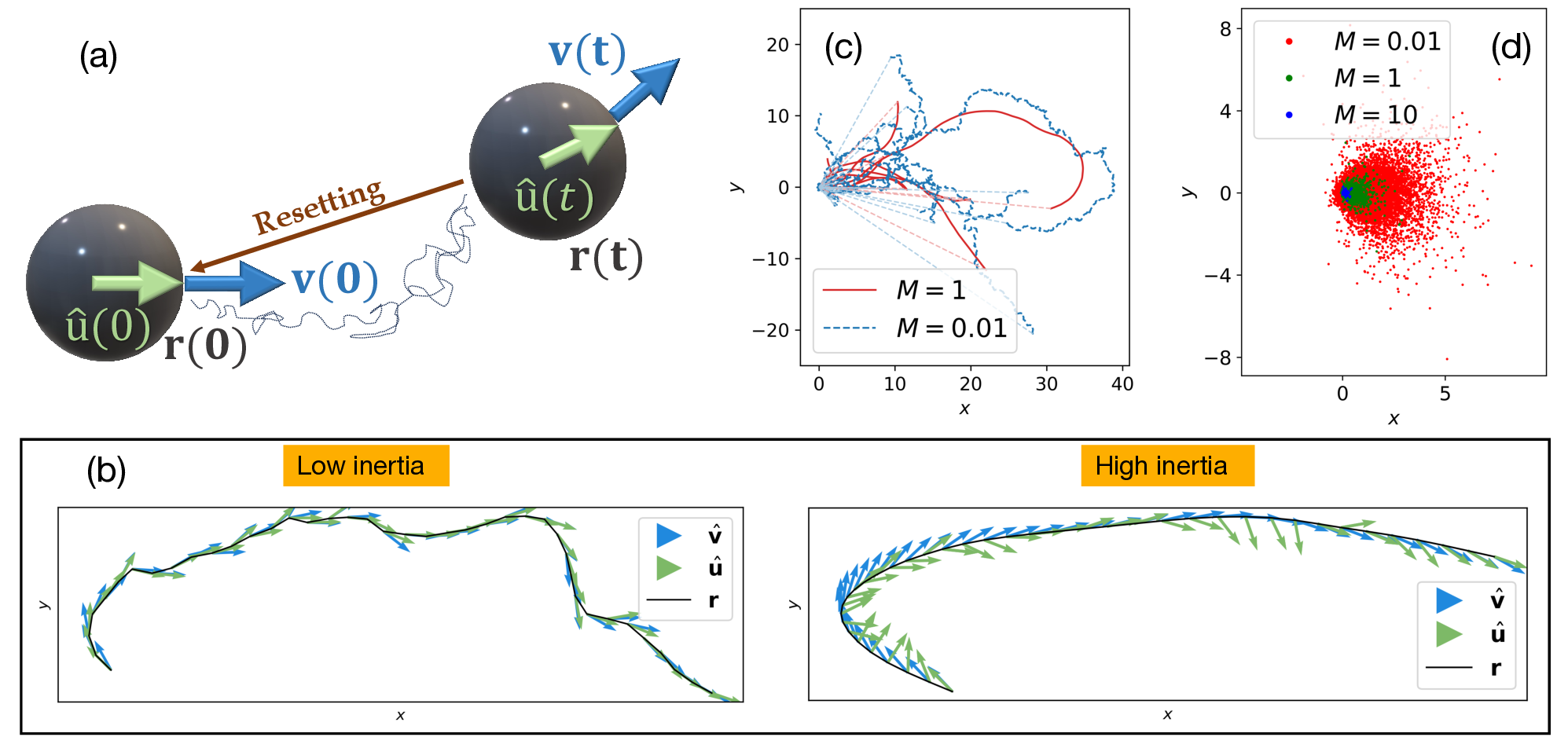} 
\caption{
Inertial effects in active Brownian particles under complete stochastic resetting. (a) Schematic diagram of the inertial active Brownian particle position $\mathbf{r}$, velocity $\mathbf{v}$, and orientation $\hat{\mathbf{u}}$ evolve(see Eqs.~\eqref{eom:disp}--\eqref{eom:resetting}), with intermittent complete resetting to the initial state $(\mathbf{r}(0),\mathbf{v}(0),\hat{\mathbf{u}}(0))\equiv (\mathbf{r}_0,\mathbf{v}_0,\hat{\mathbf{u}}_0)$.
(b) In the overdamped regime (low inertia), the particle exhibits an instantaneous orientational response: velocity direction $\hat{\vv}$ is immediately slaved to the propulsion direction $\uv$. In the underdamped regime (high inertia), the particle shows a delayed response, where velocity direction $\hat{\mathbf{v}}$ relaxes over a finite time due to inertia.
(c) Comparison of steady state single particle trajectory for inertia $M=0.01$ and $1$ at reset rate $r = 1$ and activity $\Pe = 10$.
(d) Steady state positions snapshot of $2 \times 10^4$ particles for inertia $M=0.01$, $1$, and $10$ at reset rate $r = 10$ and activity $\Pe = 10$.
We consider the initial(reset) state $\rv_{0}= 0$, $\vv_{0} = 0$, and $\uv_{0} = (1,0)$.
} 
\label{fig1}
\end{center}
\end{figure*}
%====== End of Figure 1 ========

\section{Model}
\label{sec:model}

We study an inertial active Brownian particle in two dimensions
under complete and partial stochastic resetting of its position,
velocity, and orientation. We first define complete resetting,
which serves as the reference protocol; the full family of
partial protocols is analyzed in Sec.~\ref{sec:partial}. It is characterized by its position $\mathbf{r} = (x, y)$, velocity $\vv=(v_x,v_y)$, and its orientation (or heading direction) $\uv = (u_x, u_y)$, which is a unit vector in two dimensional plane where $u_x=\cos(\theta)$ and $u_y=\sin(\theta)$, evolving over time $t$ in the presence of thermal noise at temperature $T$ satisfying the fluctuation-dissipation relation. We define the translational diffusion coefficient as $D \equiv k_B T/\g$. For the reference protocol, we impose complete stochastic resetting on $\rv$, $\vv$, and $\uv$, restoring them to $(\rv_0,\vv_0,\uv_0)$ at Poisson rate $r_0$.  Between reset events, the particle self-propels along $\uv$ while its velocity relaxes on a finite momentum–relaxation time (Fig.~\ref{fig1}(a)). 
This protocol establishes a nonequilibrium steady state for $r_0>0$ in which excursions are repeatedly “restarted,” enabling direct control of localization and excursions through $r_0$.

The position $\rv$, velocity $\vv$, and orientation angle $\theta$ evolution in $2$-dimensions can be written as 
\bea
&& d\rv = \vv \, dt\,, \label{eom:disp} \\
&& m \,d\vv = -\gamma (\vv - v_0 \uv)\,dt + \sqrt{2 \gamma k_B T} \,{\mathbf{dB}}^t(t)\,, \label{eom:vel} \\
&& 
d\theta = \sqrt{2 D_r}\, dB^r(t) \label{eom:rot_active}\\
&& (\rv, \vv, \theta) \xrightarrow{r_0} (\rv_0, \vv_0, \theta_0)\,.
\label{eom:resetting}
\eea
The dynamics is illustrated schematically in Fig.~\ref{fig1}(a).
The noise terms ${\mathbf{dB}
}^t$ and $dB^r$ are modeled as Gaussian white noise with zero mean and variances given by $\la dB_i^tdB_j^t\ra =\delta_{ij} dt$ and $\la dB^rdB^r\ra = dt$, respectively.
The dynamics of the inertial active Brownian particle under stochastic resetting is governed by Eqs.~(\ref{eom:disp})-(\ref{eom:resetting}). We assume the origin of the active orientational dynamics to be independent of inertia~\cite{Patel2023, Patel2024}.
Eq.~\eqref{eom:resetting} implements complete resetting of position, velocity, and orientation.
This inertial coupling implies that $\vv$ does not instantaneously align with $\uv$ (Fig.~\ref{fig1}(b)): after a change of heading the velocity responds with a finite delay set by inertia.
Beyond complete resetting, we analyze all six partial protocols obtained by resetting a nonempty proper subset of $(\rv,\vv,\uv)$ while leaving the remaining variables unchanged. Whenever orientation is reset, we consider both restoration to a fixed direction and independent sampling of a uniformly random direction, giving eleven protocol--ensemble combinations. In Sec.~\ref{sec:partial}, we obtain exact steady-state mean-squared velocities for all seven protocols and exact steady-state mean-squared displacements for the four protocols that reset position. The corresponding fourth-order moments and non-Gaussian parameters are analyzed in Appendix~\ref{sec:partial_fourth}. Among the protocols considered, only those that reset position admit a finite stationary mean-squared displacement; the remaining protocols have unbounded displacement moments, although their velocity statistics approach stationary limits.

We analyze the model in dimensionless units by measuring time in units of the rotational persistence time $\tau_r=1/D_r$ and length in units of $\ell=\sqrt{D/D_r}$. We present all results in dimensionless form by rescaling position as 
$\rv \to \rv/\ell$, 
time as $t \to t/\tau_r$, 
and velocity as $\vv \to \vv/\sqrt{DD_r}$. Three dimensionless parameters fully control the dynamics: (i) the \emph{inertia} $M=\tau/\tau_r$ with $\tau=m/\gamma$ the momentum–relaxation time; (ii) the \emph{activity} (Péclet number) $\Pe=v_0/\sqrt{DD_r}$, which sets the strength of self-propulsion; and (iii) the \emph{reset rate} $r=r_0\tau_r$, which sets the average duration of each excursion.
In this reduced description, “overdamped” motion corresponds to $M\to 0$ (instantaneous velocity response), while $M \gg 1$ represents the strongly inertial regime with pronounced velocity–orientation lag(Fig.~\ref{fig1}(b)). 
This delay directly impacts single--particle trajectories under resetting (Fig.~\ref{fig1}(c)). At fixed $r=1$ and $\Pe=10$, weakly inertial particles ($M=0.01$) closely track the propulsion direction, whereas increasing inertia ($M=1$) induces relatively smooth (less fluctuating; Fig.~\ref{fig1}(c)) trajectories between resets.
Figure~\ref{fig1}(d) illustrates that inertia qualitatively reshapes the steady-state spatial distribution under stochastic resetting: the density becomes increasingly concentrated near the reset position as $M$ increases.

The velocity dynamics of a free inertial ABP~\cite{Patel2023} can be viewed as mathematically equivalent to the position dynamics of an overdamped ABP in a harmonic trap~\cite{Chaudhuri2021}, with momentum relaxation playing the role of trap stiffness(see Appendix~\ref{app:mapping}). As a result, under stochastic resetting the velocity statistics of the inertial particle follow those of a trapped overdamped active particle with the appropriate parameter mapping~\cite{Shee2025}. In contrast, the position of an inertial active particle does not admit such a simple correspondence as it remains coupled to velocity, leading to novel reset-induced behavior.

We use the initial condition $(x,y,v_x,v_y,\theta)=(0,0,0,0,0)$. When selected for resetting, position and velocity are restored to zero. Orientation is restored to $\theta_0=0$ in the fixed-direction ensemble, or independently redrawn uniformly on $[-\pi,\pi]$ in the random-direction ensemble. The stochastic dynamics defined by Eqs.~\eqref{eom:disp}–\eqref{eom:rot_active} are simulated by discretizing the inertial Langevin equations with a time step $\Delta t$. The velocity update follows an underdamped Langevin scheme, while the particle orientation evolves according to rotational diffusion. We employ stochastic Euler-Maruyama for orientation and velocity–Verlet algorithm for position and velocity, which provides superior numerical stability and accuracy for inertial active systems. At each time step, a reset occurs with probability $r\Delta t$. For complete resetting, all three state variables are reset. For partial resetting, only the selected variables are reset, with the orientation reset ensemble specified in Sec.~\ref{sec:partial}. We derive exact steady-state scalar moments up to fourth order using the moment-generating framework~\cite{Hermans1952, Shee2020, Chaudhuri2021, Patel2023}. Appendices~\ref{app:moments_generator_framework}--\ref{app:fourth_moments} develop the complete-resetting results; Sec.~\ref{sec:partial} and Appendix~\ref{sec:partial_fourth} treat the partial protocols. In the main text, we analyze these results to characterize the steady-state fluctuation statistics.

%====== Figure 2 ========
\begin{figure*}[!t]
\begin{center}
\includegraphics[width=\linewidth]{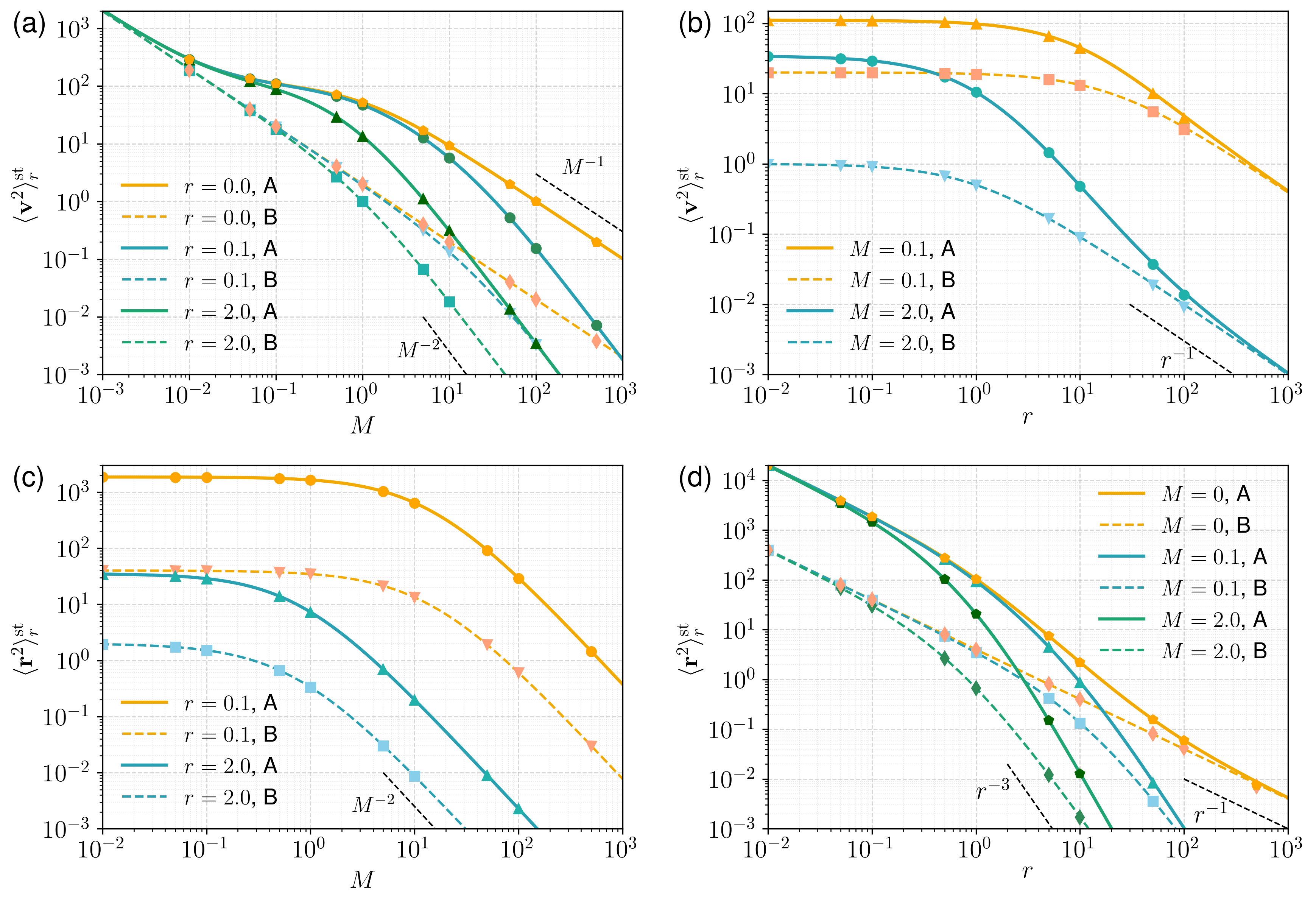} 
\caption{Steady-state second moments under complete stochastic resetting. Solid lines labelled A are active particles at $\Pe=10$; dashed lines labelled B are passive(i.e., underdamped Brownian) particles at $\Pe=0$. Symbols are simulations. (a,b) Mean-squared velocity. Lines are Eq.~\eqref{eq:MSV}. (a) Versus $M$ at $r=0$, $0.1$, and $2.0$; black dashed guides are $M^{-1}$ for $r=0$ and $M^{-2}$ at finite reset rate. (b) Versus $r$ at $M=0.1$ and $2.0$; the black dashed guide is $r^{-1}$. (c,d) Mean-squared displacement. Lines are Eq.~\eqref{eq:MSD}. (c) Versus $M$ at $r=0.1$ and $2.0$; the black dashed guide is $M^{-2}$. (d) Versus $r$ at $M=0$ (overdamped), $0.1$, and $2.0$; black dashed guides are $r^{-1}$ for $M=0$ and $r^{-3}$ at finite $M$.}
\label{fig2}
\end{center}
\end{figure*}
%====== End of Figure 2 ========

% === Table 1 ===
\begin{table*}[!t]
\centering
\caption{Limiting cases and asymptotic scaling of the steady-state mean-squared displacement (MSD) 
$\langle \rv^2\rangle_{r}^{\rm st}$ and mean-squared velocity (MSV) 
$\langle \vv^2\rangle_{r}^{\rm st}$ for different limits of inertia $M$ and reset rate $r$. “NO SS”: no steady state; “NA”: not defined in the overdamped limit.
Scaling statements refer to asymptotic limits at fixed remaining parameters.}
\label{tab:MSD_MSV_scaling}
\setlength{\tabcolsep}{6pt}
\begin{tabular}{@{} l l l @{}}
\toprule
\textbf{Limit} & \textbf{MSD} $\langle \rv^2\rangle_{r}^{\rm st}$ & \textbf{MSV} $\langle \vv^2\rangle_{r}^{\rm st}$ \\
\midrule
No reset ($r=0$)~\cite{Patel2023}
& NO SS
& $\displaystyle \frac{2}{M} + \frac{\Pe^2}{1+M}$ \\

Overdamped Brownian ($M=0$, $\Pe=0$)~\cite{Shee2025}
& $\displaystyle \frac{4}{r}$
& NA \\

Overdamped ABP ($M=0$)~\cite{Shee2025}
& $\displaystyle \frac{4}{r} + \frac{2\Pe^2}{r(1+r)}$
& NA \\

Underdamped Brownian ($\Pe=0$)
& $\displaystyle \frac{8}{r(1+Mr)(2+Mr)}$
& $\displaystyle \frac{4}{M(2+Mr)}$ \\

Our model: low reset rate
& $\sim r^{-1}$ \, (as in overdamped)
& $\displaystyle \la \vv^2\ra_{r=0}^{\rm st}$~\cite{Patel2023} \\

Our model: high reset rate
& $\sim r^{-3}$ \, (as in passive Brownian)
& $\sim r^{-1}$ \, (as in passive Brownian) \\

Our model: low inertia
& $\displaystyle \la\rv^2\ra_{r}^{\rm st}(M=0)$~\cite{Shee2025}
& $\sim M^{-1}$ \, (as in $r=0$ limit) \\

Our model: high inertia
& $\sim M^{-2}$ \, (as in passive Brownian)
& $\sim M^{-2}$ \, (as in passive Brownian) \\
\bottomrule
\end{tabular}
\end{table*}
% === End of Table 1 ===

\section{Second order moments}
\label{sec:second_moments}

\noindent
\textbf{Mean-squared velocity (MSV).~}The steady-state mean-squared velocity (MSV) of an inertial active Brownian particle under complete stochastic resetting is given by (see Appendix~\ref{app:second_moments} for derivation)
\bea
\la \vv^2 \ra_{r}^{\rm st} &=& \frac{2}{2 + M r} \left( \frac{2}{M} + \frac{\l^2}{1 + M(1+r)} \right)\,.
\label{eq:MSV}
\eea 
This expression is formally identical to the steady-state MSD of an overdamped ABP under resetting~\cite{Shee2025} via the mapping   discussed in Appendix~\ref{app:mapping}.
In the absence of resetting ($r=0$), one recovers the known result for free inertial ABPs~\cite{Patel2023},
$\la \vv^2 \ra^{\rm st} = 2/M + \Pe^2/(1+M)$, which reduces to the Brownian limit $\la \vv^2 \ra^{\rm st}=2/M$ for $\Pe=0$~\cite{Uhlenbeck1930} (Fig.~\ref{fig2}(a,b)).
For weak resetting, the MSV is nearly independent of $r$ for both active and passive particles(Fig.~\ref{fig2}(b)).  
In the passive limit ($\Pe=0$), Eq.~\eqref{eq:MSV} reduces to
\bea 
\la \vv^2 \ra_{r}^{\rm st} = \f{4}{M(2+Mr)}\,,
\eea
showing that stochastic resetting steepens the large-$M$ scaling from $M^{-1}$ (free inertial Brownian motion~\cite{Uhlenbeck1930}) to $M^{-2}$, reflecting enhanced suppression of velocity fluctuations (Fig.~\ref{fig2}(a)).  
At large reset rates, the MSV decays universally as $\la \vv^2 \ra_{r}^{\rm st}\sim r^{-1}$, independent of activity (Fig.~\ref{fig2}(b)). In the passive limit ($\Pe=0$), $\tau_r$ acts only as a reference time scale for nondimensionalization; all dimensional observables are independent of $\tau_r$. The dimensional MSV in passive limit is $\la \tilde\vv^2 \ra_r^{\rm st} = (\ell/\tau_r)^2 \la \vv^2 \ra_{r}^{\rm st} = 4 D/[\tau(2 + \tau r_0)]$.

\medskip

\noindent
\textbf{Mean-squared displacement (MSD).~}
The steady-state mean-squared displacement (MSD) of an inertial active Brownian particle under complete stochastic resetting is (see Appendix~\ref{app:second_moments} for derivation)
\bea
\la \rv^2 \ra_{r}^{\rm st} &=& \frac{2}{r(1+ Mr)(2 + Mr)}\nonumber\\
&&\times \left[ 4 + \frac{\Pe^2(2 + 2 M + 3 M r)} {(r+1)(1 + M + Mr)} \right]\,.
\label{eq:MSD}
\eea
In the passive limit ($\Pe=0$), Eq.~\eqref{eq:MSD} reduces to the exact result for underdamped Brownian motion under resetting,
\bea
\la \rv^2 \ra_{r}^{\rm st} &=& \frac{8}{r(1+ Mr)(2 + Mr)}\,.
\label{eq:MSD_Brownian}
\eea
For overdamped Brownian motion, the steady-state MSD decays as $\langle \rv^2 \rangle_{r}^{\rm st} = 4/r$~\cite{Shee2025}. The presence of inertia suppresses the MSD for all $r>0$ and leads to a markedly faster asymptotic decay at large reset rates, $\langle \rv^2 \rangle_{r}^{\rm st} \sim 8/(M^2 r^3)$ as $r\to\infty$ (Fig.~\ref{fig2}(c,d)).  
Thus, inertia enhances localization under resetting while simultaneously limiting long excursions and large-scale exploration.
The asymptotic scaling of the second moments in the relevant limiting regimes is summarized in Table~\ref{tab:MSD_MSV_scaling}.
To construct the phase diagrams, we compute the steady-state fourth-order moments(see Appendix Fig.~\ref{app_fig:fourth}) and quantify deviations from Gaussian statistics using the excess kurtosis. 
To connect these phase diagrams to the underlying probability distributions, we compare the measured distributions with the corresponding Gaussian distributions constructed from the exact second-order moments.

% === FIGURE 3 ===
\begin{figure*}[!t]
\begin{center}
\includegraphics[width=\linewidth]{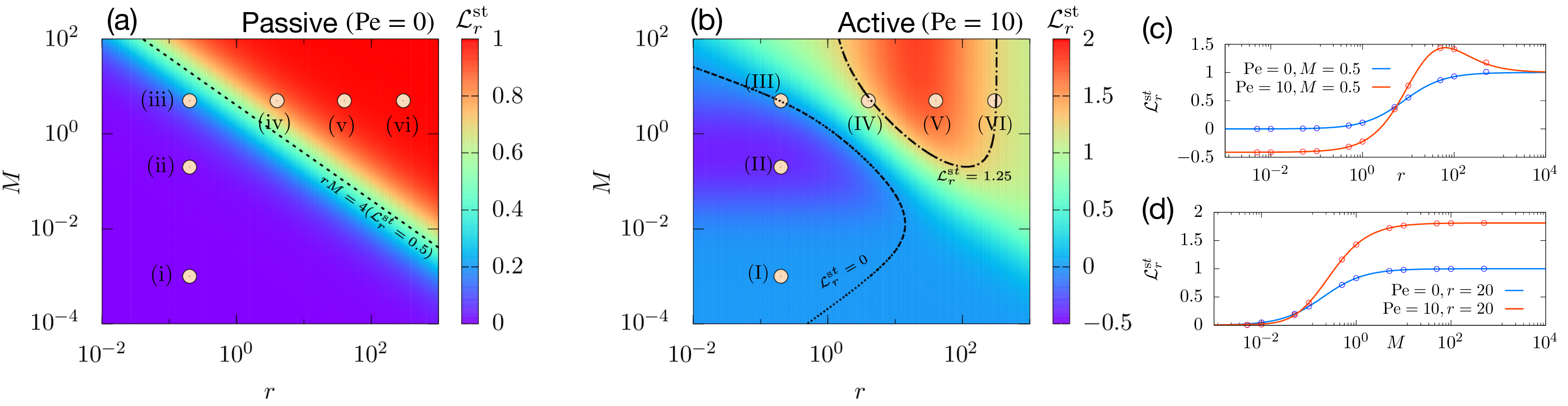} 
\includegraphics[width=1\linewidth]{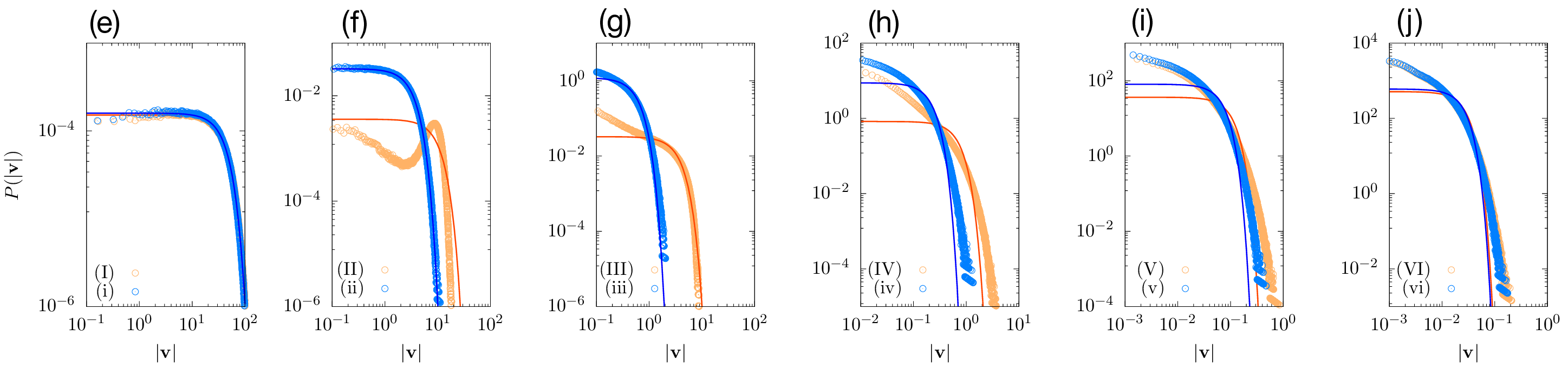}
\caption{Steady-state velocity excess kurtosis ${\cal L}_r^{\rm st}$. Color in (a) and (b) is the analytic result of Eq.~\eqref{eq:excess_kurtosis}, from the steady-state MSV, Eq.~\eqref{eq:MSV}, and fourth moment, Eq.~\eqref{eq:v4avg_ss}. (a) Passive case, $\Pe=0$. The black dashed line is $Mr=4$, where ${\cal L}_r^{\rm st}=1/2$. (b) Active case, $\Pe=10$. Black dashed contours are ${\cal L}_r^{\rm st}=0$ and $1.25$. Both panels are in the $r$--$M$ plane. Marked points (i)--(vi) and (I)--(VI) are the parameter values used in (e)--(j). (c) ${\cal L}_r^{\rm st}$ versus $r$ at $M=0.5$, for $\Pe=0$ and $\Pe=10$. (d) ${\cal L}_r^{\rm st}$ versus $M$ at $r=20$, for $\Pe=0$ and $\Pe=10$. In (c) and (d), lines are Eq.~\eqref{eq:excess_kurtosis} and symbols are simulations. (e)--(j) Steady-state distributions of $|\vv|$, pairing (i) with (I) in (e) through (vi) with (VI) in (j). Symbols are the simulated distributions; solid lines are the Gaussian reference of Eq.~\eqref{eq:Gaussian}, built from the steady-state second moment.}
\label{fig3}
\end{center}
\end{figure*}
% === End of Figure 3 ===

\section{Phase diagrams}
\label{sec:phase_diagrams}

\begin{figure*}[!t]
\begin{center}
\includegraphics[width=\linewidth]{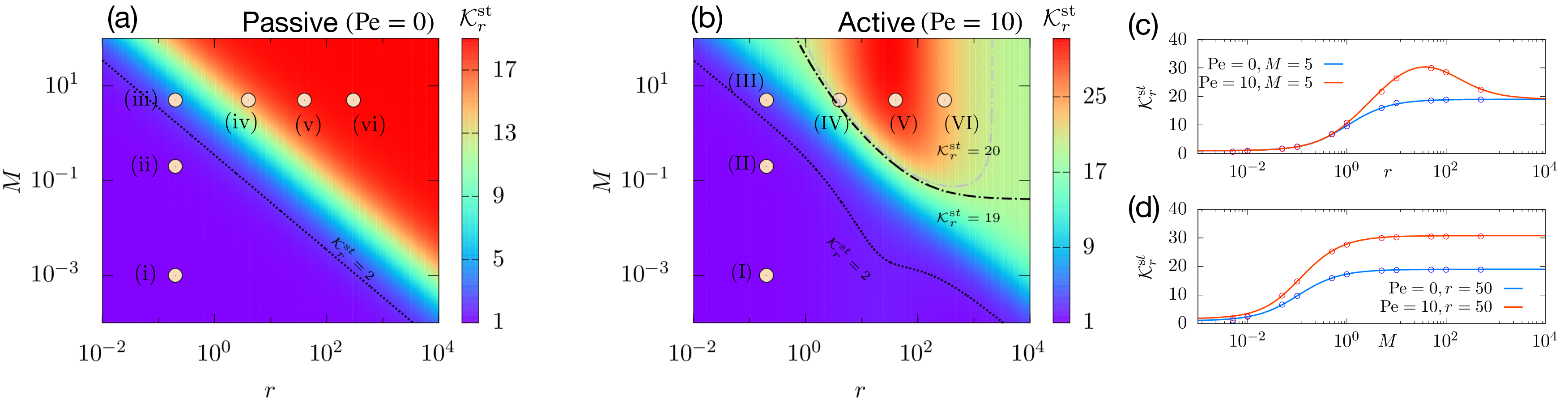} 
\includegraphics[width=\linewidth]{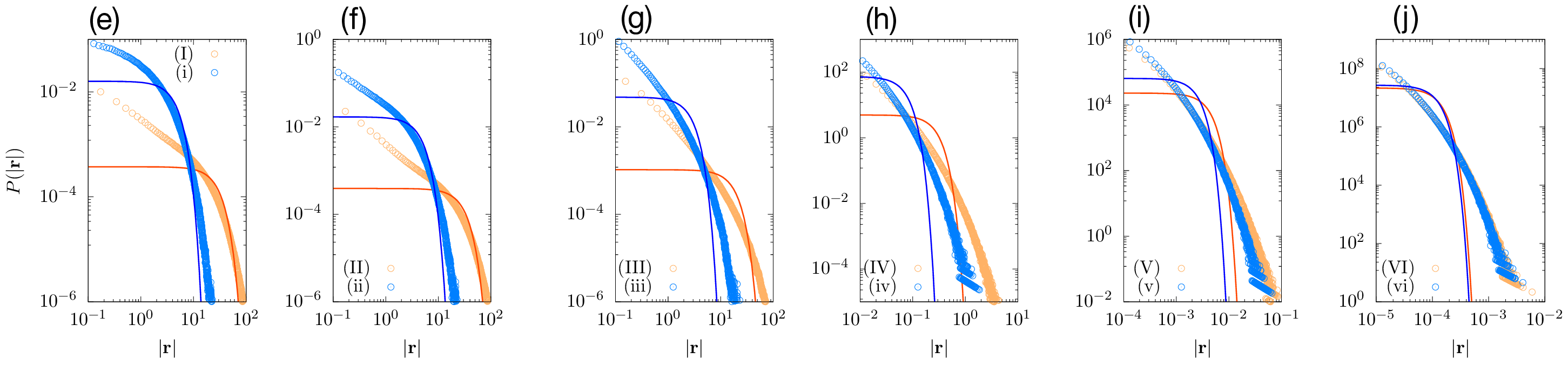}
\caption{Steady-state position excess kurtosis ${\cal K}_r^{\rm st}$. Color in (a) and (b) is the analytic result of Eq.~\eqref{eq:excess_kurtosis}, from the steady-state MSD, Eq.~\eqref{eq:MSD}, and fourth moment, Eq.~\eqref{eq:r4avg_ss}. (a) Passive case, $\Pe=0$. (b) Active case, $\Pe=10$. Both panels are in the $r$--$M$ plane. Black dashed contours are ${\cal K}_r^{\rm st}=2$ in (a), and ${\cal K}_r^{\rm st}=2$, $19$, and $20$ in (b). Marked points (i)--(vi) and (I)--(VI) are the parameter values used in (e)--(j). (c) ${\cal K}_r^{\rm st}$ versus $r$ at $M=5$, for $\Pe=0$ and $\Pe=10$. (d) ${\cal K}_r^{\rm st}$ versus $M$ at $r=50$, for $\Pe=0$ and $\Pe=10$. In (c) and (d), lines are Eq.~\eqref{eq:excess_kurtosis} and symbols are simulations. (e)--(j) Steady-state distributions of $|\rv|$, pairing (i) with (I) in (e) through (vi) with (VI) in (j). Symbols are the simulated distributions; solid lines are the Gaussian reference of Eq.~\eqref{eq:Gaussian}, built from the steady-state second moment.}
\label{fig4}
\end{center}
\end{figure*}

\noindent
\textbf{Excess kurtosis as primary metric.}
We use the steady-state excess kurtosis as the primary metric to construct phase diagrams for both velocity and position statistics. 
For a zero-mean Gaussian process, $\langle \rv \rangle_{r}^{\rm st}=\langle \vv \rangle_{r}^{\rm st}=0$, the fourth moments satisfy 
$(\mu^{r}_{4})_{\rv}^{\rm st}=2(\langle \rv^2 \rangle_{r}^{\rm st})^2$ and $(\mu^{r}_{4})_{\vv}^{\rm st}=2(\langle \vv^2 \rangle_{r}^{\rm st})^2$. 
We therefore define
\bea
\mathcal{K}_{r}^{\rm st}=\frac{\langle \rv^4 \rangle_{r}^{\rm st}}{(\mu^{r}_{4})^{\rm st}_{\rv}} - 1 \quad , \quad \mathcal{L}_{r}^{\rm st}=\frac{\langle \vv^4 \rangle_{r}^{\rm st}}{(\mu^{r}_{4})^{\rm st}_{\vv}} - 1\,,
\label{eq:excess_kurtosis}
\eea
which quantify departures from Gaussian steady-state statistics for position and velocity, respectively. 
We evaluate $\mathcal{K}_{r}^{\rm st}$ and $\mathcal{L}_{r}^{\rm st}$ analytically by combining the exact steady-state second moments (Eqs.~\eqref{eq:MSV} and \eqref{eq:MSD}) with the corresponding steady-state fourth moments (Eqs.~\eqref{eq:v4avg_ss} and \eqref{eq:r4avg_ss}) derived in Appendix~\ref{app:fourth_moments}. The resulting closed-form expressions for the excess kurtosis are algebraically lengthy. Instead, we use them to generate the phase-diagram heat maps and representative cuts, shown in Figs.~\ref{fig3}(a--d) and \ref{fig4}(a--d). In what follows, we summarize the limiting behaviors and highlight the qualitative regimes that emerge.

\medskip

\noindent
\textbf{Velocity excess kurtosis.~} In the zero activity limit, the steady-state velocity excess kurtosis reduces to the underdamped Brownian result
\bea
\lim_{\Pe \to 0}{\cal L}_{r}^{\rm st} = \frac{Mr}{4 + Mr}\,,
\label{eq:vel_excess_kurt_underdamped_Brownian}
\eea
which is strictly positive (Fig.~\ref{fig3}(a)). Equation~\eqref{eq:vel_excess_kurt_underdamped_Brownian} shows that non-Gaussianity is controlled solely by the dimensionless product $Mr$, which compares the momentum relaxation time to the mean resetting time.
 The product $Mr$ recurs throughout: it controls the position
excess kurtosis of Eq.~\eqref{eq:position_excess_kurtosis_underdamped_Brownian} and determines when velocity resetting begins to
dominate the choice of protocol in Sec.~\ref{sec:partial}.
In the absence of resetting ($r=0$), ${\cal L}_{r}^{\rm st}=0$ and the velocity distribution is Gaussian. In the overdamped limit ($M\to 0$), the resetting protocol does not affect velocity statistics, consistent with the rapid equilibration of $\vv$ relative to the other dynamical time scales. Increasing either inertia or reset rate enhances deviations from Gaussianity, with ${\cal L}_{r}^{\rm st}$ rising monotonically and saturating to unity as $Mr\to\infty$ (Fig.~\ref{fig3}(a)). Figure~\ref{fig3}(a) presents the corresponding phase diagram in the $(r,M)$ plane. The dashed line $Mr=4$ marks the contour ${\cal L}_{r}^{\rm st}=1/2$, separating weakly and strongly non-Gaussian regimes. Equation~\eqref{eq:vel_excess_kurt_underdamped_Brownian} agrees with the $\Pe=0$ curves in Fig.~\ref{fig3}(c,d). The passive distributions (i)--(vi) in Fig.~\ref{fig3}(e--j) have heavier tails as $Mr$ increases.

% === FIGURE 4 ===
% === End of Figure 4 ===

%
\noindent
For finite activity ($\Pe = 10$), the phase diagram reveals two re-entrant regimes: one emerging at low reset rates and the other in the strongly inertial limit(Fig.~\ref{fig3}b). In the small resetting limit($r\to 0$), the steady-state velocity excess kurtosis reduces to the known result for free inertial active Brownian particles~\cite{Patel2023}:
\bea
{\label{eq:noreset_kv}}
\lim_{r \to 0} {\cal L}_r^{\rm st} = - \frac{\Pe^4 M^2 (3+7M)}{2(3+M)(1+2M)[2+(2 + \Pe^2)M]^2}\,.
\eea
This expression is strictly negative(Fig.~\ref{fig3}b), consistent with the non-Gaussian velocity statistics of free active Brownian particles discussed in Ref.~\cite{Patel2023}. As a function of inertia $M$, the excess kurtosis $\mathcal{L}_{r\to 0}^{\rm st}(M,\Pe)$ exhibits a non-monotonic, re-entrant dependence.  
In the overdamped limit ($M\to 0$), the excess kurtosis vanishes, reflecting Gaussian velocity statistics.  
With increasing inertia, deviations from Gaussianity are amplified, leading to a finite negative minimum at an activity-dependent value $M=M_\star(\Pe)$. The location of this minimum follows from extremizing Eq.~\eqref{eq:noreset_kv} with respect to $M$, yielding the quartic condition
\bea
7(\Pe^{2}+2)M^{4}+(6\Pe^{2}-2)M^{3}\nonumber\\-98M^{2}-84M-18=0,
\eea
whose unique positive root defines $M_\star(\Pe)$.  
For weak activity, $M_\star(\Pe)\simeq 3.09-0.90\,\Pe^{2}+\mathcal{O}(\Pe^{4})$, whereas for strong activity it decreases asymptotically as $M_\star(\Pe)\sim 3^{1/3}\Pe^{-2/3}$.  
Thus, the inertia at which non-Gaussianity is maximal decreases monotonically with increasing activity.  
For sufficiently large inertia, the excess kurtosis relaxes back towards zero, with the asymptotic behaviour
$\mathcal{L}_{r\to 0}^{\rm st}\sim -c/M+\mathcal{O}(M^{-2})$ as $M\to\infty$, where $c=7\Pe^4/[4(2+\Pe^2)^2]$.  
Consequently, non-Gaussian effects are strongest at intermediate inertia, while both the overdamped and strongly inertial limits recover nearly Gaussian velocity statistics.

\noindent 
In the strongly inertial regime($M\to\infty$), expanding around small reset rate($r\to 0$)
\bea
\lim_{r\to 0}\lim_{M\to\infty}\mathcal{L}_r^{\rm st} &=& 1 + \f{\Pe^2(\Pe^2+16)}{4(\Pe^2+2)^2} r + \mathcal{O}(r^2)\,,\nonumber
\eea
the excess kurtosis rises above unity at small $r$, reaches a single maximum at $r=r_\star(\Pe)$, and then decreases monotonically back to $1$ as $r\to\infty$ with the asymptotic scaling 
\bea
\lim_{r\to \infty}\lim_{M\to\infty}\mathcal{L}_r^{\rm st} &=& 1+\f{\Pe^{2}}{r}+\mathcal{O}(r^{-2})\,.\nonumber
\eea
Physically, optimizing over $r$ balances frequent resets (which suppress heavy tails) against activity-induced drive; the balance shifts to larger $r$ as activity increases, but the excess kurtosis remains bounded, saturating below $2$.
Thus, as a function of resetting rate, the large--inertia excess kurtosis exhibits a re--entrant trend(Fig.~\ref{fig3}(b)).
The analytical prediction (Eq.~\eqref{eq:excess_kurtosis}, lines) is in excellent agreement with simulations (symbols) (Fig.~\ref{fig3}(c,d)), and the distributions of $|\vv|$ confirm re-entrant transitions (Fig.~\ref{fig3}(e--j)). We summarize different limits and their velocity excess kurtosis behavior in Table~\ref{tab:kurtosis_scaling}. 

% === Table 2 ===
\begin{table*}[!t]
\centering
\caption{Limiting behavior and asymptotic scaling of the steady-state excess kurtosis of velocity 
$\mathcal{L}_{r}^{\rm st}$ and position $\mathcal{K}_{r}^{\rm st}$ for different limits of inertia $M$ 
and reset rate $r$. “NO SS”: no steady state. “NA”: undefined in overdamped limit. 
Scaling refers to asymptotic limits at fixed remaining parameters.
 In the high-reset-rate limit both quantities are approached from
above for $\Pe^{2}>4/M$ and from below for $\Pe^{2}<4/M$.
}
\label{tab:kurtosis_scaling}
\setlength{\tabcolsep}{6pt}
\begin{tabular}{@{} l l l @{}}
\toprule
\textbf{Limit} & \textbf{Velocity excess kurtosis} $\boldsymbol{\mathcal{L}_{r}^{\rm st}}$ & 
\textbf{Position excess kurtosis} $\boldsymbol{\mathcal{K}_{r}^{\rm st}}$ \\
\midrule

No reset ($r=0$)~\cite{Patel2023}
& $\mathcal{L}_{r}^{\rm st} \le 0$ (Eq.~\eqref{eq:noreset_kv}) 
& NO SS \\

Overdamped Brownian ($M=0$, $\Pe=0$)~\cite{Shee2025}
& NA 
& $1$ \\

Overdamped ABP ($M=0$)~\cite{Shee2025}
& NA
& $1 \le \mathcal{K}_{r}^{\rm st} < 2$
(Eq.~\eqref{eq:position_excess_kurtosis_overdamped_ABP}) \\

Underdamped Brownian ($\Pe=0$)
& $\displaystyle \frac{Mr}{Mr+4}$
& $1<\mathcal{K}_{r}^{\rm st}<19$ 
(Eq.~\eqref{eq:position_excess_kurtosis_underdamped_Brownian}) \\

Our model: Low reset rate
& $\mathcal{L}_{r}^{\rm st} \leq 0$
& $\mathcal{K}_{r}^{\rm st} \to 1$ \\

Our model: High reset rate
& $\mathcal{L}_{r}^{\rm st} \to 1^{\pm}$ for $\Pe^2 \gtrless 4/M$
& $\mathcal{K}_{r}^{\rm st} \to 19^{\pm}$ for $\Pe^2 \gtrless 4/M$ \\

Our model: Low inertia
& $\mathcal{L}_{r}^{\rm st} \to 0$
& $\mathcal{K}_{r}^{\rm st} \to \mathcal{K}_{r}^{\rm st}(M=0)$ \\

Our model: High inertia
& $1<\mathcal{L}_{r}^{\rm st}<2$; re-entrant in $r$
& finite constant; for $\Pe=0$: $\mathcal{K}_{r}^{\rm st} \to 19$ \\

\bottomrule
\end{tabular}
\end{table*}
% === End of Table 2 ===

\medskip

\noindent
\textbf{Position excess kurtosis.~}In the zero activity limit, the excess kurtosis of position reduces to
\bea
\lim_{\Pe \to 0}  {\cal K}_r^{\rm st} &=& 19 + \frac{54}{3 + M r} - \frac{144}{4 + M r}\,.
\label{eq:position_excess_kurtosis_underdamped_Brownian}
\eea
which is strictly positive for all physical values of $M$ and $r$. In the weak-resetting or low-inertia limit ($Mr \to 0$), the excess kurtosis approaches $\mathcal{K}_{r}^{\rm st} \to 1$, recovering the overdamped passive Brownian result~\cite{Shee2025}(Fig.~\ref{fig4}).
In contrast, for strong resetting or large inertia ($Mr \to \infty$),
it saturates to a finite value $\mathcal{K}_{r}^{\rm st} \to 19$.
This value has a direct physical reading: at large $Mr$ each excursion is
short compared with the momentum-relaxation time, so a particle restarted
from rest rebuilds its velocity diffusively, giving
$\langle \rv^2\rangle \propto t^3$ at fixed excursion time $t$.
Averaging over exponentially distributed $t$ then gives
$\mathcal{K}_{r}^{\rm st} \to \langle t^6\rangle/\langle t^3\rangle^2 - 1
= 6!/[r^6\,(6/r^3)^2] - 1 = 20 - 1 = 19$
(see Appendix~\ref{sec:partial_fourth} for the formal derivation).
Thus, the product $Mr$ controls the position excess kurtosis
$\mathcal{K}_{r}^{\rm st}$, which interpolates smoothly between
$1$ and $19$ (Fig.~\ref{fig4}(a)).

For any finite activity and in the limit of vanishing inertia($M\to 0$), the excess kurtosis reduces to the overdamped ABP under position-orientation stochastic resetting result~\cite{Shee2025}:
\bea
\lim_{M \to 0}  {\cal K}_r^{\rm st} &=& \frac{1 }{(4+r)(2+\Pe^2 + 2r)^2} [2 \Pe^4 (2+r) \nn\\
&& + 4 (1+r)^2 (4+r) + 4 \Pe^2 (4+r)(1+2r)]\,. \nn\\
\label{eq:position_excess_kurtosis_overdamped_ABP}
\eea
We analyze Eq.~\eqref{eq:position_excess_kurtosis_overdamped_ABP} further on the $\Pe-r$ plane. For $\Pe \to 0$, one finds $\lim_{\Pe \to 0} \mathcal{K}_r^{\rm st} \simeq 1$, recovering the overdamped Brownian particle under stochastic resetting~\cite{Shee2025}. In contrast, for $\Pe \to \infty$, the limiting value is $\lim_{\Pe\to\infty} \mathcal{K}_r^{\rm st} \to 2(2+r)/(4+r)$ which increases monotonically from $1$ as $r \to 0^+$ to $2$ as $r \to \infty$.
In the overdamped active regime the position excess kurtosis is bounded within $1 \leq \mathcal{K}_r^{\rm st} < 2$.

At finite activity and large inertia, the position excess kurtosis exhibits a pronounced re-entrant dependence on the reset rate (Fig.~\ref{fig4}(b)). 
The analytical predictions (Eq.~\eqref{eq:excess_kurtosis}; lines) are in excellent agreement with simulations (symbols) (Fig.~\ref{fig4}(c,d)), and the corresponding distributions of $|\rv|$ explicitly reveal the emergence of heavy-tailed fluctuations (Fig.~\ref{fig4}(e--j)). 
The limiting behaviors across the different regimes are summarized in Table~\ref{tab:kurtosis_scaling}.

% === Non-monotonicity in $r$ at large inertia. ===
The expansion of excess kurtosis of position at small reset rate leads to
\bea
\lim_{r \to 0} {\cal K}_r^{\rm {st}} &=& 1 + c_1(M,\Pe) r + {\cal {O}}(r^2)\,,\nonumber\\
c_1(M,\Pe) &=& 3 M + \frac{\Pe^4 (1-3M) + 8 \Pe^2 (2+M)}{4 (2 + \Pe^2)^2 (1+M)}\,,\nonumber\\
\eea
showing that deviations from the overdamped Brownian baseline ${\cal K}_r^{\rm st}=1$ enter linearly in the reset rate.  
The coefficient $c_1(M,\Pe)$ quantifies the joint influence of inertia and activity. The purely inertial contribution $3M$ is strictly positive, implying that inertia enhances non-Gaussianity in the position statistics by introducing short-time ballistic displacements within each excursion. The activity-dependent correction vanishes as ${\cal O}(\Pe^2)$ for $\Pe\ll1$, recovering the passive limit ${\cal K}_r^{\rm st}=1+3Mr+{\cal O}(r^2)$. For finite activity it may either enhance or partially offset the inertial contribution, but the full coefficient remains positive in the explored parameter range, so the excess kurtosis initially increases above unity as the reset rate is increased from very small values.
At large reset rate, the position excess kurtosis reads
\bea
\lim_{r \to \infty}{\cal K}_r^{\rm {st}} = 19 + \frac{45}{2} \left( \Pe^2 - \frac{4}{M} \right) \frac{1}{r} + \mathcal{O}(r^{-2})\,,
\label{eq:position_excess_kurtosis_large_r}
\eea
which shows that in the strong-resetting regime the position excess kurtosis saturates to the universal underdamped-resetting value $19$, with a leading correction $\mathcal{O}(1/r)$ whose sign depends on the competition between activity and inertia. In particular, for $\Pe^2>4/M$ the kurtosis approaches $19$ from above, whereas for $\Pe^2<4/M$ it approaches from below.
The crossover condition $\Pe^2 M > 4$ has a natural physical interpretation:
in dimensional units it is equivalent to $mv_0^2/2 > 2k_BT$, i.e.,
the self-propulsion kinetic energy exceeds the full 2D thermal kinetic
energy, marking the boundary between thermally- and
activity-dominated velocity fluctuations.
Together, these two asymptotic limits at small and large reset rates explain the emergence of non-monotonic (re-entrant) dependence on the reset rate at sufficiently large inertia and finite activity (Fig.~\ref{fig4}(b,c)).
The phase diagrams in the $r-\Pe$ and $M-\Pe$ planes reveal a crossover from weakly non-Gaussian behavior at small reset rates or low inertia to strongly non-Gaussian, heavy-tailed regimes at large reset rates and large inertia, with non-monotonic dependence on the reset rate emerging only at sufficiently large activity (Appendix Fig.~\ref{app_fig:kurtosis_rpe_mpe}).

Non-monotonicity persists under several partial protocols
(Appendix~\ref{sec:partial_fourth}). $\{\rv,\vv\}$, $\{\vv,\uv\}$ and
$\{\vv\}$ inherit $\mathcal{L}_r^{\rm st}$ of complete resetting, and
$\{\rv,\vv\}$ also inherits $\mathcal{K}_r^{\rm st}$. Without velocity
resetting, $\mathcal{L}_r^{\rm st}$ is non-monotonic in $r$ only for
fixed-orientation resetting, and $\mathcal{K}_r^{\rm st}$ only under
$\{\rv,\uv\}$.

\section{Transient dynamics towards steady state}
\label{sec:dynamics}

\begin{figure*}
    \centering
    \includegraphics[width=1\linewidth]{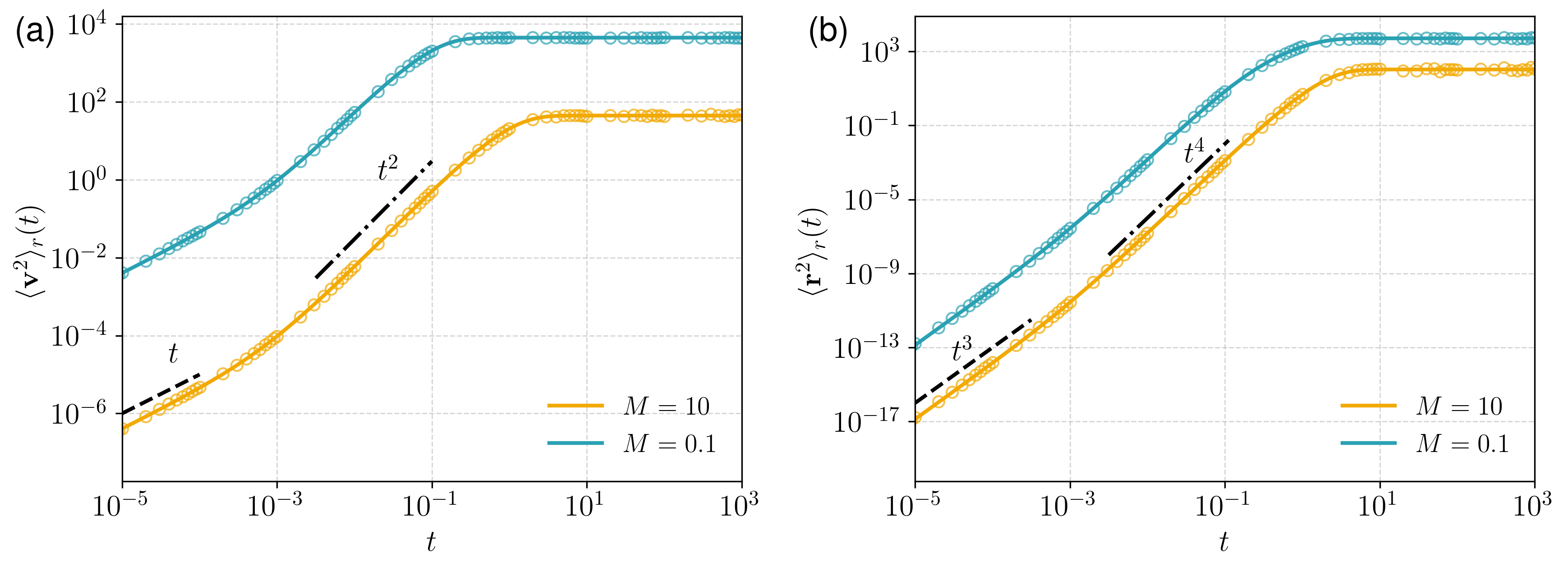}
\caption{Time evolution of the MSV $\la \vv^2 \ra_r(t)$ in (a) and the MSD $\la \rv^2 \ra_r(t)$ in (b) toward the steady state, for $M=0.1$ and $M=10$ at $r=1$ and $\Pe=75$. Symbols are simulation data. Lines are the analytic results of Eq.~\eqref{eq:msv_dynamics} in (a) and Eq.~\eqref{eq:msd_dynamics} in (b). Black dashed guides mark the short-time expansion: $t$ and $t^{2}$ in (a), where the earliest growth is $\la \vv^2 \ra_r \simeq 4t/M^{2}$, and $t^{3}$ and $t^{4}$ in (b), where the earliest growth is $\la \rv^2 \ra_r \simeq 4t^{3}/(3M^{2})$. The leading term is independent of activity and reset rate; activity and resetting enter at the next order and produce the steeper guide. The plateaus are the steady-state MSV, Eq.~\eqref{eq:MSV}, and MSD, Eq.~\eqref{eq:MSD}.}
\label{fig:dynamics}
\end{figure*}

The renewal structure of Eq.~\eqref{eq:moment_resetting} determines not only the steady state but also the full relaxation towards it. Applying the renewal equation to the exact time-dependent second moments of the reset-free dynamics~\cite{Patel2023}, or equivalently inverting the Laplace-space relation $\la \psi \ra_{r}(s) = (1 + r/s)\la \psi \ra_{s+r}$, we obtain closed-form expressions for the approach of the MSV and the MSD to their steady-state values under complete stochastic resetting.
The time evolution of the MSV reads
\bea
\label{eq:msv_dynamics}
\la \vv^2 \ra_{r} (t) &=& \frac{4+2M(2+2r+\Pe^2)}{M(2+Mr)(1+M+Mr)} + e^{-(r+2/M)t} \times \nn\\
&&\left[ \frac{4-2M(2 + \Pe^2)}{M(M-1)(2+Mr)} \right] + e^{-(r+1+1/M)t} \times \nn\\
&&\left[\frac{2\Pe^2}{(M-1)(1+M+Mr)} \right]\,.
\eea
The constant term reproduces the steady-state MSV of Eq.~\eqref{eq:MSV}, and the two decaying modes carry rates $r+2/M$ and $r+1+1/M$. Resetting thus shifts every relaxation rate of the free inertial ABP uniformly by $r$, so the MSV relaxes on a time scale $\tau_v = [r + \min(2/M,\, 1+1/M)]^{-1}$(for $\Pe=0$ the second mode has vanishing amplitude and
$\tau_v = (r+2/M)^{-1}$ for all $M$).
At large inertia and weak resetting this time grows as $\tau_v \simeq M/2$, so inertia delays velocity equilibration. The series expansion around $t = 0$ gives
\bea
\la \vv^2 \ra_r = \frac{4}{M^2}t + \frac{\left\{ \left( \Pe^2 - 2r \right) M -4 \right\}}{M^3} t^2 + {\cal O}(t^3)\,.
\eea
The leading linear growth $4t/M^2$ reflects thermal velocity diffusion from the reset state $\vv_0=\mathbf{0}$ and is independent of both activity and resetting. These enter only at ${\cal O}(t^2)$ through the combination $\Pe^2 - 2r$, which makes their competition explicit: activity accelerates the early growth of velocity fluctuations, whereas resetting suppresses it.
The MSV $\la \vv^2 \ra_r$ therefore crosses over from noise-dominated to propulsion-dominated growth at
$t_{I} = 4M/[(\Pe^{2}-2r)M-4]$, provided $\Pe^{2} > 2r + 4/M$; below this threshold resetting and damping suppress the early growth and no crossover
occurs. For $\Pe^{2}M \gg 4$ the crossover time saturates at $t_{I} \to 4/\Pe^{2}$, independent of inertia. Figure~\ref{fig:dynamics}(a) illustrates both features: $t_I = 7.16\times 10^{-4}$ at $M = 0.1$ and $t_I = 7.11\times 10^{-4}$ at $M = 10$, for $r = 1$ and $\Pe = 75$, in agreement with the asymptotic value $4/\Pe^{2} = 7.11 \times 10^{-4}$.

The time evolution of the MSD reads
\bea
\label{eq:msd_dynamics}
&&\la \rv^2 \ra_r (t) =\nonumber\\
&& \frac{2}{r(1+ Mr)(2 + Mr)}\times \left[ 4 + \frac{\Pe^2(2 + 2 M + 3 M r)} {(r+1)(1 + M + Mr)} \right]\nn\\
&& - \frac{2(2+\Pe^2)}{r} e^{-rt} - \frac{2 \Pe^2}{(1+r)(M-1)}e^{-(1+r)t} \nn\\
&& + \frac{2M\Pe^2}{(M-1)(1+M+rM)}e^{-(1+r+1/M)t} \nn\\
&&+ \frac{2M[-4 -\Pe^2 + 2M(2+ \Pe^2)]}{(M-1)(1+rM)} e^{-(r+1/M)t} \nn\\
&& - \frac{2M(-2+M(2+\Pe^2))}{(M-1)(2+rM)}e^{-(r+2/M)t}\,,
\eea
with the constant term equal to the steady-state MSD of Eq.~\eqref{eq:MSD}.
At $M=1$ the momentum and orientational relaxation rates coincide and the
amplitudes in Eqs.~\eqref{eq:msv_dynamics} and \eqref{eq:msd_dynamics} become
singular. The limit is finite: the degenerate modes combine into terms of
the form $t\,e^{-(r+2)t}$, and both moments remain smooth functions of $M$.

At $t=0$ the exponential amplitudes cancel the constant exactly, consistent with the reset condition $\rv_0=\mathbf{0}$.
In contrast to the MSV, the slowest mode of the MSD decays at exactly the bare reset rate $r$, with the $M$-independent amplitude $-2(2+\Pe^{2})/r$. The late-time approach to the steady state is therefore $e^{-rt}$ for every inertia: resetting alone fixes the terminal relaxation rate, and the position statistics equilibrate on the mean reset time $1/r$ once the faster orientational and momentum modes have decayed. Inertia enters only through those faster modes, at rates $r+1+1/M$, $r+1/M$ and $r+2/M$, and hence sets the shape of the transient rather than its terminal rate. The same bound applies to the MSV, whose growth to $\tau_v \simeq M/2$ is confined to the weak-resetting regime $Mr \lesssim 1$.
The series expansion around $t= 0$ gives
\bea
\la \rv^2 \ra_r = \frac{4}{3M^2}t^3 + \frac{\left\{ \left( \Pe^2 -4r \right)M - 4 \right\}}{4 M^3}t^4 + {\cal {O}}(t^5)\,.
\eea
The cubic short-time growth follows from the diffusive build-up of velocity from rest, $\la \vv^2 \ra_r \simeq 4t/M^2$, integrated twice in time. As for the velocity dynamics, activity and resetting appear only in the subleading term, now through the combination $\Pe^2-4r$.
The MSD $\la \rv^2 \ra_r$ crosses over from $t^{3}$ to $t^{4}$ at $t_{I} = 16M/[3\{(\Pe^{2}-4r)M-4\}]$, which requires $\Pe^{2} > 4r + 4/M$; this generalizes the activity-dominance
condition $\Pe^{2}M>4$ of Sec.~\ref{sec:phase_diagrams} by the additional
resetting-induced decorrelation $4r$, and saturates at $16/(3\Pe^{2})$ for $\Pe^{2}M \gg 4$.
Figure~\ref{fig:dynamics}(b) gives $t_I = 9.56\times 10^{-4}$ at $M = 0.1$
and $t_I = 9.49 \times 10^{-4}$ at $M = 10$, against the asymptotic value
$9.48 \times 10^{-4}$.

Figure~\ref{fig:dynamics} compares the analytical predictions of
Eqs.~\eqref{eq:msv_dynamics} and \eqref{eq:msd_dynamics} with simulations
at $r=1$ and $\Pe=75$ for $M=0.1$ and $M=10$, showing excellent agreement over the full time range.
Increasing inertia lowers the steady-state plateaus and strongly reduces the short-time amplitudes, which scale as $M^{-2}$ in both moments. Only $M=10$ satisfies $Mr \gtrsim 1$ at this reset rate, so the asymptotic $M^{-2}$ suppression of Sec.~\ref{sec:second_moments} applies at that value alone; $M=0.1$ still lies in the low-inertia regime. Velocity relaxation is visibly slower at $M=10$, whereas the MSD reaches its
plateau at $t\sim 1/r$ for both inertia values, as expected from the mode structure of Eq.~\eqref{eq:msd_dynamics}.

\section{Partial resetting protocols}
\label{sec:partial}

\begin{figure*}
\centering
\includegraphics[width=\linewidth]{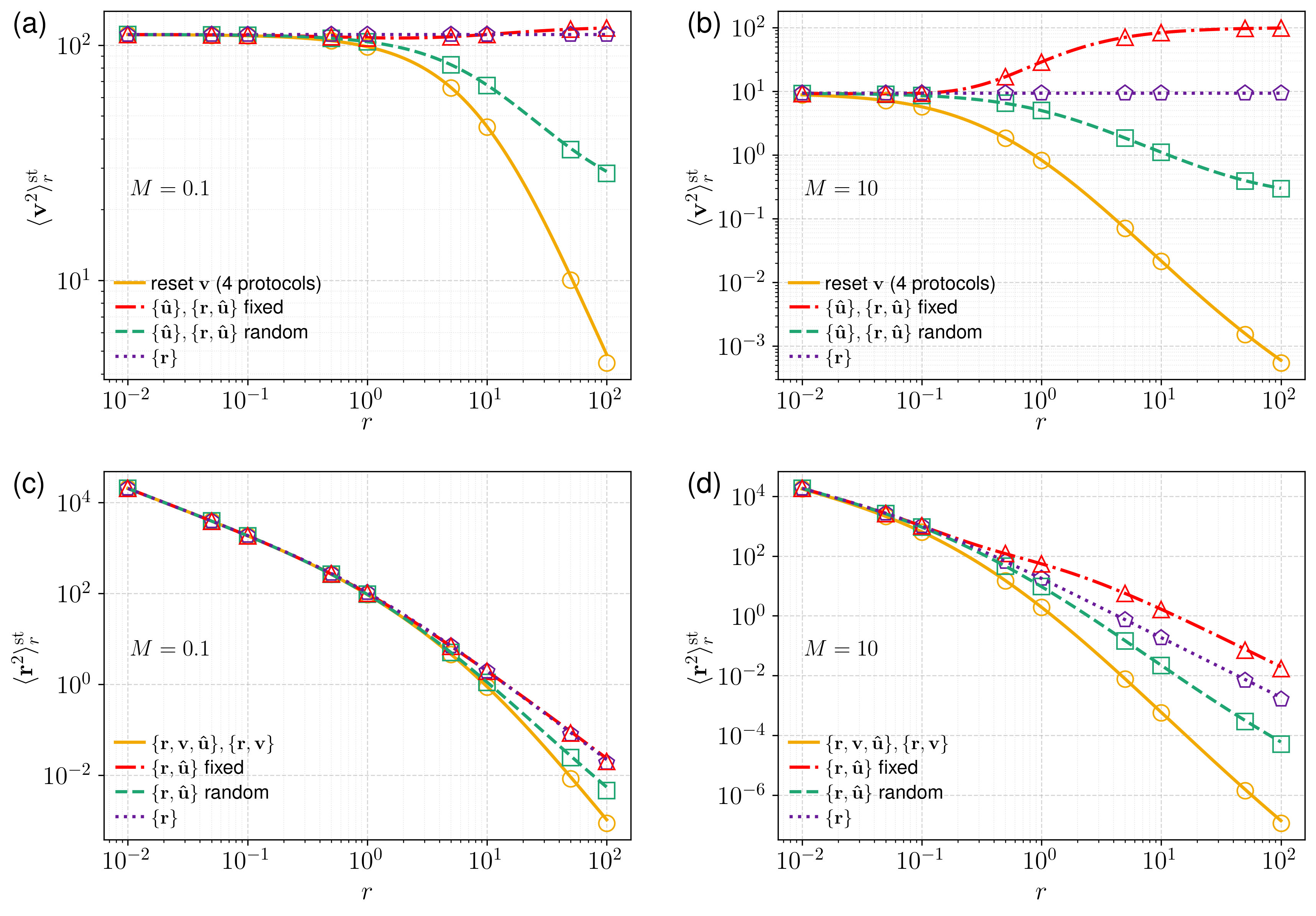}
\caption{Comparison of resetting protocols at fixed activity $\Pe=10$, for the two
orientation reset ensembles: fixed, with $\uv$ restored to $\uv_0=(1,0)$, and random,
with $\uv_0$ drawn uniformly on $[-\pi,\pi]$ at each reset event.
(a,b) Steady-state MSV $\la \vv^2 \ra_r^{\rm st}$ as a function of reset rate $r$ for all
seven protocols at (a) low inertia $M=0.1$ and (b) high inertia $M=10$.
(c,d) Steady-state MSD $\la \rv^2 \ra_r^{\rm st}$ for the four position-resetting
protocols at (c) $M=0.1$ and (d) $M=10$.
Symbols denote simulation data. In (a,b), solid lines give the complete-resetting result
Eq.~\eqref{eq:MSV}, dash-dotted lines the fixed-orientation result
Eq.~\eqref{eq:msv_fixed_u} for $\{\rv,\uv\}$ and $\{\uv\}$, dashed lines the
random-orientation result Eq.~\eqref{eq:msv_random_u} for the same two protocols, and
dotted lines the free-particle value $\la \vv^2 \ra^{\rm st}=2/M+\Pe^2/(1+M)$ retained
under $\{\rv\}$.
In (c,d), solid lines give Eq.~\eqref{eq:MSD}, dash-dotted lines the fixed-orientation result
Eq.~\eqref{eq:msd_ru_fixed} for $\{\rv,\uv\}$, dashed lines the
random-orientation result Eq.~\eqref{eq:msd_ru_random} for the same protocol of $\{\rv,\uv\}$ and dotted lines the
position-only result Eq.~\eqref{eq:msd_pos_only}.
The four velocity-resetting protocols collapse onto Eq.~\eqref{eq:MSV} in (a,b), and
$\{\rv,\vv,\uv\}$ and $\{\rv,\vv\}$ onto Eq.~\eqref{eq:MSD} in (c,d): resetting $\vv$ to
$\mathbf{0}$ removes every $\uv_0$-dependent source term, so these curves are common to
both ensembles. Under fixed-orientation resetting the MSV of $\{\rv,\uv\}$ and $\{\uv\}$
rises with $r$ towards $2/M+\Pe^2$, since each reset restores a preferred heading,
whereas under random-orientation resetting it decays towards the passive value $2/M$.
The branches separate only once $Mr \gtrsim 1$, reached at $r \gtrsim 10$ in (a,c) but at
modest reset rates in (b,d). Under the position-preserving protocols $\{\vv,\uv\}$,
$\{\vv\}$ and $\{\uv\}$ the MSD grows without bound and is omitted from (c,d).}
\label{fig:partial_reset}
\end{figure*}

% === Table: protocol summary ===
\begin{table}[!b]
\centering
\caption{Steady-state second moments across the seven resetting protocols and
the two orientation reset ensembles. ``NO SS'': the MSD grows without bound.
The four velocity-resetting protocols are insensitive to the reset ensemble,
since $\vv_0=\mathbf{0}$ removes every $\uv_0$-dependent source term.
The ensemble matters only for $\{\uv\}$ and $\{\rv,\uv\}$, whose $r\to\infty$
MSV limits differ by the entire activity contribution $\Pe^{2}$, and whose
MSD differs at large inertia according to whether the reset direction
sustains a mean velocity.
}
\label{tab:protocol_msv}
\setlength{\tabcolsep}{5pt}
\begin{tabular}{@{} l l l l @{}}
\toprule
\textbf{Protocol} & \textbf{Reset $\uv$ to} &
\textbf{MSV} $\boldsymbol{\la \vv^2\ra_{r}^{\rm st}}$ &
\textbf{MSD} $\boldsymbol{\la \rv^2\ra_{r}^{\rm st}}$ \\
\midrule
$\{\rv,\vv,\uv\}$, $\{\rv,\vv\}$ & fixed/random
& Eq.~\eqref{eq:MSV} & Eq.~\eqref{eq:MSD} \\[2pt]
$\{\vv,\uv\}$, $\{\vv\}$ & fixed/random
& Eq.~\eqref{eq:MSV} & NO SS \\[4pt]
$\{\rv\}$ & not reset
& $\displaystyle \f{2}{M}+\f{\Pe^{2}}{1+M}$
& Eq.~\eqref{eq:msd_pos_only} \\[6pt]
$\{\rv,\uv\}$ & fixed $\uv_0$ & Eq.~\eqref{eq:msv_fixed_u} & Eq.~\eqref{eq:msd_ru_fixed} \\[2pt]
$\{\rv,\uv\}$ & random & Eq.~\eqref{eq:msv_random_u} & Eq.~\eqref{eq:msd_ru_random} \\[4pt]
$\{\uv\}$ & fixed $\uv_0$ & Eq.~\eqref{eq:msv_fixed_u} & NO SS \\[2pt]
$\{\uv\}$ & random & Eq.~\eqref{eq:msv_random_u} & NO SS \\
\bottomrule
\end{tabular}
\end{table}

The complete resetting protocol in Eq.~\eqref{eom:resetting} belongs to a family of seven protocols, corresponding to different choices of the state variables $(\rv,\vv,\uv)$ to be reset. These comprise complete resetting $\{\rv,\vv,\uv\}$, the partial-resetting protocols $\{\rv,\vv\}$, $\{\rv,\uv\}$ and $\{\rv\}$, and the position-preserving
protocols $\{\vv,\uv\}$, $\{\vv\}$ and
$\{\uv\}$~\cite{Kumar2020, Baouche2024, Olsen2024, Baouche2025}. Only the protocols that reset $\uv$ admit two variants based on whether the orientation is restored to the fixed value $\uv_0=(1,0)$ or drawn uniformly on $[-\pi,\pi]$ at each reset event.
We compare the steady-state MSV across all seven and the steady-state MSD across the four position-resetting protocols, at low inertia ($M=0.1$) and high inertia ($M=10$) with the activity fixed at $\Pe=10$ (Fig.~\ref{fig:partial_reset}).
The simulations follow the scheme of Sec.~\ref{sec:model}, with only the selected components restored at each reset. Table~\ref{tab:protocol_msv} collects the results.

The eleven protocol-ensemble combinations are far from independent. They
collapse onto four distinct MSV curves and four distinct MSD curves, all of which we obtain in closed form. Two structural results account for the
collapse.

First, a positional steady state requires resetting the position. Under $\{\vv,\uv\}$, $\{\vv\}$ and $\{\uv\}$, the MSD grows without bound. Protocols with zero stationary mean velocity exhibit long-time diffusion, whereas fixed-orientation resetting can sustain a nonzero mean velocity and hence a ballistic contribution to the uncentered MSD. The MSV, by contrast, is stationary under every protocol, since the velocity of a free inertial ABP
is itself stationary~\cite{Patel2023}.

Second, resetting the velocity to $\vv_0=\mathbf{0}$ erases the reset orientation from the moment hierarchy. 
The source terms carrying $\uv_0$ all
appear through the products $\uv_0\cdot\vv_0$ and $\vv_0^2$ (Appendix~\ref{app:second_moments}), which vanish, and the orientational correlation within an excursion is $e^{-\tau}$, independent of reset value.
All four velocity-resetting protocols therefore share the MSV of Eq.~\eqref{eq:MSV}, and $\{\rv,\vv,\uv\}$ and $\{\rv,\vv\}$ share the MSD of Eq.~\eqref{eq:MSD}, identically in the fixed and the random variant.

Only $\{\rv,\uv\}$ and $\{\uv\}$ depend on the reset orientation $\uv_0$. The protocol $\{\rv\}$ leaves $\uv$ untouched. The residual correlation $\la\vv\cdot\uv\ra$ survives in all three because the velocity is not reset.
The family thus splits according to whether the velocity is reset, and the separation is controlled by $Mr$, as it was for the excess kurtosis in Sec.~\ref{sec:phase_diagrams}.
For these three velocity-preserving protocols the stationary velocity is
\bea
\vv(t) = \f{1}{M}\int_{-\infty}^{t}\! e^{-(t-s)/M}
\left[\Pe\,\uv(s)\,ds + \sqrt{2}\,{\bf{dB}}^t(s)\right],
\label{eq:v_stationary}
\eea
so that
\bea
\la \vv^2 \ra_{r}^{\rm st} = \f{2}{M}
+ \f{\Pe^2}{M^2}\int_0^\infty\!\!\int_0^\infty\! da\,da'\,
e^{-(a+a')/M}\,C_u(|a-a'|), \nn\\
\label{eq:msv_from_C}
\eea
with $C_u(\tau)=\la \uv(t)\cdot\uv(t+\tau)\ra_{r}^{\rm st}$ as the steady-state
orientation correlation. Equation~\eqref{eq:msv_from_C} is a filter
relation: momentum relaxation passes the orientational memory through a low-pass filter of width $1/M$, and the velocity statistics depend on the protocol only through $C_u$ sampled on that scale. Two subsets of fixed or random-orientation resetting shares the same form for $C_u(\tau)=A\,e^{-\lambda\tau}+B$, the double integral of which evaluates to
\bea
\la \vv^2 \ra_{r}^{\rm st} = \f{2}{M}
+ \Pe^2\left[\f{A}{1+M\lambda} + B\right]\,.
\label{eq:msv_master}
\eea
The plateau $B=|\la\uv\ra_{r}^{\rm st}|^{2}$ is the squared mean orientation. 
It is precisely this quantity that separates the two variants, and it is nonzero only when the reset direction is prescribed. 
We discuss these in the next subsection.

\subsection{Fixed-orientation resetting}
\label{sec:partial_fixed}

Under resetting to a fixed $\uv_0$ the mean orientation becomes nonzero, $\la\uv\ra_{r}^{\rm st}=r\uv_0/(1+r)$, and the correlation retains a constant offset.
The orientation correlation can then be decomposed according to whether at least one resetting event occurs during the time interval $\tau$: $C_u(\tau)=P_0(\tau)\,\left\langle\uv(t+\tau)\cdot\uv(t)\right\rangle_{0}+\bigl[1-P_0(\tau)\bigr]\,\left\langle\uv(t+\tau)\cdot\uv(t)\right\rangle_{R}$, where $P_0(\tau)$ denotes the probability of no resetting event during the interval $\tau$, while the subscripts $0$ and $R$ denote averages over trajectories with no resetting and with at least one resetting event, respectively.
This leads to the orientation correlation under fixed-orientation resetting as
\bea
C_u(\tau) = e^{-(1+r)\tau}
+ \f{r^{2}}{(1+r)^{2}}\left[1-e^{-(1+r)\tau}\right]\,,
\label{eq:C_fixed}
\eea
where we identify $\lambda=1+r$, $A=(1+2r)/(1+r)^{2}$ and $B=r^{2}/(1+r)^{2}$.
Equation~\eqref{eq:msv_master} then yields the following results for the $\{ {\uv} \}$ and $\{{\rv,\uv} \}$ protocols under fixed-orientation resetting:
\bea
\la \vv^2 \ra_{r}^{\rm st} = \f{2}{M}
+ \Pe^2\,\f{(1+r) + M r^{2}}{(1+r)\left[1+M(1+r)\right]}\,.
\label{eq:msv_fixed_u}
\eea
This reduces to the free inertial ABP result $2/M+\Pe^2/(1+M)$ at $r=0$ and rises to the frozen-orientation value $2/M+\Pe^2$ as $r\to\infty$, but the approach is not monotonic.
Expanding at small reset rate, MSV yields
\bea
\la \vv^2 \ra_{r}^{\rm st} = \f{2}{M}
+ \f{\Pe^{2}}{1+M}\left[1 - \f{Mr}{1+M} + {\cal O}(r^{2})\right]\,,
\label{eq:msv_fixed_smallr}
\eea
where the leading correction in $r$ is negative for every $M>0$. The two effects of an
orientational reset enter at different orders. Interrupting the current
heading adds a decorrelation channel, $\lambda=1+r$, at ${\cal O}(r)$,
whereas the alignment bias towards $\uv_0$ builds up only through
$B={\cal O}(r^{2})$. Decorrelation therefore dominates at small $r$ and
alignment at large $r$. Extremizing Eq.~\eqref{eq:msv_fixed_u} locates the
minimum at
\bea
r_{\star}(M) = \f{1}{2}\left[\sqrt{1+2/M}-1\right]\,,
\label{eq:rstar_fixed}
\eea
and the MSV returns to its free-particle value at exactly $r=1/M$, so that
\bea
\la \vv^2 \ra_{r}^{\rm st}\gtrless\la \vv^2 \ra_{r=0}^{\rm st}
\quad\text{for}\quad Mr\gtrless1\,.
\eea
The same product that governs the excess kurtosis therefore fixes the sign of the orientational reset as well: $Mr_{\star}\to1/2$ at large inertia, and the crossover sits at $Mr=1$ for all $M$. The dip is shallow, of relative depth ${\cal O}(M/(1+M)^{2})$ at its deepest, and the dominant feature of Fig.~\ref{fig:partial_reset}(a,b) remains the rise above the free-particle value at $Mr\gtrsim1$: orientational resetting acts on the velocity statistics opposite to velocity resetting, raising the MSV rather than
decaying as $r^{-1}$.

For the resetting protocol of $\{\rv\}$ alone, the orientation is untouched, $C_u(\tau)=e^{-\tau}$, and the MSV stays at its free-particle value $2/M+\Pe^2/(1+M)$ for all $r$.

\subsection{Random-orientation resetting}
\label{sec:partial_random}

If the orientation is instead redrawn uniformly at each reset, the mean orientation vanishes, the plateau in Eq.~\eqref{eq:C_fixed} is absent, and $C_u(\tau)=e^{-(1+r)\tau}$, with $\lambda=1+r$, $A=1$ and $B=0$.
Thus for the random-orientation resetting in $\{\uv\}$ and $\{\rv,\uv\}$, Eq.~\eqref{eq:msv_master} gives,
\bea
\la \vv^2 \ra_{r}^{\rm st} = \f{2}{M} + \f{\Pe^2}{1+M(1+r)}\,.
\label{eq:msv_random_u}
\eea
The reset rate now acts purely as an added rotational decorrelation channel, shifting the persistence rate from $1$ to $1+r$. 
The MSV decreases monotonically to the passive value $2/M$: frequent randomization destroys the directed component of the propulsion before the velocity, which
responds on the time scale $M$, can follow it. 
The suppression is set by $M r$ once $M\gtrsim1$, so inertia and randomization reinforce one another.

The contrast with Eq.~\eqref{eq:msv_fixed_u} is complete. The same reset rate applied to the same variable drives the MSV to $2/M+\Pe^2$ under fixed-orientation resetting and to $2/M$ under random-orientation resetting, the two limits differing by the entire activity contribution.
Orientational resetting is thus not a single control knob but two opposite ones, and its sign is set by whether the restart direction is prescribed or
drawn afresh. The distinction is directly relevant to experiment: a robotic agent programmed to reorient along a fixed heading and one that reorients randomly implement opposite limits of the same protocol.

\subsection{Displacement statistics under partial resetting}
\label{sec:partial_msd}

The MSD admits a closed form under every position-resetting protocol. 
We treat $\{\rv\}$ first, where a renewal argument applies directly. Under $\{\rv\}$ the velocity and orientation are never reset, so the velocity remains stationary across reset events and the position statistics follow from the renewal equation~\eqref{eq:moment_resetting} applied to the free displacement built from a stationary velocity, $\la \rv_p^2(t)\ra = 2\int_0^t (t-\tau)\,C_v(\tau)\,d\tau$, with $C_v(\tau)=\la\vv(t)\cdot\vv(t+\tau)\ra^{\rm st}$ evaluated from
Eq.~\eqref{eq:v_stationary},
\bea
C_v(\tau) = \f{2}{M}e^{-\tau/M}
+ \f{\Pe^{2}}{M^{2}-1}\left(M e^{-\tau/M} - e^{-\tau}\right)\,.
\label{eq:Cv}
\eea
The double time integral becomes a factor $s^{-2}$ in Laplace space, so the
Final Value Theorem gives $\la\rv^2\ra_{r}^{\rm st} = 2\hat C_v(r)/r$. The
apparent poles at $M=1$ cancel identically, leaving
\bea
\la \rv^2 \ra_{r}^{\rm st} = \f{2}{r(1+Mr)}
\left[\,2 + \f{\Pe^{2}\,(1+M+Mr)}{(1+M)(1+r)}\right]\,.
\label{eq:msd_pos_only}
\eea
In the overdamped limit this reduces to $4/r + 2\Pe^{2}/[r(1+r)]$, the known
result for an overdamped ABP under position-orientation
resetting~\cite{Shee2025}, as it must, since $\vv$ is then slaved to $\uv$
and carries no memory across the reset. In the passive limit it gives
$4/[r(1+Mr)]$, larger than the complete-resetting result
Eq.~\eqref{eq:MSD_Brownian} by exactly the factor $(2+Mr)/2$, the reciprocal
of the velocity-suppression factor in Eq.~\eqref{eq:MSV}.

Comparing Eq.~\eqref{eq:msd_pos_only} with Eq.~\eqref{eq:MSD} isolates the
effect of velocity resetting on localization. At large reset rate the
position-only MSD decays as
$\la\rv^2\ra_{r}^{\rm st}\simeq 2[2+M\Pe^{2}/(1+M)]/(Mr^{2})$, one power
slower than the $r^{-3}$ of complete resetting; at large inertia it decays
as $M^{-1}$, again one power slower than the $M^{-2}$ of
Eq.~\eqref{eq:MSD}. Velocity resetting therefore contributes exactly one
additional power of suppression in each of $r$ and $M$. The mechanism is
transparent: a particle restarted with $\vv_0=\mathbf{0}$ must rebuild its
velocity over the momentum-relaxation time $M$ before executing an
appreciable excursion, and it spends a fraction $\sim Mr$ of its time doing
so, whereas one restarted at full speed loses nothing. This is the
separation between the two branches in
Fig.~\ref{fig:partial_reset}(c,d), which opens at $Mr\sim1$ and grows
without bound.

The MSD of $\{\rv\}$ and $\{\rv,\vv\}$ is trivially ensemble independent,
since neither resets $\uv$. Under $\{\rv,\vv,\uv\}$ the reset
$\vv_0=\mathbf{0}$ removes every $\uv_0$-dependent source term, so
$\{\rv,\vv\}$ and $\{\rv,\vv,\uv\}$ share the MSD of Eq.~\eqref{eq:MSD},
consistent with the simulations. Only $\{\rv,\uv\}$ distinguishes the two
ensembles, through the velocity carried across the reset.
There the position and orientation resets occur at the same instants,
so the excursions are not statistically independent of the orientation
history and the renewal argument leading to Eq.~\eqref{eq:msd_pos_only} does
not close. The stationary hierarchy closes regardless. Under any protocol
that resets the position, every observable of degree at least one in $\rv$ is
annihilated by the reset, since $\psi(\rv=\mathbf{0},\vv,\uv)=0$ whatever is
done to $\uv$ at the same instant, and each such observable therefore
acquires the single extra term $-r\la\psi\ra_{r}^{\rm st}$ in its stationary
equation. The position sector obeys the Laplace-space relations of
Appendix~\ref{app:second_moments} evaluated at $s\to r$, with vanishing
initial terms for $\rv_{0}=\mathbf{0}$,
\bea
(1+r)\la \uv\cdot\rv\ra_{r}^{\rm st} &=& \la \uv\cdot\vv\ra_{r}^{\rm st}\,,\nn\\
\left(r+\f{1}{M}\right)\la \vv\cdot\rv\ra_{r}^{\rm st} &=&
\la \vv^{2}\ra_{r}^{\rm st}
+ \f{\Pe}{M}\la \uv\cdot\rv\ra_{r}^{\rm st}\,,\nn\\
r\la \rv^{2}\ra_{r}^{\rm st} &=& 2\la \vv\cdot\rv\ra_{r}^{\rm st}\,,
\label{eq:pos_second_generic}
\eea
so that the protocol enters through two velocity averages only,
\bea
\la \rv^{2}\ra_{r}^{\rm st} = \f{2}{r(1+Mr)}
\left[\,M\la \vv^{2}\ra_{r}^{\rm st}
+ \f{\Pe\,\la \uv\cdot\vv\ra_{r}^{\rm st}}{1+r}\right]\,.
\label{eq:msd_master}
\eea
Equation~\eqref{eq:msd_master} holds for all four position-resetting
protocols. It returns Eq.~\eqref{eq:MSD} and Eq.~\eqref{eq:msd_pos_only} on
inserting $\la\uv\cdot\vv\ra_{r}^{\rm st}=\Pe/(1+M+Mr)$ for complete
resetting and $\Pe/(1+M)$ for $\{\rv\}$. For the velocity-preserving
protocols the two averages are not independent,
$\la\vv^{2}\ra_{r}^{\rm st}=2/M+\Pe\la\uv\cdot\vv\ra_{r}^{\rm st}$, so
Eqs.~\eqref{eq:msv_fixed_u} and \eqref{eq:msv_random_u} already fix the MSD
of $\{\rv,\uv\}$,
\bea
\la \rv^{2}\ra_{r}^{\rm st}\Big|_{\rm fixed} &=& \f{2}{r(1+Mr)}
\left[\,2+\f{\Pe^{2}\left[(1+r)+Mr^{2}\right]}{(1+r)^{2}}\right]\,, \nn\\
\label{eq:msd_ru_fixed}\\
\la \rv^{2}\ra_{r}^{\rm st}\Big|_{\rm random} &=& \f{2}{r(1+Mr)}
\left[\,2+\f{\Pe^{2}}{1+r}\right]\,,
\label{eq:msd_ru_random}
\eea
which account for the corresponding branches in
Fig.~\ref{fig:partial_reset}(c,d).

Both reduce to $4/r+2\Pe^{2}/[r(1+r)]$ as $M\to0$, the overdamped result of
Ref.~\cite{Shee2025}, and to $4/[r(1+Mr)]$ at $\Pe=0$, where a prescribed
reset direction is immaterial. Both decay as $r^{-2}$ at large reset rate,
one power slower than the $r^{-3}$ of the complete protocol, in agreement
with the counting above: the extra power is supplied by velocity resetting
alone. Their behaviour at large inertia differs. The random ensemble decays
as $\la\rv^{2}\ra_{r}^{\rm st}\simeq2(2+2r+\Pe^{2})/[(1+r)Mr^{2}]$, again one
power slower than the $M^{-2}$ of Eq.~\eqref{eq:MSD}, whereas the fixed
ensemble saturates,
\bea
\lim_{M\to\infty}\la \rv^{2}\ra_{r}^{\rm st}\Big|_{\rm fixed}
= \f{2\Pe^{2}}{(1+r)^{2}}\,,
\label{eq:msd_ru_fixed_largeM}
\eea
and is not suppressed by inertia at all. A prescribed reset direction
sustains the mean velocity $\la\vv\ra_{r}^{\rm st}=\Pe\,r\,\uv_{0}/(1+r)$,
and at large $M$ the fluctuations about it no longer contribute, leaving the
drift alone: $|\la\vv\ra_{r}^{\rm st}|^{2}\la t^{2}\ra$ with
$\la t^{2}\ra=2/r^{2}$ reproduces the limit exactly. This drift survives the
momentum-relaxation bottleneck that suppresses every other protocol at large
inertia. The two ensembles separate in the same way under strong resetting,
\bea
\la \rv^{2}\ra_{r}^{\rm st}\Big|_{\rm fixed} \simeq
\f{4}{Mr^{2}}+\f{2\Pe^{2}}{r^{2}}\,,\qquad
\la \rv^{2}\ra_{r}^{\rm st}\Big|_{\rm random} \simeq
\f{4}{Mr^{2}}\,, \nn\\
\eea
the fixed-ensemble activity term being the ballistic displacement
$\Pe^{2}\la t^{2}\ra$ accumulated along $\uv_{0}$ over an excursion of
exponentially distributed duration $t$. The same substitution $s\to r$
closes the position hierarchy at fourth order, treated in
Appendix~\ref{sec:partial_fourth}.

\subsection{Experimental realizations}
\label{sec:partial_expt}

The seven protocols are not equally easy to implement, and the distinctions that matter theoretically map onto operations of very different cost.
Resetting the orientation is the cheapest. It requires only a torque, leaves the particle where it is, and is realized by aligning a magnetic colloid with an external field, by reprogramming the heading of a self-propelled robot~\cite{Deblais2018, Paramanick2024}, or, in the random variant, by an engineered tumble event.
The two ensembles correspond to two equally accessible instructions, reorient along a prescribed laboratory direction or draw a new heading at random, and Eqs.~\eqref{eq:msv_fixed_u} and
\eqref{eq:msv_random_u} show that they drive the velocity statistics to opposite limits.
Resetting the velocity requires dissipating the kinetic energy $M\la\vv^{2}\ra^{\rm st}_{r}/2$ at every reset event, a real actuation cost for a vibrated granular particle~\cite{Koumakis2016,
Scholz2018} or a robot, and it is this operation that Sec.~\ref{sec:partial_msd} identifies as the one controlling localization.
Resetting the position is the most demanding, since the agent must physically return to the origin, as in the homing protocols of Refs.~\cite{Altshuler2024, Paramanick2024}.

An important consequence is that the overdamped colloidal realizations of stochastic resetting~\cite{Tal-Friedman2020Experimental, Faisant2021OptimalMFPTResetting} cannot distinguish most of these protocols.
When $M\to0$ the velocity is slaved to the orientation and carries no memory across a reset, so $\{\rv\}$, $\{\rv,\vv\}$ and $\{\rv,\vv,\uv\}$ collapse
onto a single curve, as Fig.~\ref{fig:partial_reset}(c) confirms at $M=0.1$ for $r\lesssim1/M$. The hierarchy becomes observable only once
$Mr\gtrsim1$, which is why inertial platforms are required to test it.
Vibrated granular disks, low-damping colloids and robotic agents all reach
this regime at accessible reset rates: an agent with momentum-relaxation
time $\tau$ restarted every $\tau_{\rm reset}\lesssim\tau$ sits where the
branches of Fig.~\ref{fig:partial_reset} separate by orders of magnitude,
and where the choice between braking and merely reorienting decides whether
the steady state is localized or exploratory.

Velocity resetting thus controls localization at all inertia, through the
single combination $Mr$, and complete resetting is the most strongly
localizing protocol, although for the MSD it is exactly degenerate with
$\{\rv,\vv\}$.
Together with Eqs.~\eqref{eq:msv_fixed_u}, \eqref{eq:msv_random_u}, \eqref{eq:msd_pos_only}, \eqref{eq:msd_ru_fixed} and \eqref{eq:msd_ru_random}, the results of Sec.~\ref{sec:second_moments} give the stationary MSV for every protocol and the stationary MSD for every position-resetting protocol in closed form. The full transient MSV and MSD under complete resetting are given in Sec.~\ref{sec:dynamics}.
The corresponding
analysis at fourth order, including a master relation for the velocity
excess kurtosis of the velocity-preserving protocols, is given in
Appendix~\ref{sec:partial_fourth}.

\section{Conclusions}
\label{sec:conclusions}

We have demonstrated that inertia dramatically alters the steady–state statistics of active particles under stochastic resetting, giving rise to enhanced typical localization, strong non-Gaussian fluctuations, and distinct scaling regimes in both displacement and velocity statistics. Whereas overdamped active particles under resetting in confinement exhibits modest deviations from Gaussianity, under complete resetting the inclusion of inertia produces qualitatively new behavior: the mean-squared displacement decays as $r^{-3}$ at high reset rates and as $M^{-2}$ at large inertia, and the excess kurtosis exhibits pronounced re-entrant trends reflecting competition between persistence, relaxation, and resetting. These results establish inertia as a key control variable in reset-driven nonequilibrium steady states, bridging the overdamped active matter framework and the largely unexplored inertial regime.

The systematic comparison of resetting protocols shows that position resetting establishes localization, whereas velocity resetting controls its strength. Resetting position and velocity together reproduces the scalar displacement moments of complete resetting, including the $r^{-3}$ suppression of the stationary MSD at large reset rates. When velocity is preserved, the stationary MSD instead decays as $r^{-2}$. The orientation reset ensemble provides a further distinction: under position--orientation resetting, a fixed reset direction sustains a drift contribution and a nonzero large-inertia MSD, whereas uniformly random orientation resetting yields an MSD that decreases as $M^{-1}$.

\noindent
The exact moment framework yields closed-form expressions for steady-state fluctuations up to fourth order, enabling a quantitative characterization of non-Gaussian statistics. Kurtosis-based phase diagrams demonstrate that inertia and resetting produce heavy-tailed velocity and displacement distributions, with a re-entrant dependence that identifies inertial momentum relaxation as a key control parameter for rare fluctuations.
These predictions are experimentally accessible in platforms ranging from inertial colloids and optically trapped active particles to programmable robotic agents, where reset protocols can be used to obtain steady states in which both small- and large-amplitude displacements become more probable.

More broadly, this work establishes an analytically tractable framework connecting inertia, stochastic resetting, and extreme fluctuations in active systems, with implications for active matter, search processes, and autonomous agents.
By showing that inertia can be harnessed to amplify rare events and modulate spatial statistics, our results suggest new strategies for controlling active transport in synthetic and biological systems. Extensions incorporating rotational inertia~\cite{Caprini2022, Lisin2022} or external potentials~\cite{Patel2024} provide a natural route to broaden the framework toward active suspensions, granular systems, and mobile robotics.
 The analytical characterization of the probability distributions in inertial active systems remains highly challenging and can provide important insights for resetting protocols.

Our results show that inertia qualitatively reshapes active dynamics under stochastic resetting, enhancing rare fluctuations and producing strongly non-Gaussian steady states.
These predictions are directly testable in robotic and colloidal platforms, where re-entrant kurtosis behavior and distinct scaling regimes at large inertia or reset rates can be probed~\cite{Tal-Friedman2020Experimental, Faisant2021OptimalMFPTResetting}.
The exact results for complete and partial resetting provide a basis for optimizing the choice of reset variables and reset ensembles. Adaptive protocols and optimization for exploration or sampling efficiency remain directions for future work~\cite{Olsen2024, Kumar2020, Baouche2024, Baouche2025, Keidar2025}.
First-passage observables quantify the efficiency of locating specific targets~\cite{Baouche2025}, whereas steady-state spatial statistics (e.g., MSD and kurtosis) characterize the global structure and the exploration process~\cite{Shee2025, Olsen2025Optimal}. These two aspects are related but not equivalent: heavy-tailed steady states can coexist with poor target-finding efficiency when large excursions are rare.  A dedicated study of first-passage statistics is therefore an important future direction, lying beyond the scope of the present work.
 Partial resetting protocols, such as velocity-only~\cite{Olsen2024} or orientation-only resetting~\cite{Baouche2024}, are treated in Sec.~\ref{sec:partial}; using them to choose optimal strategies remains open.
In summary, our results establish inertial active matter under resetting as a fertile framework to connect active matter, search theory, and autonomous systems, with clear implications for future theory and experiment.

\begin{acknowledgments}
We gratefully acknowledge Debasish Chaudhuri and Priyo Shankar Pal for their insightful feedback and discussions. The numerical simulations were performed using SAMKHYA, the High-Performance Computing Facility provided by the Institute of Physics, Bhubaneswar.
\end{acknowledgments}

\section*{Author declarations}
\subsection*{Conflict of Interest}
The authors have no conflicts to disclose.
\subsection*{Author Contributions}
\textbf{Manish Patel:} Conceptualization (equal); Formal analysis (equal); Investigation (equal); Methodology (equal); Software (equal); Visualization (equal); Writing -- original draft (equal); Writing -- review \& editing (equal).
\textbf{Amir Shee:} Conceptualization (equal); Formal analysis (equal); Investigation (equal); Methodology (equal); Software (equal); Visualization (equal); Writing -- original draft (equal); Writing -- review \& editing (equal).

\section*{Data availability}
The data that support the findings of this study are available within the article. The code that supports this study is available from the corresponding author upon reasonable request.

\appendix

\section{Mapping to overdamped ABP in a harmonic trap}
\label{app:mapping}

It is instructive to note that the dimensionless velocity dynamics of a free inertial ABP (the dimensionless form of Eq.~\eqref{eom:vel})
\begin{equation}
M\, d\vv = - \vv\,dt+ \Pe\, \uv \,dt+ \sqrt{2} \,{\bf{dB}}^t,
\end{equation}
can be mapped exactly onto the position dynamics of an overdamped ABP in a harmonic potential of stiffness $\beta$~\cite{Chaudhuri2021},
\begin{equation}
d \rv = -\beta \rv \,dt + v_a \uv \,dt + \sqrt{2}/M \,{\bf{dB}}^{t}.
\end{equation}
Identifying coefficients yields the correspondence
\begin{equation}
\beta = \frac{1}{M}, \qquad v_a = \frac{\Pe}{M},
\end{equation}
with noise strength $2/M^{2}$, so that the inertial velocity process is mathematically equivalent to the position of an overdamped ABP confined in a harmonic trap with parameters $(\beta, v_a)$.

Under stochastic resetting, this mapping implies that the velocity statistics of an inertial particle mirror those of a harmonically trapped overdamped ABP with the above parameter substitution~\cite{Shee2025}. By contrast, the \emph{position} of an inertial ABP under resetting does not follow this simple mapping, as it couples to velocity and therefore exhibits rich complex behavior.

\section{Moments generator framework}
\label{app:moments_generator_framework}
The probability distribution $P(\rv, \vv, \uv, t)$ of the position $\rv$, the velocity $\vv$ and the heading (or orientation) $\uv$ of the particle  follows the Fokker-Planck equation~\cite{Hermans1952, Patel2023}
\bea
\p_t P &=& -\nabla \cdot(\vv\, P) - (1/M)\nabla_v \cdot ((\l \, \uv - \vv)P)\nonumber\\
&& + ( 1/M^2) \nabla_v^2P + \nabla_u^2 P\,,
\label{eq:F-P}
\eea
where $\nabla$, $\nabla_{v}$, and $\nabla_u$ is the gradient on the position, velocity, and orientation space, respectively. 

Utilizing the Laplace transform $\tilde P(\rv, \vv, \uv, s) = \int_0^\infty dt\, e^{-s t}\, P(\rv, \vv, \uv, t) $ and defining the mean of an observable $\la \psi \ra_s = \int d\rv \, d\vv \, d\uv\, \psi(\rv, \vv, \uv ) \tilde P(\rv, \vv, \uv, s)$, multiplying by $\psi(\rv, \vv, \uv)$ and integrating over all possible $(\rv, \vv, \uv)$, we find,
\bea \label{observable}
     -\la \psi \ra_0 + s \la \psi \ra_s &=& \la \vv \cdot \nabla \psi \ra_s + (\l/M) \la \uv \cdot \nabla_v \psi \ra_s \nonumber\\
&&- (1/M) \la \vv \cdot \nabla_v \psi \ra_s + (1/M^2) \la \nabla_v^2 \psi \ra_s \nonumber\\
&&+  \la \nabla_u^2 \psi \ra_s,
\label{eq:moment}
\eea
where the initial condition sets $\la \psi \ra_0 = \int d\rv \, d\vv\, d\uv\,  \psi(\rv, \vv, \uv) P(\rv, \vv, \uv, 0)$. Without any loss of generality, we consider the initial condition to follow $P(\rv, \vv, \uv, 0) = \d(\rv-\rv_0) \d(\vv - \vv_0) \d(\uv - \uv_0)$, where $\rv_0$, $\vv_0$, and $\uv_0$ are the initial position, velocity, and orientation respectively. Equation~(\ref{eq:moment}) can be utilize to compute exact moments as a function of time without stochastic resetting; see Patel {\em et al}~\cite{Patel2023}.

Under stochastic resetting of position, velocity, and orientation, the moments satisfy the renewal equation~\cite{Kumar2020, Shee2025}
\bea
\la \psi (t)\ra_r &=& e^{-r t} \la\psi (t)\ra + r \int_{0}^{t} dt^{\prime} e^{-rt^{\prime}} \la\psi (t^{\prime})\ra\,.
\label{eq:moment_resetting}
\eea
This formalism allows us to compute exact expressions for moments up to fourth order and to analyze their long-time limit ($t\to\infty$), which defines the non-equilibrium steady state.
Alternatively, the steady state under stochastic resetting can be obtained directly from the Final Value Theorem (FVT) applied to Eq.~\eqref{eq:moment}
\bea
\la\psi\ra^{\rm st}_{r} &=& r \la\psi\ra_s(s=r)\,.
\label{eq:moment_resetting_SS}
\eea
We present the primary results in the main text and provide full derivations below.

\section{Derivation of second order moments}
\label{app:second_moments}
\noindent
{\bf{Orientation: }}
 Without stochastic resetting, the Laplace space orientation $\la \uv \ra_s$ can be obtain from Eq.~\eqref{eq:moment} using $\la \nabla_u^2 (\uv)\ra = - \uv$, while $\vv \cdot \nabla (\uv) = \uv \cdot \nabla_v (\uv) = \vv \cdot \nabla_v (\uv) = \nabla_v^2 (\uv) = 0$ as 
 \bea
 \la \uv \ra_s = \frac{\uv_0}{s+1}. 
 \eea
The inverse Laplace transform of the above equation gives the full dynamics in absence of stochastic resetting as $\la \uv \ra (t) = \uv_0 e^{-t}$.
The full orientation evolution under stochastic resetting can be obtained by using $\la \uv \ra (t)$ in Eq.~\eqref{eq:moment_resetting}. One can directly obtain the steady state orientation under stochastic resetting using Eq.~\eqref{eq:moment_resetting_SS} as 
\bea
\la \uv \ra_{r}^{\rm st} = \frac{r \, \uv_0}{r + 1}.
\eea

\noindent
{\bf{Second moment of Velocity:}} We set $\psi = \vv^2 $ in Eq.~(\ref{eq:moment}) and use $\la \psi \ra_0 = \vv_0^2$, 
$\vv \cdot \nabla (\vv^2) = 0$, $\la \vv \cdot \nabla_v (\vv^2 ) \ra_s = 2 \la \vv^2 \ra_s$, $\la \uv \cdot \nabla_v (\vv^2) \ra_s = 2 \la \uv \cdot \vv \ra_s$, $\nabla_v^2(\vv^2) = 4/s$ and $ \nabla_u^2 (\vv^2) = 0$ to obtain moments in absence of resetting as 
\bea
\la \vv^2 \ra_s = \frac{1}{s+2/M} \left( \vv_0^2 + \frac{2 \Pe \la \uv \cdot \vv \ra_s}{M} + \frac{4 }{s M^2} \right).
\eea
We again set $\psi = \uv \cdot \vv$ in Eq.~(\ref{eq:moment}) to get $\la \uv \cdot \vv \ra_s = [\uv_0 \cdot \vv _0 + \Pe/(sM) ]/[s + 1+ 1/M ]$. Following Eq.~(\ref{eq:moment_resetting_SS}), we use $\la \vv^2 \ra_s$ and $\la \uv \cdot \vv \ra_s$ to get the steady state second moment of velocity under stochastic resetting as Eq.~\eqref{eq:MSV} in the main text, where we have used the initial condition as $\rv_0 = {\bf{0}}$ and $\vv_0 = {\bf{0}}$.

\noindent
{\bf{Second moment of Displacement:}} We follow the same step as for velocity dynamics to get the second moment of displacement in absence of stochastic resetting using Eq.~(\ref{eq:moment}) as s $\la \rv^2 \ra_s = \la \rv^2 \ra_0 + 2 \la \vv \cdot \rv \ra_s$. 
We further use $\psi = \vv \cdot \rv$ and $\psi = \uv \cdot \rv$ in Eq.~(\ref{eq:moment}) to get $(s+1/M) \la \vv \cdot \rv \ra_s = \la \vv \cdot \rv \ra_0 + \la \vv^2 \ra_s + \Pe \la \uv \cdot \rv \ra_s/M$ and $(s+1) \la \uv \cdot \rv \ra_s = \la \uv \cdot \rv \ra_0 +  \la \uv \cdot \vv \ra_s$, respectively. 
We use these moments in Eq.~(\ref{eq:moment_resetting_SS}) to get the steady state mean squared displacement under stochastic resetting for initial condition $\rv_0 = {\bf{0}}$, $\vv_0 = {\bf{0}}$ as Eq.~\eqref{eq:MSD} presented in main text.

\noindent
\textbf{Gaussian from second order moments:~}
Given the exact steady--state second moments derived above, we write the corresponding Gaussian. 
Writing the variances as
\bea
\s_r^2 \;\equiv\; \frac{1}{2}\,\la \rv^2 \ra_{r}^{\rm st},\quad \s_v^2 \;\equiv\; \frac{1}{2}\,\la \vv^2 \ra_{r}^{\rm st},
\label{eq:vars}
\eea
the Gaussian distributions of position and velocity reads
\bea
\mathcal{P}_{\rm G}(|\rv|)=\frac{1}{2\pi\s_r^2}\exp \left(-\frac{|\rv|^2}{2\s_r^2}\right)\,,\\
\mathcal{P}_{\rm G}(|\vv|) =\frac{1}{2\pi\s_v^2}\exp \left(-\frac{|\vv|^2}{2\s_v^2}\right)\,.
\label{eq:Gaussian}
\eea
We use this Gaussian distribution as a reference to visually quantify deviations of the numerically computed distributions.

\noindent
\textbf{Transient dynamics:~}
The Laplace-space moments $\la \vv^2 \ra_s$ and $\la \rv^2 \ra_s$ derived
above give the steady state through the Final Value Theorem,
Eq.~\eqref{eq:moment_resetting_SS}, but they also determine the full time
dependence. Using the renewal relation
$\la \psi \ra_{r}(s) = (1 + r/s)\,\la \psi \ra_{s+r}$ and inverting the
Laplace transform, we obtain $\la \vv^2 \ra_r(t)$ and $\la \rv^2 \ra_r(t)$
in closed form. Resetting shifts every relaxation rate of the free
inertial ABP uniformly by $r$ and adds a mode at the bare rate $r$ in the
MSD. These results are presented and analyzed in
Sec.~\ref{sec:dynamics}, Eqs.~\eqref{eq:msv_dynamics} and
\eqref{eq:msd_dynamics}.

\onecolumngrid
\section{Derivation of fourth order moments} 
\label{app:fourth_moments}
\begin{figure*}[!t]
\begin{center}
\includegraphics[width=\linewidth]{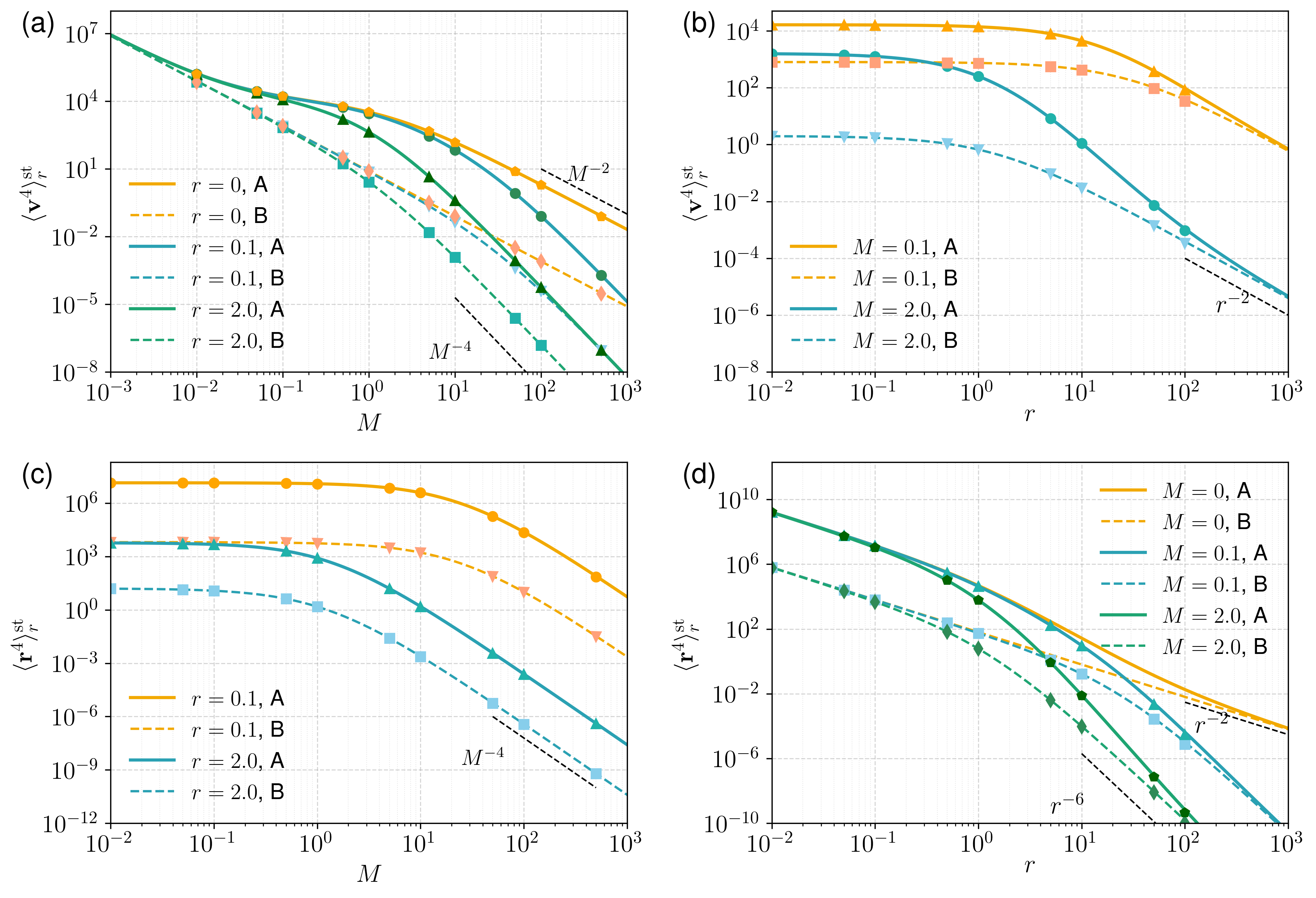} 
\caption{Steady-state fourth moment of velocity $\la \vv^4 \ra_{r}^{\rm st}$ in (a,b)
and of position $\la \rv^4 \ra_{r}^{\rm st}$ in (c,d), under complete resetting.
Symbols are simulation data.
Lines are the analytic results: Eq.~\eqref{eq:v4avg_ss} in (a,b) and Eq.~\eqref{eq:r4avg_ss} in (c,d).
Active particles ($\Pe=10$, solid, labeled A) are compared with passive particles ($\Pe=0$, dashed, labeled B).
(a) Velocity fourth moment versus inertia $M$ at $r=0$, $0.1$, and $2.0$.
(b) The same versus reset rate $r$ at $M=0.1$ and $2.0$.
(c) Position fourth moment versus $M$ at $r=0.1$ and $2.0$.
(d) The same versus $r$ at $M=0$ (overdamped), $0.1$, and $2.0$.
Black dashed lines are the asymptotic scalings derived in this appendix:
$M^{-2}$ and $M^{-4}$ in (a), $r^{-2}$ in (b), $M^{-4}$ in (c),
and in (d) $r^{-2}$ for $M=0$ together with $r^{-6}$ at finite $M$.}
\label{app_fig:fourth}
\end{center}
\end{figure*}

We set $\psi = \mathbf{v}^4$ and $\psi = \mathbf{r}^4$ in Eq.~\eqref{eq:moment}, respectively, to obtain the fourth moments of the velocity and displacement vectors, in the absence of stochastic resetting, in Laplace space:
\bea
(s+4/M) \la \vv^4 \ra_s &=& \vv^4_0 + \frac{4 \Pe}{M} \la (\uv \cdot \vv) \vv^2 \ra_s + \frac{16}{M^2} \la \vv^2 \ra_s, \\
(s+3/M + 1) \la (\uv \cdot \vv) \vv^2 \ra_s &=& (\uv_0 \cdot \vv_0) \vv_0^2 + \frac{\Pe}{M} \la \vv^2 \ra_s + \frac{2 \Pe}{M} \la (\uv \cdot \vv)^2 \ra_s + \frac{8}{M^2} \la \uv \cdot \vv \ra_s, \\
(s+2/M+4) \la (\uv \cdot \vv)^2 \ra_s &=& (\uv_0 \cdot \vv_0 )^2 + \frac{2 \Pe}{M} \la \uv \cdot \vv \ra_s + \frac{2}{sM^2} + 2 \la \vv^2 \ra_s, \\
s \la \rv^4 \ra_s &=& \rv_0^4 + 4 \la (\vv \cdot \rv) \rv^2 \ra_s, \\
(s+1/M) \la (\vv \cdot \rv) \rv^2 \ra_s &=& (\vv_0 \cdot \rv_0) \rv_0^2 + \la \vv^2 \rv^2 \ra_s + 2 \la (\vv \cdot \rv)^2 \ra_s + \frac{\Pe}{M} \la (\uv \cdot \rv) \rv^2 \ra_s, \\
(s+2/M) \la \vv^2 \rv^2 \ra_s &=& \vv_0^2 \rv_0^2 +  2 \la (\vv \cdot \rv) \vv^2 \ra_s + \frac{2 \Pe}{M} \la (\uv \cdot \vv) \rv^2 \ra_s + \frac{4}{M^2} \la \rv^2 \ra_s, \\
(s + 3/M) \la (\vv \cdot \rv) \vv^2 \ra_s &=& (\vv_0 \cdot \rv_0 ) \vv_0^2 + \la \vv^4 \ra_s + \frac{\Pe}{M} \la (\uv \cdot \rv) \vv^2 \ra_s + \frac{2 \Pe}{M} \la (\uv \cdot \vv) (\vv \cdot \rv) \ra_s + \frac{8}{M^2} \la \vv \cdot \rv \ra_s, \\
(s+2/M + 1) \la (\uv \cdot \vv ) (\vv \cdot \rv) \ra_s &=& (\uv_0 \cdot \vv_0)(\vv_0 \cdot \rv_0) + \la (\uv \cdot \vv ) \vv^2 \ra_s + \frac{\Pe}{M} \la \vv \cdot \rv \ra_s + \frac{\Pe}{M} \la (\uv \cdot \vv) (\uv \cdot \rv) \ra_s  + \frac{2}{M^2} \la \uv \cdot \rv \ra_s, \nn\\ 
(s + 2/M + 1) \la (\uv \cdot \rv) \vv^2 \ra_s &=& (\uv_0 \cdot \rv_0) \vv_0^2 +  \la (\uv \cdot \vv) \vv^2 \ra_s + \frac{2 \Pe}{M} \la (\uv \cdot \vv) (\uv \cdot \rv) \ra_s + \frac{4}{M^2} \la \uv \cdot \rv \ra_s, \\
(s + 1/M  + 4) \la (\uv \cdot \vv) (\uv \cdot \rv) \ra_s &=& (\uv_0 \cdot \vv_0) (\uv_0 \cdot \rv_0) + \la (\uv \cdot \vv)^2 \ra_s + \frac{\Pe}{M} \la \uv \cdot \rv \ra_s + 2 \la \vv \cdot \rv \ra_s, \\
(s + 1/M + 1) \la (\uv \cdot \vv) \rv^2 \ra_s &=& (\uv_0 \cdot \vv_0) \rv_0^2 + 2 \la (\uv \cdot \vv)(\vv \cdot \rv) \ra_s + \frac{\Pe}{M} \la \rv^2 \ra_s, \\
(s+ 2/M) \la (\vv \cdot \rv)^2 \ra_s &=& (\vv_0 \cdot \rv_0)^2 + 2 \la (\vv \cdot \rv) \vv^2 \ra_s + \frac{2 \Pe}{M} \la (\uv \cdot \rv)(\vv \cdot \rv) \ra_s + \frac{2}{M^2} \la \rv^2 \ra_s, \\
(s + 1) \la (\uv \cdot \rv) \rv^2 \ra_s &=& (\uv_0 \cdot \rv_0) \rv_0^2 + \la (\uv \cdot \vv) \rv^2 \ra_s + 2 \la (\uv \cdot \rv) (\vv \cdot \rv) \ra_s, \\
(s+1/M+1) \la (\uv \cdot \rv) (\vv \cdot \rv) \ra_s &=& (\uv_0 \cdot \rv_0) (\vv_0 \cdot \rv_0) + \la (\uv \cdot \vv) (\vv \cdot \rv) \ra_s + \la (\uv \cdot \rv) \vv^2 \ra_s + \frac{\Pe}{M} \la (\uv \cdot \rv)^2 \ra_s, \\
(s + 4) \la (\uv \cdot \rv)^2 \ra_s &=& (\uv_0 \cdot \rv_0)^2 + 2 \la (\uv \cdot \vv) (\uv \cdot \rv) \ra_s + 2 \la \rv^2 \ra_s.
\eea
The inverse Laplace transformation of $\langle \mathbf{v}^4 \rangle_s$ and $\langle \mathbf{r}^4 \rangle_s$ leads to the full time evolution of the fourth moments, $\langle \mathbf{v}^4 \rangle(t)$ and $\langle \mathbf{r}^4 \rangle(t)$, respectively. These results correspond to inertial active Brownian particles in the absence of stochastic resetting and were discussed in detail by Patel {\em et al.}~\cite{Patel2023}.
We may use Eq.~(\ref{eq:moment_resetting}) to obtain exact expressions for the fourth moments, $\langle \mathbf{v}^4 \rangle_{r}(t)$ and $\langle \mathbf{r}^4 \rangle_{r}(t)$, in the presence of full stochastic resetting.
 The same inversion yields the time-dependent fourth moments, although the resulting expressions are algebraically unwieldy; we restrict the transient analysis of Sec.~\ref{sec:dynamics} to the second moments.
Since our primary focus in this article is the steady state, we directly employ the Final Value Theorem in Eq.~\eqref{eq:moment_resetting_SS}, which leads to
\bea
\la \vv^4 \ra_{r}^{\rm st} = \frac{{\cal A}_1}{{\cal B}_1},
\label{eq:v4avg_ss}
\eea
where 
\bea
{\cal A}_1 &=&
64 M^{3} r^{3} \nonumber\\
&&+ \big(96 M^{3}\Pe^{2} + 384 M^{3} + 384 M^{2}\big)\, r^{2} \nonumber\\
&&+ \big(24 M^{3}\Pe^{4} + 448 M^{3}\Pe^{2} + 576 M^{3} 
     + 384 M^{2}\Pe^{2} + 1536 M^{2} + 704 M\big)\, r \nonumber\\
&&+ \big(64 M^{3}\Pe^{4} + 256 M^{3}\Pe^{2} + 256 M^{3} 
     + 48 M^{2}\Pe^{4} + 896 M^{2}\Pe^{2} + 1152 M^{2} 
     + 384 M\Pe^{2} + 1280 M + 384\big)\,,\nonumber\\
{\cal B}_1 &=& M^{2}\,(Mr+2)\,(Mr+4)\,(Mr+M+1)\,(Mr+M+3)\,(Mr+4M+2)\,.\nonumber
\eea
The above steady state fourth moment of velocity agrees well with numerical simulation, as shown in Fig.~\ref{app_fig:fourth}(a,b). For zero reset rate $r = 0$, $\la \vv^4 \ra_{r}^{\rm st}$ goes as $\sim 8/M^2$ in small inertia limit and $\sim (8 + 8 \Pe^2 + 2 \Pe^4)/M^2$ in the large inertia limit, as shown in Fig.~\ref{app_fig:fourth}(a). At any finite reset rate $r \neq 0$, $\la \vv^4 \ra_{r}^{\rm st}$ show same scaling of $\sim 8/M^2$ at small inertia. However, at large inertia the behavior is $\sim 8[8(1+r)^2(4+r)+4(4+r)(2+3r)\Pe^2 + (8 + 3r) \Pe^4]/[r^2 (1+r)^2(4+r) M^4]$, as shown in Fig.~\ref{app_fig:fourth}(a).
At any fix inertia, $\la \vv^4 \ra_r^{\rm st}$ starts from a fixed value at small reset rate, which is governed by inertia. With increase in reset rate $\la \vv^4 \ra_r^{\rm st}$ decreases and goes as $\sim 64/(M^4 r^2)$ at large reset rate, as shown in Fig.~\ref{app_fig:fourth}(b).

Again using the Eq.~\eqref{eq:moment_resetting_SS}, we can get the steady state fourth moment of position as
\bea
\la \rv^4 \ra_{r}^{\rm st} &=& \frac{256 (12 + 5rM(5 + 2rM) )}{r^2 (1+rM)^2(2+rM)^2(3+rM)(4+rM)}+ \frac{{\cal A}_2}{{\cal B}_2} \Pe^2  + \frac{{\cal A}_3}{{\cal B}_2(4+r)(1+M(4+r))(2+M(4+r))} \Pe^4\,, \nn\\
\label{eq:r4avg_ss}
\eea
with 
\bea
{\cal B}_2 &=& r^2 (1+r)^2 (1+rM)^2 (2+rM)^2 (3+rM)(4+rM) (1+M+rM)^2 (2+M+rM) (3+M+rM)\,, \nn\\
{\cal A}_2 &=& 64\left(105 M^6 r^7+14 M^5 (29 M+65) r^6+7 M^4 (M (93 M+431)+443) r^5+5 M^3 (M (M (108 M+805)+1725)+1064) r^4 \right. \nn\\
&&\left. +10 M^2 (M (M (M (23 M+267)+926)+1198)+486) r^3+M (M (M (M (5 M (8 M+171)+4628)+9773)+8372) \right. \nn\\
&& \left. +2268) r^2+4 (M+1) (M (M (M (25 M+219)+602)+570)+108) r+48 (M+1)^2 (M+2) (M+3)\right)\,,\nn\\
{\cal A}_3 &=& 8 \left(315 M^8 r^9+693 M^7 (6 M+5) r^8+63 M^6 (M (349 M+653)+255) r^7+M^5 (2 M (M (29512 M+96487)+84558) \right. \nn\\
 && \left. +41055) r^6+4 M^2 \left(M \left(M \left(M \left(M \left(7328 M^2+88388 M+334233\right)+520114\right)+352254\right)+99437\right)+8976\right) r^3 \right. \nn\\
 && \left. +M^4 \left(M \left(M \left(85856 M^2+449710 M+689867\right)+376869\right)+63420\right) r^5+2 M^3 \left(M \left(M \left(M \left(34584 M^2+274438  \right. \right. \right. \right. \right. \nn\\  
&& \left. \left. \left. \left. \left. M+678807\right)+651656\right)+248685\right)+30480\right) r^4+8 M (M (M (M (M (8 M (16 M (5 M+109)+10081)+193597)+ \right. \nn\\
&& \left. 214555)+109643)+23467)+1494) r^2+32 (M+1) (M (M (M (4 M (M (100 M+933)+2935)+15153)+7750)  \right. \nn\\
&& \left. +1419)+54) r+768 (M+1)^2 (M+2) (M+3) (2 M+1) (4 M+1)\right)\,.\nonumber
\eea
The above expression for steady state fourth moment of position agrees well with numerical simulation results, as shown in Fig.~\ref{app_fig:fourth}(c,d). At any finite reset rate, $\la \rv^4 \ra_{r}^{\rm st}$ starts from a saturation value $8[8(1+r)^2(4+r)+4(4+r)(2+3r)\Pe^2 + (8 + 3r) \Pe^4]/[r^2 (1+r)^2(4+r)]$ at small inertia and decreases with increase in inertia. It vanishes in large limit of inertia as $\sim 1/M^4$, as shown in Fig.~\ref{app_fig:fourth}(c).  In the limit of vanishing inertia $M = 0$, $\la \rv^4 \ra_{r}^{\rm st}$ behaves as $\sim 16(2+\Pe^2)^2/r^2$ at small reset rate and decrease with increase in reset rate. It vanishes as $\sim 64 /r^2 $ at large reset rate, as shown in Fig.~\ref{app_fig:fourth}(d). For any finite inertia, the small-reset-rate behavior of $\la \rv^4 \ra_{r}^{\rm st}$ is same as in overdamped case. In contrast, in the large–reset-rate limit, $\la \rv^4 \ra_{r}^{\rm st}$ vanishes as $\sim 2560/(M^4r^6)$, as shown in Fig.~\ref{app_fig:fourth}(d).

\begin{figure*}
\centering
\includegraphics[width = \linewidth]{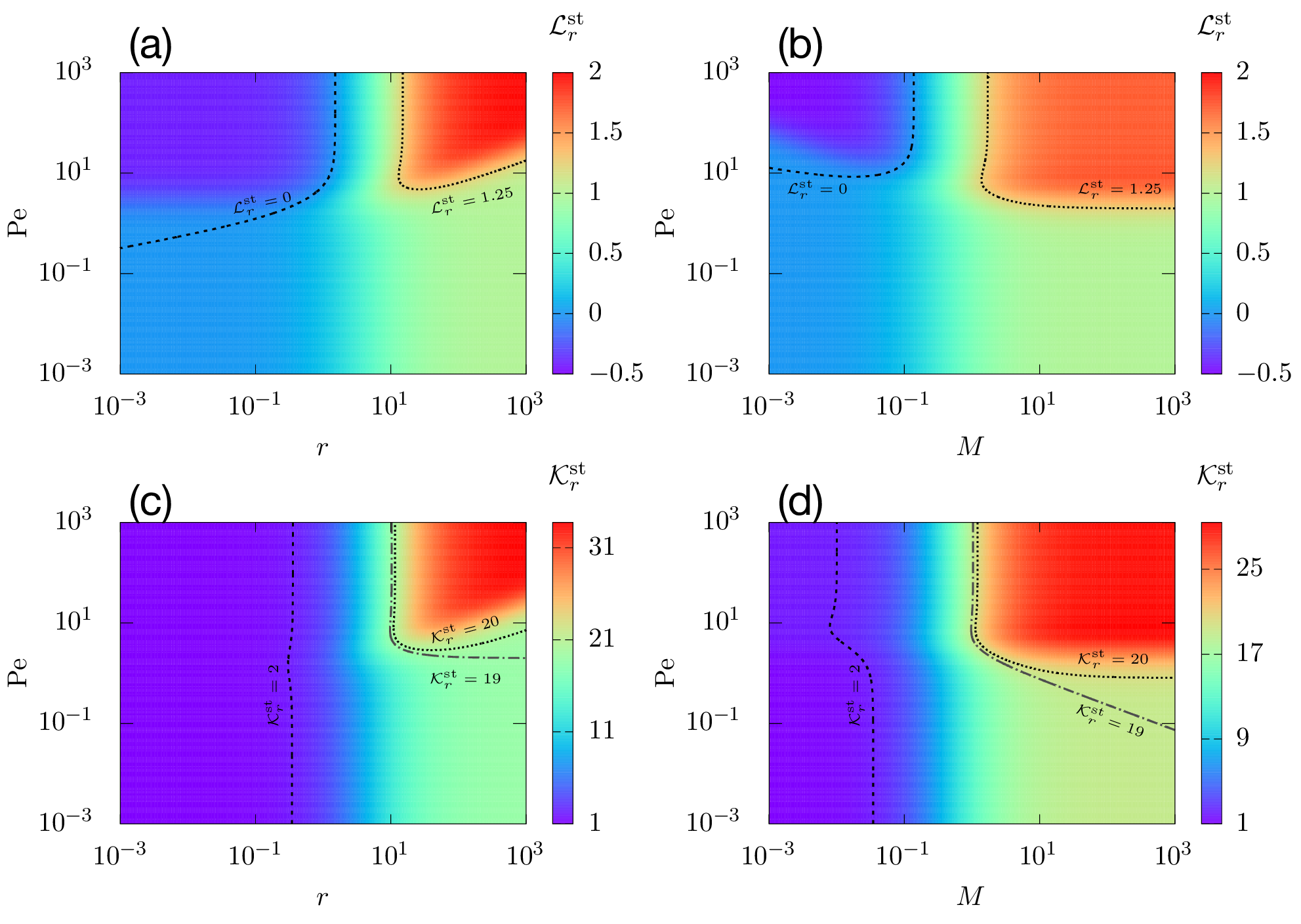}
\caption{Activity-controlled phase diagrams of the steady-state excess kurtosis under complete resetting.
Color is the analytic result of Eq.~\eqref{eq:excess_kurtosis}:
$\mathcal{L}_{r}^{\rm st}$ in (a,b), from Eqs.~\eqref{eq:MSV} and \eqref{eq:v4avg_ss},
and $\mathcal{K}_{r}^{\rm st}$ in (c,d), from Eqs.~\eqref{eq:MSD} and \eqref{eq:r4avg_ss}.
(a,c) show the $r$--$\Pe$ plane at fixed inertia $M=1$;
(b,d) show the $M$--$\Pe$ plane at fixed reset rate $r=10$.
Curves are contours of constant excess kurtosis:
$\mathcal{L}_{r}^{\rm st}=0$ and $1.25$ in (a,b),
and $\mathcal{K}_{r}^{\rm st}=1$, $10$ and $20$ in (c,d).
Increasing activity drives a crossover from weakly non-Gaussian steady states to strongly non-Gaussian regimes.}
\label{app_fig:kurtosis_rpe_mpe}
\end{figure*}

\begin{figure*}
    \centering
    \includegraphics[width=1.0\linewidth]{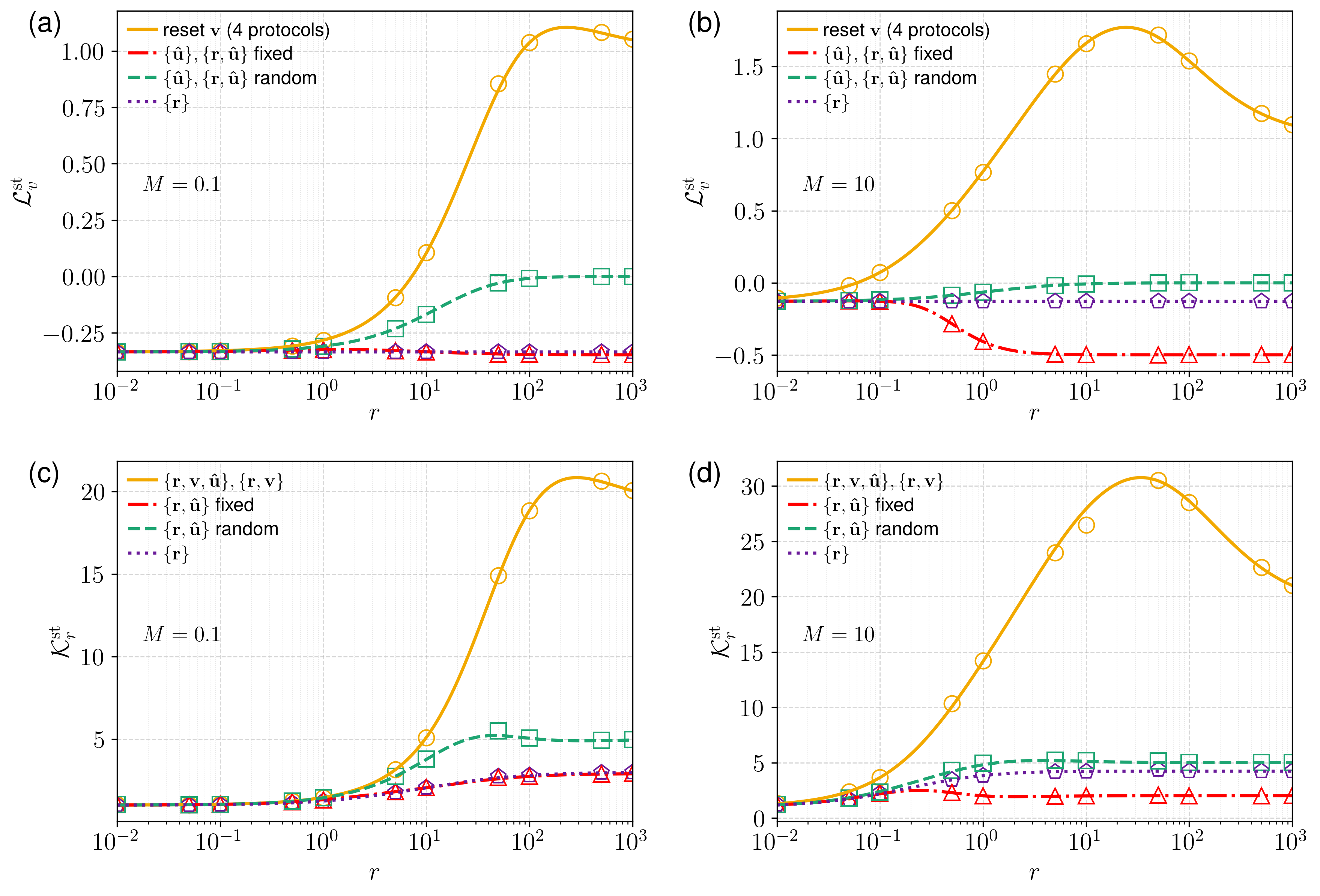}
    \caption{ Steady state excess kurtosis at fixed activity $\Pe=10$  under all possible resetting protocols; for the two orientation reset ensembles: fixed, with $\uv$ restored to $\uv_0=(1,0)$, and random, with $\uv_0$ drawn uniformly on $[-\pi,\pi]$ at each reset event.
Symbols denote simulation data.
Lines are the analytic steady-state expressions of Eq.~\eqref{eq:excess_kurtosis};
the second and fourth moments of each curve are given in Table~\ref{tab:fig9_curves}.
(a,b) Velocity excess kurtosis ${\cal L}_v^{\rm st}$ as a function of reset rate $r$
for all seven protocols at (a) low inertia $M=0.1$ and (b) high inertia $M=10$.
Solid lines are the result shared by the four velocity-resetting protocols
$\{\rv,\vv,\uv\}$, $\{\rv,\vv\}$, $\{\vv,\uv\}$ and $\{\vv\}$,
from Eqs.~\eqref{eq:MSV} and \eqref{eq:v4avg_ss}.
Dash-dotted lines are the fixed-orientation result Eq.~\eqref{eq:kurtosis_fixed}
for $\{\uv\}$ and $\{\rv,\uv\}$,
dashed lines the random-orientation result Eq.~\eqref{eq:kurtosis_random}
for the same two protocols,
and dotted lines the reset-rate-independent value Eq.~\eqref{eq:noreset_kv}
retained under $\{\rv\}$.
(c,d) Position excess kurtosis ${\cal K}_r^{\rm st}$ for the four position-resetting
protocols at (c) $M=0.1$ and (d) $M=10$.
Solid lines are the result shared by $\{\rv,\vv,\uv\}$ and $\{\rv,\vv\}$,
from Eqs.~\eqref{eq:MSD} and \eqref{eq:r4avg_ss}.
Dash-dotted lines are the fixed-orientation result for $\{\rv,\uv\}$,
dashed lines the random-orientation result for the same protocol,
both obtained from Eqs.~\eqref{eq:pos4_start}--\eqref{eq:pos4_end}
with Eqs.~\eqref{eq:msd_ru_fixed} and \eqref{eq:msd_ru_random},
and dotted lines the position-only result with Eq.~\eqref{eq:msd_pos_only}.
Protocols that preserve the position are omitted from (c,d).}
    \label{fig:kurtosis_partial_reset}
\end{figure*}

% === Table: analytic curves in Fig. 9 ===
\begin{table*}[!b]
\centering
\caption{Analytic expressions plotted as lines in Fig.~\ref{fig:kurtosis_partial_reset}.
Excess kurtosis is evaluated from Eq.~\eqref{eq:excess_kurtosis} in every case.
The three velocity-preserving position-resetting protocols require the reduced
fourth-order position system of Eqs.~\eqref{eq:pos4_start}--\eqref{eq:pos4_end},
whose five velocity inputs, Eq.~\eqref{eq:vel_inputs}, are protocol dependent.}
\label{tab:fig9_curves}
\setlength{\tabcolsep}{6pt}
\begin{tabular}{@{} l l l l @{}}
\toprule
\textbf{Panel} & \textbf{Protocols (one curve)} & \textbf{Second moment} &
\textbf{Fourth moment} \\
\midrule
(a,b) & $\{\rv,\vv,\uv\}$, $\{\rv,\vv\}$, $\{\vv,\uv\}$, $\{\vv\}$
& Eq.~\eqref{eq:MSV} & Eq.~\eqref{eq:v4avg_ss} \\
(a,b) & $\{\uv\}$, $\{\rv,\uv\}$: fixed
& Eq.~\eqref{eq:msv_fixed_u} & Eq.~\eqref{eq:v4_fixed};
$\mathcal{L}_r^{\rm st}$ in closed form, Eq.~\eqref{eq:kurtosis_fixed} \\
(a,b) & $\{\uv\}$, $\{\rv,\uv\}$: random
& Eq.~\eqref{eq:msv_random_u} & Eq.~\eqref{eq:v4_random};
$\mathcal{L}_r^{\rm st}$ in closed form, Eq.~\eqref{eq:kurtosis_random} \\
(a,b) & $\{\rv\}$ ($r$-independent)
& $2/M+\Pe^{2}/(1+M)$ & Eq.~\eqref{eq:v4_master} at $\lambda_1=1$, $\lambda_2=4$;
$\mathcal{L}_r^{\rm st}$ = Eq.~\eqref{eq:noreset_kv} \\
\midrule
(c,d) & $\{\rv,\vv,\uv\}$, $\{\rv,\vv\}$
& Eq.~\eqref{eq:MSD} & Eq.~\eqref{eq:r4avg_ss} \\
(c,d) & $\{\rv,\uv\}$: fixed
& Eq.~\eqref{eq:msd_ru_fixed}
& Eqs.~\eqref{eq:pos4_start}--\eqref{eq:pos4_end} with inputs from
Eqs.~\eqref{eq:fix_level1}--\eqref{eq:v4_fixed} \\
(c,d) & $\{\rv,\uv\}$: random
& Eq.~\eqref{eq:msd_ru_random}
& Eqs.~\eqref{eq:pos4_start}--\eqref{eq:pos4_end} with inputs from
Eqs.~\eqref{eq:second_master}--\eqref{eq:v4_master} at
$\lambda_1=1+r$, \\$$ & $$ & $$ & $\lambda_2=4+r$ \\
(c,d) & $\{\rv\}$
& Eq.~\eqref{eq:msd_pos_only}
& Eqs.~\eqref{eq:pos4_start}--\eqref{eq:pos4_end} with the same inputs at
$\lambda_1=1$, $\lambda_2=4$ \\
\bottomrule
\end{tabular}
\end{table*}

\noindent
\textbf{Position hierarchy, Eqs.~\eqref{eq:pos4_start}--\eqref{eq:pos4_end}.}
The list below closes the steady-state moments of positive degree in $\rv$ up
to fourth order, for any protocol that resets the position. Each such
observable vanishes at the reset state $\rv_0=\mathbf{0}$, whatever is done to
$\vv$ and $\uv$, so a reset contributes the single term $-r\la\psi\ra$ to its
stationary equation. Each line therefore follows from the corresponding free
identity $(s+\Gamma)\la\psi\ra_s=\la\mathcal{S}\ra_s+\la\psi\ra_0$ at the
beginning of this appendix by dropping the initial term and setting $s=r$,
which gives $\la\psi\ra=\la\mathcal{S}\ra/(r+\Gamma)$. For complete resetting
this coincides with the Final Value Theorem, Eq.~\eqref{eq:moment_resetting_SS}.
The protocol enters only through the five $\rv$-independent averages
$\la\uv\cdot\vv\ra$, $\la\vv^{2}\ra$, $\la(\uv\cdot\vv)^{2}\ra$,
$\la(\uv\cdot\vv)\vv^{2}\ra$ and $\la\vv^{4}\ra$, collected in
Eq.~\eqref{eq:vel_inputs}. They are given in Appendices~\ref{app:second_moments}
and \ref{app:fourth_moments} for complete resetting and in
Secs.~\ref{sec:pf_pos}--\ref{sec:pf_fixed} for the partial protocols.
All averages below are steady-state averages; the subscript $r$ is kept only
on $\la\rv^{4}\ra_{r}^{\rm st}$. The list is triangular.
Each source $\la\mathcal{S}\ra$ contains only the five inputs or a moment already obtained on an earlier line, so the values follow by direct substitution, with no linear system left to solve.

\bea
\la\uv\cdot\rv\ra &=& \f{\la\uv\cdot\vv\ra}{1+r}\,,\qquad
\la\vv\cdot\rv\ra = \f{\la\vv^{2}\ra+(\Pe/M)\la\uv\cdot\rv\ra}{r+1/M}\,,\qquad
\la\rv^{2}\ra = \f{2\la\vv\cdot\rv\ra}{r}\,,
\label{eq:pos4_start}
\eea

Eq.~\eqref{eq:pos4_start} rebuilds, in order, the mean orientation along the displacement, the velocity--displacement correlation, and the mean-squared displacement.
Inserting the five inputs of complete resetting recovers Eq.~\eqref{eq:MSD}.
The same three relations with the partial-protocol inputs recover Eqs.~\eqref{eq:msd_pos_only}, \eqref{eq:msd_ru_fixed}, and \eqref{eq:msd_ru_random}.
The remaining lines repeat the substitution at fourth order and end at $\la\rv^{4}\ra_{r}^{\rm st}$.

\bea
\la(\uv\cdot\vv)(\uv\cdot\rv)\ra &=&
\f{\la(\uv\cdot\vv)^{2}\ra+(\Pe/M)\la\uv\cdot\rv\ra+2\la\vv\cdot\rv\ra}{r+1/M+4}\,,\\
\la(\uv\cdot\vv)(\vv\cdot\rv)\ra &=&
\f{\la(\uv\cdot\vv)\vv^{2}\ra+(\Pe/M)\left[\la\vv\cdot\rv\ra
+\la(\uv\cdot\vv)(\uv\cdot\rv)\ra\right]+(2/M^{2})\la\uv\cdot\rv\ra}{r+2/M+1}\,,\\
\la(\uv\cdot\rv)\vv^{2}\ra &=&
\f{\la(\uv\cdot\vv)\vv^{2}\ra+(2\Pe/M)\la(\uv\cdot\vv)(\uv\cdot\rv)\ra
+(4/M^{2})\la\uv\cdot\rv\ra}{r+2/M+1}\,,\\
\la(\uv\cdot\vv)\rv^{2}\ra &=&
\f{2\la(\uv\cdot\vv)(\vv\cdot\rv)\ra+(\Pe/M)\la\rv^{2}\ra}{r+1/M+1}\,,\qquad
\la(\uv\cdot\rv)^{2}\ra =
\f{2\la(\uv\cdot\vv)(\uv\cdot\rv)\ra+2\la\rv^{2}\ra}{r+4}\,,\\
\la(\uv\cdot\rv)(\vv\cdot\rv)\ra &=&
\f{\la(\uv\cdot\vv)(\vv\cdot\rv)\ra+\la(\uv\cdot\rv)\vv^{2}\ra
+(\Pe/M)\la(\uv\cdot\rv)^{2}\ra}{r+1/M+1}\,,\\
\la(\vv\cdot\rv)\vv^{2}\ra &=&
\f{\la\vv^{4}\ra+(\Pe/M)\la(\uv\cdot\rv)\vv^{2}\ra
+(2\Pe/M)\la(\uv\cdot\vv)(\vv\cdot\rv)\ra+(8/M^{2})\la\vv\cdot\rv\ra}{r+3/M}\,,\\
\la(\uv\cdot\rv)\rv^{2}\ra &=&
\f{\la(\uv\cdot\vv)\rv^{2}\ra+2\la(\uv\cdot\rv)(\vv\cdot\rv)\ra}{r+1}\,,\\
\la(\vv\cdot\rv)^{2}\ra &=&
\f{2\la(\vv\cdot\rv)\vv^{2}\ra+(2\Pe/M)\la(\uv\cdot\rv)(\vv\cdot\rv)\ra
+(2/M^{2})\la\rv^{2}\ra}{r+2/M}\,,\\
\la\vv^{2}\rv^{2}\ra &=&
\f{2\la(\vv\cdot\rv)\vv^{2}\ra+(2\Pe/M)\la(\uv\cdot\vv)\rv^{2}\ra
+(4/M^{2})\la\rv^{2}\ra}{r+2/M}\,,
\eea

The closing relation is the steady fourth moment of the displacement.
With the inputs of complete resetting it reproduces Eq.~\eqref{eq:r4avg_ss}.
With the inputs of a partial protocol it is the moment that enters the excess kurtosis, Eq.~\eqref{eq:excess_kurtosis}, plotted in Fig.~\ref{fig:kurtosis_partial_reset}.

\bea
\la(\vv\cdot\rv)\rv^{2}\ra &=&
\f{\la\vv^{2}\rv^{2}\ra+2\la(\vv\cdot\rv)^{2}\ra
+(\Pe/M)\la(\uv\cdot\rv)\rv^{2}\ra}{r+1/M}\,,\qquad
\la\rv^{4}\ra_{r}^{\rm st} = \f{4\la(\vv\cdot\rv)\rv^{2}\ra}{r}\,.
\label{eq:pos4_end}
\eea

\section{Fourth moments and non-Gaussianity under partial resetting}
\label{sec:partial_fourth}

The exact fourth moments of Sec.~\ref{sec:phase_diagrams} were derived for
complete resetting. We now ask which of the seven protocols admit the same
treatment. The answer is governed by the same criterion that organized the
second moments, namely whether the velocity is reset, together with one new
distinction: whether the reset ensemble leaves a nonzero mean orientation.
Table~\ref{tab:fig9_curves} collects the resulting expressions and
identifies the curve each one plots in Fig.~\ref{fig:kurtosis_partial_reset}.

\subsection{Protocols that inherit the complete-resetting result}
\label{sec:pf_inherit}

For $\{\rv,\vv,\uv\}$, $\{\rv,\vv\}$, $\{\vv,\uv\}$ and $\{\vv\}$ the
velocity is restored to $\vv_0=\mathbf{0}$ at every reset, so every
$\uv_0$-dependent source term in the hierarchy of
Appendix~\ref{app:fourth_moments} vanishes and the renewal structure is
unchanged. All four therefore share the velocity fourth moment of
Eq.~\eqref{eq:v4avg_ss}, and $\{\rv,\vv,\uv\}$ and $\{\rv,\vv\}$ share the
position fourth moment of Eq.~\eqref{eq:r4avg_ss}, in both the fixed and the
random variant. The velocity kurtosis phase diagrams of Fig.~\ref{fig3} therefore apply to all four velocity-resetting protocols, whereas the position kurtosis phase diagrams of Fig.~\ref{fig4} apply to the two protocols that reset both position and velocity.

\subsection{ Master relation for velocity preserving protocols with zero-mean orientation}
\label{sec:pf_master}

For the protocols that preserve the velocity, $\{\uv\}$, $\{\rv,\uv\}$ and
$\{\rv\}$, the velocity is stationary and Eq.~\eqref{eq:v_stationary}
applies. Decomposing $\vv=\vv_{\rm th}+\vv_a$ into the thermal and
propulsion-driven parts, which are statistically independent, and using the
isotropy of $\vv_{\rm th}$ with $\la \vv_{\rm th}^2\ra = 2/M$, the odd terms
drop and
\bea
\la \vv^4 \ra_{r}^{\rm st} = \f{8}{M^{2}}
+ \f{8}{M}\,\la \vv_a^{2}\ra + \la \vv_a^{4}\ra\,.
\label{eq:v4_decomposition}
\eea
Both remaining averages are determined by the orientational statistics
alone. Whenever the reset ensemble has zero mean orientation,
$\la\uv\ra_{r}^{\rm st}=\mathbf{0}$, the angular harmonics of $\uv$ decay as
pure exponentials and the process is characterized by just two rates: the
polar rate $\lambda_{1}$ governing $\la\uv(t)\cdot\uv(t+\tau)\ra$ and the
nematic rate $\lambda_{2}$ governing $\la\cos 2[\theta(t+\tau)-\theta(t)]\ra$.
The stationary moment hierarchy then closes on
$\la\uv\cdot\vv\ra$, $\la\vv^{2}\ra$, $\la(\uv\cdot\vv)^{2}\ra$ and
$\la(\uv\cdot\vv)\vv^{2}\ra$, with the relaxation rates
$\lambda_{1}+1/M$, $2/M$, $\lambda_{2}+2/M$, $\lambda_{1}+3/M$ and $4/M$.
Introducing
\bea
D_{1}=1+M\lambda_{1}\,,\quad
D_{2}=2+M\lambda_{2}\,,\quad
D_{3}=3+M\lambda_{1}=D_{1}+2\,,
\label{eq:Ddefs}
\eea
the hierarchy is solved by
\bea
\la\uv\cdot\vv\ra_{r}^{\rm st} &=& \f{\Pe}{D_{1}}\,,\qquad
\la\vv^{2}\ra_{r}^{\rm st} = \f{2}{M}+\f{\Pe^{2}}{D_{1}}\,,
\label{eq:second_master}\\
\la(\uv\cdot\vv)^{2}\ra_{r}^{\rm st} &=& \f{1}{M}
+ \f{\Pe^{2}(D_{2}+2)}{2D_{1}D_{2}}\,,
\label{eq:uv2_master}\\
\la(\uv\cdot\vv)\vv^{2}\ra_{r}^{\rm st} &=& \f{4\Pe}{M D_{1}}
+ \f{2\Pe^{3}(D_{2}+1)}{D_{1}D_{2}D_{3}}\,,
\label{eq:uvv2_master}
\eea
and Eq.~\eqref{eq:v4_decomposition} becomes
\bea
\la \vv^4 \ra_{r}^{\rm st} = \f{8}{M^{2}}
+ \f{8\Pe^{2}}{M D_{1}}
+ \f{2\Pe^{4}\,(D_{2}+1)}{D_{1}D_{2}D_{3}}\,.
\label{eq:v4_master}
\eea
Combining Eqs.~\eqref{eq:second_master} and \eqref{eq:v4_master}, the
excess kurtosis of Eq.~\eqref{eq:excess_kurtosis} collapses to a single
ratio, the terms of order $\Pe^{0}$ and $\Pe^{2}$ cancelling identically
between numerator and denominator:
\bea
\mathcal{L}_{r}^{\rm st} =
\f{M^{2}\Pe^{4}\,(D_{1}-2D_{2})}
{D_{2}D_{3}\left(2D_{1}+M\Pe^{2}\right)^{2}}\,.
\label{eq:kurtosis_master}
\eea
Since $D_{1}-2D_{2}=-3+M(\lambda_{1}-2\lambda_{2})$ is negative whenever
$\lambda_{2}>\lambda_{1}/2$, which holds for every case below, the velocity
statistics of the velocity-preserving protocols are non-Gaussian with
strictly negative excess kurtosis. This is the opposite sign to the
velocity-resetting protocols, where $\mathcal{L}_{r}^{\rm st}$ is positive
and saturates at unity for $Mr\gg1$ (Fig.~\ref{fig3}).

Setting $\lambda_{1}=1$ and $\lambda_{2}=4$ in
Eqs.~\eqref{eq:v4_master}--\eqref{eq:kurtosis_master} recovers the free
inertial ABP results of Ref.~\cite{Patel2023}, including
Eq.~\eqref{eq:noreset_kv}, which fixes the construction.

\subsection{Position-only resetting}
\label{sec:pf_pos}

Under $\{\rv\}$ the orientation is never reset, so $\lambda_{1}=1$ and
$\lambda_{2}=4$ irrespective of $r$, and Eqs.~\eqref{eq:v4_master} and
\eqref{eq:kurtosis_master} reduce to the free-particle expressions. The
velocity fourth moment and its excess kurtosis are therefore exactly
independent of the reset rate,
\bea
\mathcal{L}_{r}^{\rm st}\big|_{\{\rv\}} =
-\f{\Pe^{4}M^{2}(3+7M)}{2(3+M)(1+2M)\left[2+(2+\Pe^{2})M\right]^{2}}\,,
\eea
consistent with the corresponding statement for the second moment in
Sec.~\ref{sec:partial_fixed}. Position resetting alone leaves every velocity
observable untouched, at all orders.

\subsection{Random-orientation resetting}
\label{sec:pf_random}

If the orientation is redrawn uniformly at each reset, a reset severs the
orientational memory completely, so each angular harmonic acquires the reset
rate additively,
\bea
\lambda_{1}=1+r\,,\qquad \lambda_{2}=4+r\,.
\label{eq:lambda_random}
\eea
Equation~\eqref{eq:v4_master} then gives, for $\{\uv\}$ and $\{\rv,\uv\}$,
\bea
\la \vv^4 \ra_{r}^{\rm st} &=& \f{8}{M^{2}}
+ \f{8\Pe^{2}}{M\left[1+M(1+r)\right]} \nn\\
&&+ \f{2\Pe^{4}\left[3+M(4+r)\right]}
{\left[1+M(1+r)\right]\left[2+M(4+r)\right]\left[3+M(1+r)\right]}\,,\nn\\
\label{eq:v4_random}
\eea
and Eq.~\eqref{eq:kurtosis_master}, using
$D_{1}-2D_{2}=-\left[3+M(7+r)\right]$, yields the closed form
\bea
\mathcal{L}_{r}^{\rm st} = -\,
\f{M^{2}\Pe^{4}\left[3+M(7+r)\right]}
{\left[2+M(4+r)\right]\left[3+M(1+r)\right]
\left[2+M(2+2r+\Pe^{2})\right]^{2}}\,.\nn\\
\label{eq:kurtosis_random}
\eea
Three features follow directly. At $r=0$ this reduces to
Eq.~\eqref{eq:noreset_kv}, as it must. It is strictly negative for all
$M$, $\Pe$ and $r$, so random-orientation resetting can never generate
heavy-tailed velocity statistics. And as a function of the reset rate it
decays monotonically to zero,
\bea
\lim_{r\to\infty}\mathcal{L}_{r}^{\rm st} = -\f{\Pe^{4}}{4Mr^{3}}
+ \mathcal{O}(r^{-4})\,,
\eea
the numerator of Eq.~\eqref{eq:kurtosis_random} growing linearly in $r$
against a quartic denominator. Frequent randomization drives the velocity
distribution towards the passive Gaussian from below, in contrast to the
velocity-resetting protocols, where $\mathcal{L}_{r}^{\rm st}$ is re-entrant
in $r$ and returns to unity (Fig.~\ref{fig3}(b)).

The dependence on inertia remains re-entrant. Equation~\eqref{eq:kurtosis_random}
vanishes as $-M^{2}\Pe^{4}/8$ for $M\to0$ and as
$-\Pe^{4}(7+r)/\{M(4+r)(1+r)\left[2+2r+\Pe^{2}\right]^{2}\}$ for
$M\to\infty$, with a negative minimum at intermediate inertia. Resetting
therefore shifts and shallows the free-particle minimum without removing it.

\subsection{Fixed-orientation resetting}
\label{sec:pf_fixed}

When the orientation is restored to a fixed $\uv_{0}$ the mean orientation
is nonzero, $\la\uv\ra_{r}^{\rm st}=r\uv_{0}/(1+r)$, the angular harmonics
no longer decay as pure exponentials, and the master relation
\eqref{eq:kurtosis_master} does not apply. The hierarchy nevertheless
closes, on an enlarged set of observables that includes the projections
onto the reset direction. Writing
\bea
a=\uv\cdot\vv\,,\qquad b=\uv_{0}\cdot\vv\,,\qquad c=\uv\cdot\uv_{0}\,,
\eea
a reset leaves $\vv$ untouched and maps $(a,b,c)\to(b,b,1)$, so that $b$
and $c$ are the images of $a$ and of unity under the reset. The
orientational generator acts as
$\mathcal{L}_{u}\psi=\nabla_{u}^{2}\psi-r[\psi(\uv)-\psi(\uv_{0})]$, with
$\nabla_u^2(\uv\cdot{\bf p})(\uv\cdot{\bf q})
=-4[(\uv\cdot{\bf p})(\uv\cdot{\bf q})-\tfrac12{\bf p}\cdot{\bf q}]$ for
any ${\bf p}$ and ${\bf q}$ independent of $\uv$. The stationary
conditions $\la\mathcal{L}\psi\ra_{r}^{\rm st}=0$ then give a triangular
system, which we list in the order in which it is solved,
\bea
\la c\ra &=& \f{r}{1+r}\,,\qquad
\la c^{2}\ra = \f{2+r}{4+r}\,,\qquad
\la b\ra = \Pe\la c\ra\,,
\label{eq:fix_level1}\\
D_{1}\la a\ra &=& \Pe + Mr\la b\ra\,,
\label{eq:fix_a}\\
D_{1}\la bc\ra &=& \Pe\la c^{2}\ra + Mr\la b\ra\,,
\label{eq:fix_bc}\\
E_{2}\la ac\ra &=& \Pe\la c\ra + M(2+r)\la b\ra\,,
\label{eq:fix_ac}\\
\la \vv^{2}\ra &=& \f{2}{M} + \Pe\la a\ra\,,\qquad
\la b^{2}\ra = \f{1}{M} + \Pe\la bc\ra\,,
\label{eq:fix_second}\\
E_{1}\la ab\ra &=& \Pe\left[\la b\ra+\la ac\ra\right]
+ \f{2\la c\ra}{M} + Mr\la b^{2}\ra\,,
\label{eq:fix_ab}\\
E_{1}\la c\vv^{2}\ra &=& 2\Pe\la ac\ra + \f{4\la c\ra}{M}
+ Mr\la \vv^{2}\ra\,,
\label{eq:fix_cv2}\\
D_{2}\la a^{2}\ra &=& 2\Pe\la a\ra + \f{2}{M}
+ 2M\la \vv^{2}\ra + Mr\la b^{2}\ra\,,
\label{eq:fix_a2}\\
3\la b\vv^{2}\ra &=& \Pe\la c\vv^{2}\ra + 2\Pe\la ab\ra
+ \f{8\la b\ra}{M}\,,
\label{eq:fix_bv2}\\
D_{3}\la a\vv^{2}\ra &=& \Pe\left[\la \vv^{2}\ra+2\la a^{2}\ra\right]
+ \f{8\la a\ra}{M} + Mr\la b\vv^{2}\ra\,,
\label{eq:fix_av2}\\
\la \vv^{4}\ra &=& \Pe\la a\vv^{2}\ra + \f{4\la \vv^{2}\ra}{M}\,,
\label{eq:fix_v4chain}
\eea
where all averages are steady-state averages at reset rate $r$, and
$D_{1}$, $D_{2}$, $D_{3}$ are the denominators of Eq.~\eqref{eq:Ddefs}
evaluated at $\lambda_{1}=1+r$ and $\lambda_{2}=4+r$, supplemented by the
two factors generated by the projected observables,
\bea
E_{1}=D_{1}+1=2+M(1+r)\,,\qquad E_{2}=D_{2}-1=1+M(4+r)\,.
\label{eq:Edefs}
\eea
The first three levels are explicit,
\bea
\la a\ra &=& \f{\Pe\left[(1+r)+Mr^{2}\right]}{(1+r)D_{1}}\,,\qquad
\la ac\ra = \f{\Pe\,r\left[1+M(2+r)\right]}{(1+r)E_{2}}\,,\nn\\
\la bc\ra &=& \f{\Pe\left[(1+r)(2+r)+Mr^{2}(4+r)\right]}
{(1+r)(4+r)D_{1}}\,,
\label{eq:fix_explicit}
\eea
and Eq.~\eqref{eq:fix_second} with the first of these reproduces the MSV of
Eq.~\eqref{eq:msv_fixed_u}, which fixes the construction at second order.
Back substitution through Eqs.~\eqref{eq:fix_ab}--\eqref{eq:fix_v4chain}
gives the velocity fourth moment in closed form. Retaining the thermal
decomposition of Eq.~\eqref{eq:v4_decomposition},
\bea
\la \vv^{4}\ra_{r}^{\rm st} &=& \f{8}{M^{2}}
+ \f{8\Pe^{2}\left[(1+r)+Mr^{2}\right]}{M(1+r)D_{1}}\nn\\
&&+\f{\Pe^{4}\,\mathcal{N}}{(1+r)(4+r)D_{1}E_{1}D_{3}E_{2}D_{2}}\,,
\label{eq:v4_fixed}
\eea
with
\bea
\mathcal{N} &=& M^{5}r^{4}(r+4)^{3}
+ M^{4}r^{2}(r+4)^{2}\left(9r^{2}+22r+16\right)\nn\\
&&+ M^{3}(r+4)\left(31r^{4}+152r^{3}+211r^{2}+80r+32\right)\nn\\
&&+ M^{2}\left(51r^{4}+380r^{3}+889r^{2}+768r+384\right)\nn\\
&&+ M\left(40r^{3}+246r^{2}+438r+280\right)\nn\\
&&+ 12r^{2}+60r+48\,.
\eea
The fixed ensemble therefore carries the same five relaxation rates as the
random one, but the reset direction adds the factors $E_{1}$, $E_{2}$ and
$(1+r)(4+r)$ to the denominator, the last two coming from the nematic
projections $\la ac\ra$ and $\la bc\ra$.

Combining Eqs.~\eqref{eq:msv_fixed_u} and \eqref{eq:v4_fixed}, the excess
kurtosis of Eq.~\eqref{eq:excess_kurtosis} is
\bea
\mathcal{L}_{r}^{\rm st} = -\,
\f{M^{2}\Pe^{4}\,\mathcal{Q}}
{2(4+r)\,E_{1}D_{3}E_{2}D_{2}\,\mathcal{S}^{2}}\,,
\label{eq:kurtosis_fixed}
\eea
with
\bea
\mathcal{S} &=& 2(1+r)D_{1} + M\Pe^{2}\left[(1+r)+Mr^{2}\right]
\;=\;M(1+r)D_{1}\la \vv^{2}\ra_{r}^{\rm st}\,,\nn\\
\mathcal{Q} &=& M^{6}r^{9} + \left(14M^{6}+10M^{5}\right)r^{8}
+ \left(73M^{6}+125M^{5}+40M^{4}\right)r^{7}\nn\\
&&+ \left(172M^{6}+553M^{5}+441M^{4}+82M^{3}\right)r^{6}\nn\\
&&+ \left(176M^{6}+982M^{5}+1680M^{4}+789M^{3}+91M^{2}\right)r^{5}\nn\\
&&+ \left(64M^{6}+512M^{5}+2590M^{4}+2595M^{3}+757M^{2}
+52M\right)r^{4}\nn\\
&&+ \left(-32M^{5}+1623M^{4}+3669M^{3}+2103M^{2}+370M+12\right)r^{3}\nn\\
&&+ \left(636M^{4}+2697M^{3}+2619M^{2}+816M+72\right)r^{2}\nn\\
&&+ \left(816M^{3}+1710M^{2}+826M+108\right)r\nn\\
&&+ 224M^{3}+600M^{2}+328M+48\,.
\eea
Four properties follow. At $r=0$, Eq.~\eqref{eq:kurtosis_fixed} reduces to the free inertial ABP result, Eq.~\eqref{eq:noreset_kv}, and at $\Pe=0$ it vanishes, since a prescribed heading is immaterial for a passive particle.
It is strictly negative, as in the random ensemble, so no velocity-preserving protocol generates heavy-tailed velocity statistics.
In the strong-resetting limit the orientation is frozen at $\uv_{0}$
between resets and $\vv$ becomes Gaussian with mean $\Pe\uv_{0}$ and
variance $1/M$ per component, giving
\bea
\lim_{r\to\infty}\la \vv^{4}\ra_{r}^{\rm st}
&=& \Pe^{4}+\f{8\Pe^{2}}{M}+\f{8}{M^{2}}\,,\nn\\
\mathcal{L}_{r}^{\rm st} &=& -\f{M^{2}\Pe^{4}}{2\left(2+M\Pe^{2}\right)^{2}}
+ \f{4M^{2}\Pe^{4}}{r\left(2+M\Pe^{2}\right)^{3}}
+ \mathcal{O}(r^{-2})\,,
\label{eq:kurtosis_fixed_larger}
\eea
so that $\mathcal{L}_{r}^{\rm st}>-1/2$ throughout, the bound being
saturated only for $M\Pe^{2}\to\infty$. This is the opposite extreme to
the velocity-resetting protocols, where $\mathcal{L}_{r}^{\rm st}$ returns to unity (Fig.~\ref{fig3}).

Finally, the dependence on the reset rate is non-monotonic for all $M$ and
$\Pe$. Expanding Eq.~\eqref{eq:kurtosis_fixed} at small $r$,
\bea
\mathcal{L}_{r}^{\rm st} = \mathcal{L}_{0}^{\rm st}
+ \f{M^{3}\Pe^{4}\left[31M^{3}\Pe^{2}+174M^{3}+36M^{2}\Pe^{2}+574M^{2}
+9M\Pe^{2}+426M+90\right]}
{4(3+M)^{2}(1+2M)^{2}\left[2+(2+\Pe^{2})M\right]^{3}}\,r
+ \mathcal{O}(r^{2})\,,\nn\\
\eea
whose linear coefficient is manifestly positive, whereas the difference
between the two endpoints,
\bea
\mathcal{L}_{0}^{\rm st}-\mathcal{L}_{\infty}^{\rm st} =
\f{M^{3}\Pe^{4}\left[M^{3}(\Pe^{2}+2)^{2}+18M^{2}\Pe^{2}+22M^{2}
+6M\Pe^{2}+38M+12\right]}
{(3+M)(1+2M)\left[2+(2+\Pe^{2})M\right]^{2}
\left(2+M\Pe^{2}\right)^{2}}\,,\nn\\
\eea
is manifestly positive as well. The excess kurtosis therefore rises from
its free-particle value, passes through a maximum at intermediate $r$, and
falls to the frozen-heading value of
Eq.~\eqref{eq:kurtosis_fixed_larger}, which lies below the starting point.
Weak resetting to a fixed heading makes the velocity statistics more
Gaussian by interrupting the orientational memory, while strong resetting
makes them less Gaussian by imposing a persistent drift. The random
ensemble shows neither feature, decaying monotonically to zero
(Sec.~\ref{sec:pf_random}), so the two variants differ at fourth order as
sharply as they do at second.

\subsection{Position statistics}
\label{sec:pf_position}

The renewal route of Sec.~\ref{sec:partial_msd} does not extend to
$\la\rv^{4}\ra$, which would require the four-time stationary velocity
correlator. The stationary hierarchy closes regardless, by the argument given
below Eq.~\eqref{eq:msd_pos_only}: under any protocol that resets the
position, every observable of degree at least one in $\rv$ is annihilated by
the reset and acquires the single extra term $-r\la\psi\ra_{r}^{\rm st}$ in
its stationary equation. The position sector therefore obeys the
Laplace-space relations of Appendix~\ref{app:fourth_moments} evaluated at
$s\to r$, with vanishing initial terms for $\rv_{0}=\mathbf{0}$, the only
protocol dependence entering through the five $\rv$-independent averages
\bea
\left\{\la\uv\cdot\vv\ra,\ \la\vv^{2}\ra,\ \la(\uv\cdot\vv)^{2}\ra,\
\la(\uv\cdot\vv)\vv^{2}\ra,\ \la\vv^{4}\ra\right\}_{r}^{\rm st}\,,
\label{eq:vel_inputs}
\eea
which are supplied by Sec.~\ref{sec:pf_pos} for $\{\rv\}$,
Sec.~\ref{sec:pf_random} for $\{\rv,\uv\}$ with a random reset direction, and
Sec.~\ref{sec:pf_fixed} for $\{\rv,\uv\}$ with a fixed one. The resulting
linear system is triangular and is solved by back substitution. At second
order it gives Eqs.~\eqref{eq:msd_ru_fixed} and \eqref{eq:msd_ru_random} of
the main text, which closes the last gap in
Table~\ref{tab:protocol_msv}; the twelve equations at fourth order are those
of Appendix~\ref{app:fourth_moments} under the same substitution.

At fourth order, the two velocity-preserving position-resetting protocols, $\{\rv\}$ and $\{\rv,\uv\}$, share the following compact passive limit:
\bea
\la \rv^{4}\ra_{r}^{\rm st}\Big|_{\Pe=0} &=&
\f{64\left(2+3Mr\right)}{r^{2}(1+Mr)^{2}(2+Mr)}\,,\nn\\
\mathcal{K}_{r}^{\rm st}\Big|_{\Pe=0} &=& \f{2+5Mr}{2+Mr}\,,
\label{eq:K_passive_partial}
\eea
to be compared with Eq.~\eqref{eq:position_excess_kurtosis_underdamped_Brownian}
for the complete protocol. Both interpolate between the overdamped value
$1$ at $Mr\to0$ and a constant at $Mr\to\infty$, but the constant is $5$
when the velocity is preserved and $19$ when it is reset. The two values
have a direct reading. For $Mr\gg1$ an excursion is short compared with
the momentum relaxation time, so a particle that keeps its velocity moves
ballistically, $\rv\simeq\vv t$ with $\vv$ Gaussian and $t$ exponential,
and $\mathcal{K}_{r}^{\rm st}\to\la\vv^{4}\ra\la t^{4}\ra/
[2\la\vv^{2}\ra^{2}\la t^{2}\ra^{2}]-1=5$; a particle restarted from rest
instead rebuilds its velocity diffusively, $\la\rv^{2}\ra\propto t^{3}$ at
fixed $t$, and the same average gives
$\la t^{6}\ra/\la t^{3}\ra^{2}-1=19$. The contrast between these two saturation values is thus a direct signature of velocity resetting.

For $\Pe>0$ the solution of the fourth-order system is closed but
algebraically unwieldy, and we quote its limits. All four protocols share
the overdamped limit
\bea
\lim_{M\to0}\mathcal{K}_{r}^{\rm st} =
\text{Eq.~\eqref{eq:position_excess_kurtosis_overdamped_ABP}}\,,
\eea
and the coincidence is exact to all orders, not merely at fourth: when
$\vv$ is slaved to $\uv$ the displacement accumulated in an excursion is
built from orientational increments alone, whose law is independent of the
orientation at the start of the excursion by rotational invariance, so
resetting $\uv$ or leaving it untouched cannot alter the position
statistics. This strengthens the argument of
Sec.~\ref{sec:partial_expt}: overdamped realizations of stochastic
resetting are blind to the protocol hierarchy at every order, not only in
the MSD. At small reset rate all protocols approach the overdamped
Brownian baseline linearly, $\mathcal{K}_{r}^{\rm st}=1+c_{1}r
+\mathcal{O}(r^{2})$, with
\bea
c_{1}\Big|_{\{\rv,\uv\}} &=& 2M
+ \f{\Pe^{2}\left(16+\Pe^{2}\right)}{4\left(2+\Pe^{2}\right)^{2}}\,,\nn\\
c_{1}\Big|_{\{\rv\}} &=& \f{2M^{2}}{1+M}
+ \f{M\left(\Pe^{4}+32\Pe^{2}+32\right)+\Pe^{4}+16\Pe^{2}}
{4(1+M)\left(2+\Pe^{2}\right)^{2}}\,,
\eea
the first holding for both reset ensembles. The purely inertial
contribution is $2M$ here against $3M$ for complete resetting
(Sec.~\ref{sec:phase_diagrams}), the difference again being the ballistic
excursion that velocity resetting removes. At strong resetting the
saturation values separate,
\bea
\lim_{r\to\infty}\mathcal{K}_{r}^{\rm st}\Big|_{\{\rv,\uv\},\,\rm random}
&=& 5\,,\nn\\
\lim_{r\to\infty}\mathcal{K}_{r}^{\rm st}\Big|_{\{\rv,\uv\},\,\rm fixed}
&=& \f{2\left(M^{2}\Pe^{4}+10M\Pe^{2}+10\right)}
{\left(2+M\Pe^{2}\right)^{2}}\,,
\label{eq:K_larger_partial}
\eea
the latter decreasing from $5$ at $\Pe=0$ to $2$ as $M\Pe^{2}\to\infty$,
which is the value $\la t^{4}\ra/[2\la t^{2}\ra^{2}]-1$ for a purely
ballistic excursion along the prescribed heading. For $\{\rv\}$ the limit
retains a full dependence on both parameters,
\bea
&&\lim_{r\to\infty}\mathcal{K}_{r}^{\rm st}\Big|_{\{\rv\}} =
\f{2}{(3+M)(1+2M)\left[2+M(2+\Pe^{2})\right]^{2}}\nn\\
&&\times\Big[5M^{4}\Pe^{4}+20M^{4}\Pe^{2}+20M^{4}
+7M^{3}\Pe^{4}+90M^{3}\Pe^{2}\nn\\
&&+110M^{3}+3M^{2}\Pe^{4}+100M^{2}\Pe^{2}+190M^{2}
+30M\Pe^{2}\nn\\
&&+130M+30\Big]\,,
\eea
which tends to $2(5M^{2}+7M+3)/(2M^{2}+7M+3)$ at large activity and
therefore interpolates between $2$ and $5$. Every velocity-preserving
protocol is thus confined to $1\le\mathcal{K}_{r}^{\rm st}\lesssim 6$,
against the saturation value $19$ of the complete protocol
(Table~\ref{tab:kurtosis_scaling}).

Evaluation of the closed forms shows one further difference between the
protocols. Under $\{\rv\}$ the position excess kurtosis rises
monotonically with the reset rate, whereas under $\{\rv,\uv\}$ it
overshoots its saturation value before returning to it, reaching
$\mathcal{K}_{r}^{\rm st}\simeq5.8$ at $M=1$, $\Pe=50$, $r=10$ in the
random ensemble. Among the velocity-preserving protocols, re-entrance in
$r$ therefore requires an orientational reset channel, in contrast to the
complete protocol, where it survives with the orientation untouched since
$\{\rv,\vv\}$ and $\{\rv,\vv,\uv\}$ are degenerate.

\bibliography{reference}

@article{Hermans1952,
  author  = {J. J. Hermans and R. Ullman},
  title   = {The statistics of stiff chains, with applications to light scattering},
  journal = {Physica},
  volume  = {18},
  number  = {11},
  pages   = {951--971},
  year    = {1952},
  publisher = {Elsevier},
  doi     = {10.1016/S0031-8914(52)80231-9}
}

@article{Shee2025,
  author  = {Shee, Amir},
  title   = {Active Brownian particle under stochastic position and orientation resetting in a harmonic trap},
  journal = {Journal of Physics Communications},
  year    = {2025},
  volume  = {9},
  number  = {2},
  pages   = {025003},
  month   = {Feb},
  doi     = {10.1088/2399-6528/adb36e},
  url     = {https://iopscience.iop.org/article/10.1088/2399-6528/adb36e}
}

@article{Boyer2024,
  title = {Active particle in one dimension subjected to resetting with memory},
  author = {Boyer, Denis and Majumdar, Satya N.},
  journal = {Phys. Rev. E},
  volume = {109},
  issue = {5},
  pages = {054105},
  numpages = {13},
  year = {2024},
  month = {May},
  publisher = {American Physical Society},
  doi = {10.1103/PhysRevE.109.054105},
  url = {https://link.aps.org/doi/10.1103/PhysRevE.109.054105}
}

@article{Scholz2018,
  author  = {Scholz, Christoph and Jahanshahi, Soroush and Ldov, Alexander and L{\"o}wen, Hartmut},
  title   = {Inertial delay of self-propelled particles},
  journal = {Nature Communications},
  year    = {2018},
  volume  = {9},
  pages   = {5156},
  doi     = {10.1038/s41467-018-07596-x},
  publisher = {Nature Publishing Group}
}

@article{Koumakis2016,
  author       = {Koumakis, Nikolaos and Gnoli, Alessandro and Maggi, Claudio and Puglisi, Andrea and Di Leonardo, Roberto},
  title        = {Mechanism of self-propulsion in 3D-printed active granular particles},
  journal      = {New Journal of Physics},
  volume       = {18},
  number       = {11},
  pages        = {113046},
  year         = {2016},
  doi          = {10.1088/1367-2630/18/11/113046},
  url          = {https://iopscience.iop.org/article/10.1088/1367-2630/18/11/113046},
  publisher    = {IOP Publishing}
}

@article{Dauchot2019,
  title = {Dynamics of a Self-Propelled Particle in a Harmonic Trap},
  author = {Dauchot, Olivier and D\'emery, Vincent},
  journal = {Phys. Rev. Lett.},
  volume = {122},
  issue = {6},
  pages = {068002},
  numpages = {5},
  year = {2019},
  month = {Feb},
  publisher = {American Physical Society},
  doi = {10.1103/PhysRevLett.122.068002},
  url = {https://link.aps.org/doi/10.1103/PhysRevLett.122.068002}
}

@article{Palacci2013,
  author       = {Palacci, J{\'e}r{\'e}mie and Sacanna, S. and Steinberg, A.~P. and Pine, D.~J. and Chaikin, P.~M.},
  title        = {Living crystals of light-activated colloidal surfers},
  journal      = {Science},
  volume       = {339},
  number       = {6122},
  pages        = {936--940},
  year         = {2013},
  doi          = {10.1126/science.1230020},
  url          = {https://www.science.org/doi/10.1126/science.1230020},
}

@article{Caprini2022,
  title        = {Role of rotational inertia for collective phenomena in active matter},
  author       = {Caprini, Lorenzo and Gupta, Rahul Kumar and Löwen, Hartmut},
  journal      = {Physical Chemistry Chemical Physics},
  year         = {2022},
  volume       = {24},
  number       = {--},
  pages        = {24910--24916},
  doi          = {10.1039/D2CP02940E},
}

@article{Olsen2024,
  title        = {Dynamics of inertial particles under velocity resetting},
  author       = {Kristian Stølevik Olsen and Hartmut Löwen},
  journal      = {Journal of Statistical Mechanics: Theory and Experiment},
  volume       = {},
  number       = {3},
  pages        = {033210},
  year         = {2024},
  publisher    = {IOP Publishing},
  doi          = {10.1088/1742-5468/ad319a},
  url          = {https://doi.org/10.1088/1742-5468/ad319a}
}

@article{Kumar2020,
  title = {Active Brownian motion in two dimensions under stochastic resetting},
  author = {Kumar, Vijay and Sadekar, Onkar and Basu, Urna},
  journal = {Phys. Rev. E},
  volume = {102},
  issue = {5},
  pages = {052129},
  numpages = {19},
  year = {2020},
  month = {Nov},
  publisher = {American Physical Society},
  doi = {10.1103/PhysRevE.102.052129},
  url = {https://link.aps.org/doi/10.1103/PhysRevE.102.052129}
}

@article{Gupta2019,
  author       = {Gupta, Deepak},
  title        = {Stochastic resetting in underdamped Brownian motion},
  journal      = {Journal of Statistical Mechanics: Theory and Experiment},
  year         = {2019},
  number       = {3},
  pages        = {033212},
  doi          = {10.1088/1742-5468/ab054a},
}

@article{Shee2020,
  author       = {A. Shee and A. Dhar and D. Chaudhuri},
  title        = {Active Brownian particles: mapping to equilibrium polymers and exact computation of moments},
  journal      = {Soft Matter},
  year         = {2020},
  volume       = {16},
  pages        = {4776--4787},
  doi          = {10.1039/D0SM00367K},
}

@article{Uhlenbeck1930,
  author    = {Uhlenbeck, G. E. and Ornstein, L. S.},
  title     = {On the Theory of the Brownian Motion},
  journal   = {Physical Review},
  volume    = {36},
  pages     = {823--841},
  year      = {1930},
  doi       = {10.1103/PhysRev.36.823}
}

@article{Bressloff2020,
  author  = {Paul C. Bressloff},
  title   = {Modeling active cellular transport as a directed search process with stochastic resetting and delays},
  journal = {Journal of Physics A: Mathematical and Theoretical},
  volume  = {53},
  number  = {35},
  pages   = {355001},
  year    = {2020},
  doi     = {10.1088/1751-8121/ab9fb7},
  publisher = {IOP Publishing}
}

@incollection{Basu2024,
  author    = {Urna Basu and Sanjib Sabhapandit and Ion Santra},
  title     = {Target Search by Active Particles},
  booktitle = {Target Search Problems},
  editor    = {Denis Grebenkov and Ralf Metzler and Gleb Oshanin},
  publisher = {Springer},
  address   = {Cham},
  pages     = {463--487},
  year      = {2024},
  doi       = {10.1007/978-3-031-67802-8_19},
  isbn      = {978-3-031-67802-8}
}

@article{Evans2018,
  author    = {Martin R. Evans and Satya N. Majumdar},
  title     = {Run and tumble particle under resetting: a renewal approach},
  journal   = {Journal of Physics A: Mathematical and Theoretical},
  volume    = {51},
  number    = {47},
  pages     = {475003},
  year      = {2018},
  publisher = {IOP Publishing},
  doi       = {10.1088/1751-8121/aae74e}
}

@article{Ghosh2026,
  title = {Anisotropic active Brownian particle in two dimensions under stochastic resetting},
  author = {Ghosh, Anirban and Mandal, Sudipta and Chaki, Subhasish},
  journal = {Phys. Rev. E},
  volume = {113},
  issue = {1},
  pages = {014142},
  numpages = {19},
  year = {2026},
  month = {Jan},
  publisher = {American Physical Society},
  doi = {10.1103/11f6-srsx},
  url = {https://link.aps.org/doi/10.1103/11f6-srsx}
}

@article{Bressloff2020PRE,
  title = {Occupation time of a run-and-tumble particle with resetting},
  author = {Bressloff, Paul C.},
  journal = {Phys. Rev. E},
  volume = {102},
  issue = {4},
  pages = {042135},
  numpages = {12},
  year = {2020},
  month = {Oct},
  publisher = {American Physical Society},
  doi = {10.1103/PhysRevE.102.042135},
  url = {https://link.aps.org/doi/10.1103/PhysRevE.102.042135}
}

@article{Loewen2020Inertial,
  author       = {L{\"o}wen, Hartmut},
  title        = {Inertial effects of self-propelled particles: From active Brownian to active Langevin motion},
  journal      = {The Journal of Chemical Physics},
  volume       = {152},
  number       = {4},
  pages        = {040901},
  year         = {2020},
  doi          = {10.1063/1.5134455},
  url          = {https://doi.org/10.1063/1.5134455}
}

@article{Chaudhuri2021,
  title = {Active Brownian particle in harmonic trap: exact computation of moments, and re-entrant transition},
  author = {Chaudhuri, Debasish and Dhar, Abhishek},
  journal = {Journal of Statistical Mechanics: Theory and Experiment},
  volume = {2021},
  number = {1},
  pages = {013207},
  year = {2021},
  doi = {10.1088/1742-5468/abd031},
  url = {https://iopscience.iop.org/article/10.1088/1742-5468/abd031}
}

@article{Faisant2021OptimalMFPTResetting,
  author    = {Faisant, Félix and Besga, Benjamin and Petrosyan, Artyom and Ciliberto, Sergio and Majumdar, Satya N.},
  title     = {Optimal mean first-passage time of a Brownian searcher with resetting in one and two dimensions: experiments, theory and numerical tests},
  journal   = {Journal of Statistical Mechanics: Theory and Experiment},
  year      = {2021},
  number    = {11},
  pages     = {113203},
  doi       = {10.1088/1742-5468/AC2CC7},
  url       = {https://doi.org/10.1088/1742-5468/AC2CC7}
}

@misc{Sinha2025,
  title         = {Optimal navigation in a noisy environment},
  author        = {Sinha, Abhijit and Jangid, Sandeep and Sadhu, Tridib and Ghosh, Shankar},
  year          = {2025},
  eprint        = {2512.20336},
  archivePrefix = {arXiv},
  primaryClass  = {cond-mat.stat-mech},
  url           = {https://arxiv.org/abs/2512.20336}
}

@article{Reuveni2014SubstrateUnbinding,
  author  = {Reuveni, Shlomi and Urbakh, Michael and Klafter, Joseph},
  title   = {Role of substrate unbinding in Michaelis--Menten enzymatic reactions},
  journal = {Proceedings of the National Academy of Sciences of the United States of America},
  volume  = {111},
  number  = {12},
  pages   = {4391--4396},
  year    = {2014},
  doi     = {10.1073/pnas.1318122111},
  url     = {https://doi.org/10.1073/pnas.1318122111}
}

@article{Robin2018SingleMolecule,
  author  = {Robin, Tal and Reuveni, Shlomi and Urbakh, Michael},
  title   = {Single-molecule theory of enzymatic inhibition},
  journal = {Nature Communications},
  volume  = {9},
  number  = {1},
  pages   = {779},
  year    = {2018},
  doi     = {10.1038/s41467-018-02995-6},
  url     = {https://doi.org/10.1038/s41467-018-02995-6}
}

@article{Rotbart2015,
  title = {Michaelis-Menten reaction scheme as a unified approach towards the optimal restart problem},
  author = {Rotbart, Tal and Reuveni, Shlomi and Urbakh, Michael},
  journal = {Phys. Rev. E},
  volume = {92},
  issue = {6},
  pages = {060101},
  numpages = {6},
  year = {2015},
  month = {Dec},
  publisher = {American Physical Society},
  doi = {10.1103/PhysRevE.92.060101},
  url = {https://link.aps.org/doi/10.1103/PhysRevE.92.060101}
}

@article{Roldan2016,
  title = {Stochastic resetting in backtrack recovery by RNA polymerases},
  author = {Rold\'an, \'Edgar and Lisica, Ana and S\'anchez-Taltavull, Daniel and Grill, Stephan W.},
  journal = {Phys. Rev. E},
  volume = {93},
  issue = {6},
  pages = {062411},
  numpages = {10},
  year = {2016},
  month = {Jun},
  publisher = {American Physical Society},
  doi = {10.1103/PhysRevE.93.062411},
  url = {https://link.aps.org/doi/10.1103/PhysRevE.93.062411}
}

@article{Tal-Friedman2020Experimental,
  author  = {Tal-Friedman, Ofir and Pal, Arnab and Sekhon, Amandeep and Reuveni, Shlomi and Roichman, Yael},
  title   = {Experimental Realization of Diffusion with Stochastic Resetting},
  journal = {The Journal of Physical Chemistry Letters},
  volume  = {11},
  number  = {17},
  pages   = {7350–7355},
  year    = {2020},
  doi     = {10.1021/acs.jpclett.0c02122},
  url     = {https://doi.org/10.1021/acs.jpclett.0c02122}
}

@article{Altshuler2024,
  title = {Environmental memory facilitates search with home returns},
  author = {Altshuler, Amy and Bonomo, Ofek Lauber and Gorohovsky, Nicole and Marchini, Shany and Rosen, Eran and Tal-Friedman, Ofir and Reuveni, Shlomi and Roichman, Yael},
  journal = {Phys. Rev. Res.},
  volume = {6},
  issue = {2},
  pages = {023255},
  numpages = {9},
  year = {2024},
  month = {Jun},
  publisher = {American Physical Society},
  doi = {10.1103/PhysRevResearch.6.023255},
  url = {https://link.aps.org/doi/10.1103/PhysRevResearch.6.023255}
}

@article{PSPal2024,
  title = {Active motion can be beneficial for target search with resetting in a thermal environment},
  author = {Pal, P. S. and Park, Jong-Min and Pal, Arnab and Park, Hyunggyu and Lee, Jae Sung},
  journal = {Phys. Rev. E},
  volume = {110},
  issue = {5},
  pages = {054124},
  numpages = {10},
  year = {2024},
  month = {Nov},
  publisher = {American Physical Society},
  doi = {10.1103/PhysRevE.110.054124},
  url = {https://link.aps.org/doi/10.1103/PhysRevE.110.054124}
}

@article{Nguyen2022InertialAOUP,
  author  = {Nguyen, G. H. Philipp and Wittmann, René and Löwen, Hartmut},
  title   = {Active Ornstein–Uhlenbeck model for self-propelled particles with inertia},
  journal = {Journal of Physics: Condensed Matter},
  volume  = {34},
  number  = {3},
  pages   = {035101},
  year    = {2022},
  doi     = {10.1088/1361-648X/ac2c3f},
  url     = {https://doi.org/10.1088/1361-648X/ac2c3f}
}

@article{Patel2023,
  author  = {Patel, Manish and Chaudhuri, Debasish},
  title   = {Exact moments and re-entrant transitions in the inertial dynamics of active Brownian particles},
  journal = {New Journal of Physics},
  year    = {2023},
  volume  = {25},
  number  = {12},
  pages   = {123048},
  month   = dec,
  doi     = {10.1088/1367-2630/ad1538},
  url     = {https://iopscience.iop.org/article/10.1088/1367-2630/ad1538}
}

@article{Patel2024,
  author       = {Manish Patel and Debasish Chaudhuri},
  title        = {Exact moments for trapped active particles: inertial impact on steady-state properties and re-entrance},
  journal      = {New Journal of Physics},
  year         = {2024},
  volume       = {26},
  number       = {7},
  pages        = {073048},
  doi          = {10.1088/1367-2630/ad6349},
}

@article{Marchetti2013,
  title = {Hydrodynamics of soft active matter},
  author = {Marchetti, M. C. and Joanny, J. F. and Ramaswamy, S. and Liverpool, T. B. and Prost, J. and Rao, Madan and Simha, R. Aditi},
  journal = {Rev. Mod. Phys.},
  volume = {85},
  issue = {3},
  pages = {1143--1189},
  numpages = {0},
  year = {2013},
  month = {Jul},
  publisher = {American Physical Society},
  doi = {10.1103/RevModPhys.85.1143},
  url = {https://link.aps.org/doi/10.1103/RevModPhys.85.1143}
}

@article{Bechinger2016,
  title = {Active particles in complex and crowded environments},
  author = {Bechinger, Clemens and Di Leonardo, Roberto and L{\"o}wen, Hartmut and Reichhardt, Charles and Volpe, Giorgio and Volpe, Giovanni},
  journal = {Reviews of Modern Physics},
  volume = {88},
  number = {4},
  pages = {045006},
  year = {2016},
  doi = {10.1103/RevModPhys.88.045006},
  url = {https://doi.org/10.1103/RevModPhys.88.045006}
}

@article{Evans2020,
  title = {Stochastic Resetting and Applications},
  author = {Evans, M. R. and Majumdar, S. N. and Schehr, G.},
  journal = {Journal of Physics A: Mathematical and Theoretical},
  volume = {53},
  pages = {193001},
  year = {2020},
  doi = {10.1088/1751-8121/ab7cfe},
  url = {https://doi.org/10.1088/1751-8121/ab7cfe}
}

@article{Caprini2021,
  author       = {Caprini, Lorenzo and Marini Bettolo Marconi, Umberto},
  title        = {Inertial self-propelled particles},
  journal      = {The Journal of Chemical Physics},
  volume       = {154},
  number       = {2},
  pages        = {024902},
  year         = {2021},
  doi          = {10.1063/5.0030940},
}

@article{Paramanick2024,
  title = {Uncovering Universal Characteristics of Homing Paths using Foraging Robots},
  author = {Paramanick, Somnath and Biswas, Arup and Soni, Harsh and Pal, Arnab and Kumar, Nitin},
  journal = {PRX Life},
  volume = {2},
  issue = {3},
  pages = {033007},
  numpages = {13},
  year = {2024},
  month = {Aug},
  publisher = {American Physical Society},
  doi = {10.1103/PRXLife.2.033007},
  url = {https://link.aps.org/doi/10.1103/PRXLife.2.033007}
}

@article{Sandoval2020,
  title = {Pressure and diffusion of active matter with inertia},
  author = {Sandoval, Mario},
  journal = {Phys. Rev. E},
  volume = {101},
  issue = {1},
  pages = {012606},
  numpages = {8},
  year = {2020},
  month = {Jan},
  publisher = {American Physical Society},
  doi = {10.1103/PhysRevE.101.012606},
  url = {https://link.aps.org/doi/10.1103/PhysRevE.101.012606}
}

@article{Baouche2024,
  author    = {Yanis Baouche and Thomas Franosch and Matthias Meiners and Christina Kurzthaler},
  title     = {Active Brownian particle under stochastic orientational resetting},
  journal   = {New Journal of Physics},
  year      = {2024},
  volume    = {26},
  number    = {7},
  pages     = {073041},
  doi       = {10.1088/1367-2630/ad602a},
  url       = {https://doi.org/10.1088/1367-2630/ad602a}
}

@article{Pal2017,
  title = {First Passage under Restart},
  author = {Pal, Arnab and Reuveni, Shlomi},
  journal = {Phys. Rev. Lett.},
  volume = {118},
  issue = {3},
  pages = {030603},
  numpages = {6},
  year = {2017},
  month = {Jan},
  publisher = {American Physical Society},
  doi = {10.1103/PhysRevLett.118.030603},
  url = {https://link.aps.org/doi/10.1103/PhysRevLett.118.030603}
}

@article{Palagi2018,
  title = {Bioinspired microrobots},
  author = {Palagi, Stefano and Fischer, Peer},
  journal = {Nature Reviews Materials},
  volume = {3},
  number = {6},
  pages = {113--124},
  year = {2018},
  doi = {10.1038/s41578-018-0016-9},
  url = {https://www.nature.com/articles/s41578-018-0016-9}
}

@article{Volpe2017,
  author       = {Volpe, Giorgio and Volpe, Giovanni},
  title        = {The topography of the environment alters the optimal search strategy for active particles},
  journal      = {Proceedings of the National Academy of Sciences of the United States of America},
  volume       = {114},
  number       = {43},
  pages        = {11350--11355},
  year         = {2017},
  doi          = {10.1073/pnas.1711371114},
  url          = {https://doi.org/10.1073/pnas.1711371114}
}

@article{Gupta2022,
  author       = {Gupta, Shamik and Jayannavar, Arun M.},
  title        = {Stochastic Resetting: A (Very) Brief Review},
  journal      = {Frontiers in Physics},
  volume       = {10},
  pages        = {789097},
  year         = {2022},
  doi          = {10.3389/fphy.2022.789097},
  url          = {https://www.frontiersin.org/journals/physics/articles/10.3389/fphy.2022.789097/full}
}

@article{Evans2011PRL,
  title = {Diffusion with Stochastic Resetting},
  author = {Evans, Martin R. and Majumdar, Satya N.},
  journal = {Phys. Rev. Lett.},
  volume = {106},
  issue = {16},
  pages = {160601},
  numpages = {4},
  year = {2011},
  month = {Apr},
  publisher = {American Physical Society},
  doi = {10.1103/PhysRevLett.106.160601},
  url = {https://link.aps.org/doi/10.1103/PhysRevLett.106.160601}
}

@article{Evans2011JPA,
  author    = {Evans, Martin R. and Majumdar, Satya N.},
  title     = {Diffusion with optimal resetting},
  journal   = {Journal of Physics A: Mathematical and Theoretical},
  volume    = {44},
  number    = {43},
  pages     = {435001},
  year      = {2011},
  doi       = {10.1088/1751-8113/44/43/435001},
  url       = {https://doi.org/10.1088/1751-8113/44/43/435001}
}

@book{Viswanathan2011,
  title     = {The Physics of Foraging: An Introduction to Random Searches and Biological Encounters},
  author    = {Viswanathan, G. M. and da Luz, M. G. E. and Raposo, E. P. and Stanley, H. E.},
  year      = {2011},
  publisher = {Cambridge University Press},
  address   = {Cambridge, UK},
  isbn      = {9780521143693}
}

@article{Romanczuk2012,
  author    = {Pawel Romanczuk and Markus B{\"a}r and Werner Ebeling and Benjamin Lindner and Lutz Schimansky-Geier},
  title     = {Active Brownian particles: From individual to collective stochastic dynamics},
  journal   = {Eur. Phys. J. - Special Topics},
  year      = {2012},
  volume    = {202},
  number    = {1},
  pages     = {1--162},
  doi       = {10.1140/epjst/e2012-01529-y},
  url       = {https://doi.org/10.1140/epjst/e2012-01529-y}
}

@article{Olsen2025Optimal,
  title   = {{Optimal area exploration by resetting active particles}},
  author  = {Olsen, Kristian St{\o}levik and L{\"o}wen, Hartmut and Caprini, Lorenzo},
  journal = {arXiv preprint arXiv:2510.01087},
  year    = {2025},
  month   = {oct},
  doi     = {10.48550/arXiv.2510.01087},
  url     = {https://arxiv.org/abs/2510.01087},
  eprint  = {2510.01087},
  archivePrefix = {arXiv},
  primaryClass  = {cond-mat.soft}
}

@article{Keidar2025,
  author    = {Tommer D. Keidar and Ofir Blumer and Barak Hirshberg and Shlomi Reuveni},
  title     = {Adaptive resetting for informed search strategies and the design of non-equilibrium steady-states},
  journal   = {Nature Communications},
  year      = {2025},
  volume    = {16},
  pages     = {7259},
  doi       = {10.1038/s41467-025-62398-2},
  url       = {https://doi.org/10.1038/s41467-025-62398-2}
}

@article{Baouche2025,
  author    = {Yanis Baouche and Christina Kurzthaler},
  title     = {Optimal first-passage times of active Brownian particles under stochastic resetting},
  journal   = {Soft Matter},
  year      = {2025},
  volume    = {21},
  pages     = {5998--6011},
  doi       = {10.1039/D5SM00340G}
}

@article{Santra2020,
    title = {{Run-and-tumble particles in two dimensions: Marginal position distributions}},
    year = {2020},
    journal = {Physical Review E},
    author = {Santra, Ion and Basu, Urna and Sabhapandit, Sanjib},
    number = {6},
    month = {6},
    pages = {62120},
    volume = {101},
    publisher = {American Physical Society},
    url = {https://link.aps.org/doi/10.1103/PhysRevE.101.062120},
    doi = {10.1103/PhysRevE.101.062120}
}

@article{Pattanayak2024,
  author       = {Anweshika Pattanayak and Amir Shee and Debasish Chaudhuri and Abhishek Chaudhuri},
  title        = {Impact of torque on active Brownian particle: exact moments in two and three dimensions},
  journal      = {New Journal of Physics},
  volume       = {26},
  number       = {8},
  pages        = {083024},
  year         = {2024},
  month        = aug,
  doi          = {10.1088/1367-2630/ad6a32},
}

@misc{Patel2025,
      title={Crossover dynamics and non-Gaussian fluctuations in inertial active chains}, 
      author={Manish Patel and Subhajit Paul and Debasish Chaudhuri},
      year={2025},
      eprint={2511.13270},
      archivePrefix={arXiv},
      primaryClass={cond-mat.stat-mech},
      url={https://arxiv.org/abs/2511.13270}, 
}

@article{Lisin_26,
  title = {Fully inertial active Brownian particle in a harmonic potential},
  author = {Lisin, Evgeny A. and Lisina, Irina I.},
  journal = {Phys. Rev. E},
  volume = {113},
  issue = {1},
  pages = {015417},
  numpages = {13},
  year = {2026},
  month = {Jan},
  publisher = {American Physical Society},
  doi = {10.1103/9kcc-9hxm},
  url = {https://link.aps.org/doi/10.1103/9kcc-9hxm}
}

@article{GutierrezMartinez2020,
  author  = {Gutierrez-Martinez, Luis L. and Sandoval, Mario},
  title   = {Inertial effects on trapped active matter},
  journal = {The Journal of Chemical Physics},
  volume  = {153},
  number  = {4},
  pages   = {044906},
  year    = {2020},
  month   = {07},
  doi     = {10.1063/5.0011270},
  url     = {https://doi.org/10.1063/5.0011270}
}

@article{Lisin2022,
  author  = {Lisin, E. A. and Vaulina, O. S. and Lisina, I. I. and Petrov, O. F.},
  title   = {Motion of a self-propelled particle with rotational inertia},
  journal = {Physical Chemistry Chemical Physics},
  year    = {2022},
  volume  = {24},
  pages   = {14150--14158},
  doi     = {10.1039/D2CP01313D},
  publisher = {Royal Society of Chemistry}
}

@article{Martins2022,
  title = {Inertial dynamics of an active Brownian particle},
  author = {Mayer Martins, Jonas and Wittkowski, Raphael},
  journal = {Phys. Rev. E},
  volume = {106},
  issue = {3},
  pages = {034616},
  numpages = {17},
  year = {2022},
  month = {Sep},
  publisher = {American Physical Society},
  doi = {10.1103/PhysRevE.106.034616},
  url = {https://link.aps.org/doi/10.1103/PhysRevE.106.034616}
}

@article{Dombrowski2019,
  title = {Transition in swimming direction in a model self-propelled inertial swimmer},
  author = {Dombrowski, Thomas and Jones, Shannon K. and Katsikis, Georgios and Bhalla, Amneet Pal Singh and Griffith, Boyce E. and Klotsa, Daphne},
  journal = {Phys. Rev. Fluids},
  volume = {4},
  issue = {2},
  pages = {021101},
  numpages = {9},
  year = {2019},
  month = {Feb},
  publisher = {American Physical Society},
  doi = {10.1103/PhysRevFluids.4.021101},
  url = {https://link.aps.org/doi/10.1103/PhysRevFluids.4.021101}
}

@article{Antonov2024,
  title = {Inertial Active Matter with Coulomb Friction},
  author = {Antonov, Alexander P. and Caprini, Lorenzo and Ldov, Anton and Scholz, Christian and L\"owen, Hartmut},
  journal = {Phys. Rev. Lett.},
  volume = {133},
  issue = {19},
  pages = {198301},
  numpages = {6},
  year = {2024},
  month = {Nov},
  publisher = {American Physical Society},
  doi = {10.1103/PhysRevLett.133.198301},
  url = {https://link.aps.org/doi/10.1103/PhysRevLett.133.198301}
}

@article{Lisin2025,
  author  = {Evgeny A. Lisin and Irina I. Lisina},
  title   = {Active Brownian motion with inertia: The role of dimensionality},
  journal = {Physics of Fluids},
  volume  = {37},
  number  = {2},
  pages   = {021709},
  year    = {2025},
  doi     = {10.1063/5.0252359},
  publisher = {AIP Publishing}
}

@article{Deblais2018,
  title = {Boundaries Control Collective Dynamics of Inertial Self-Propelled Robots},
  author = {Deblais, A. and Barois, T. and Guerin, T. and Delville, P. H. and Vaudaine, R. and Lintuvuori, J. S. and Boudet, J. F. and Baret, J. C. and Kellay, H.},
  journal = {Phys. Rev. Lett.},
  volume = {120},
  issue = {18},
  pages = {188002},
  numpages = {5},
  year = {2018},
  month = {May},
  publisher = {American Physical Society},
  doi = {10.1103/PhysRevLett.120.188002},
  url = {https://link.aps.org/doi/10.1103/PhysRevLett.120.188002}
}

@article{Debasish2025,
  title = {Active stop and go motion: A strategy to improve spatial exploration and survival},
  author = {Peruani, Fernando and Chaudhuri, Debasish},
  journal = {Phys. Rev. E},
  volume = {112},
  issue = {3},
  pages = {L033402},
  numpages = {5},
  year = {2025},
  month = {Sep},
  publisher = {American Physical Society},
  doi = {10.1103/2m8x-g81x},
  url = {https://link.aps.org/doi/10.1103/2m8x-g81x}
}

@article{Andrey2025,
    author = {Liang, Yingjie and Wei, Qing and Wang, Wei and Cherstvy, Andrey G.},
    title = {Ultraslow diffusion processes under stochastic resetting},
    journal = {Physics of Fluids},
    volume = {37},
    number = {3},
    pages = {032014},
    year = {2025},
    month = {03},
    issn = {1070-6631},
    doi = {10.1063/5.0255601}
}

@article{Smith_2022,
doi = {10.1088/1742-5468/ac6f04},
url = {https://doi.org/10.1088/1742-5468/ac6f04},
year = {2022},
month = {jun},
publisher = {IOP Publishing and SISSA},
volume = {2022},
number = {5},
pages = {053212},
author = {Smith, Naftali R and Majumdar, Satya N},
title = {Condensation transition in large deviations of self-similar Gaussian processes with stochastic resetting},
journal = {Journal of Statistical Mechanics: Theory and Experiment}
}

@article{Santra_2025,
    author = {Santra, Ion and Stølevik Olsen, Kristian},
    title = {Brownian motion with stochastic energy renewals},
    journal = {Chaos: An Interdisciplinary Journal of Nonlinear Science},
    volume = {35},
    number = {9},
    pages = {093110},
    year = {2025},
    month = {09},
    issn = {1054-1500},
    doi = {10.1063/5.0285588},
    url = {https://doi.org/10.1063/5.0285588}
}

\end{document}